\documentclass[10pt,letterpaper]{article}

\usepackage{jheppub_mod}
\usepackage{slashed}
\usepackage{graphicx}

\makeatletter
\newsavebox\myboxA
\newsavebox\myboxB
\newlength\mylenA

\newcommand*\xoverline[2][0.75]{%
    \sbox{\myboxA}{$\m@th#2$}%
    \setbox\myboxB\null
    \ht\myboxB=\ht\myboxA%
    \dp\myboxB=\dp\myboxA%
    \wd\myboxB=#1\wd\myboxA
    \sbox\myboxB{$\m@th\overline{\copy\myboxB}$}
    \setlength\mylenA{\the\wd\myboxA}
    \addtolength\mylenA{-\the\wd\myboxB}%
    \ifdim\wd\myboxB<\wd\myboxA%
       \rlap{\hskip 0.5\mylenA\usebox\myboxB}{\usebox\myboxA}%
    \else
        \hskip -0.5\mylenA\rlap{\usebox\myboxA}{\hskip 0.5\mylenA\usebox\myboxB}%
    \fi}
\makeatother

\newcommand{\cA}{\mathcal{A}}

\newcommand{\cE}{\mathcal{E}}
\newcommand{\cF}{\mathcal{F}}

\newcommand{\cJ}{\mathcal{J}}
\newcommand{\cK}{\mathcal{K}}
\newcommand{\cL}{\mathcal{L}}

\newcommand{\cN}{\mathcal{N}}
\newcommand{\cO}{\mathcal{O}}
\newcommand{\cQ}{\mathcal{Q}}

\newcommand{\cS}{\mathcal{S}}
\newcommand{\cV}{\mathcal{V}}
\newcommand{\cW}{\mathcal{W}}

\newcommand{\uD}{\mathrm{D}}

\newcommand{\ud}{\mathrm{d}}
\newcommand{\ue}{\mathrm{e}}
\newcommand{\ui}{\mathrm{i}}

\newcommand{\us}{\mathrm{s}}

\newcommand{\IIB}{\mathrm{IIB}}

\newcommand{\Ga}{\Gamma}

\newcommand{\Sig}{\Sigma}

\newcommand{\ep}{\epsilon}
\newcommand{\ga}{\gamma}
\newcommand{\ka}{\kappa}

\newcommand{\sig}{\sigma}

\newcommand{\vp}{\varphi}

\newcommand{\vol}{\mathrm{vol}}

\newcommand{\SO}[1]{\mathrm{SO}\left(#1\right)}
\newcommand{\SU}[1]{\mathrm{SU}\left(#1\right)}
\newcommand{\U}[1]{\mathrm{U}\left(#1\right)}

\newcommand{\abs}[1]{\left\vert#1\right\rvert}
\newcommand{\ul}[1]{\underline{#1}}

\title{Falling flavors in AdS/CFT}

\author{Paul McGuirk}

\affiliation{Laboratory for Elementary-Particle Physics, Cornell
  University, Ithaca, New York, 14853, USA}

\emailAdd{mcguirk@cornell.edu}

\date{\today}

\abstract{We consider the behavior of $\uD 7$ probes of supersymmetric
  warped geometries that are perturbed by the presence of
  $\overline{\uD 3}$-branes.  Such constructions are the gravitational
  duals of non-supersymmetric states in supersymmetric flavored gauge
  theories.  Although the $\uD 7$s we consider do not feel a force
  from either $\uD 3$s or $\overline{\uD 3}$s alone, when both types
  of $3$-branes are present the $\uD 7$s deform and fall a small
  distance toward the $3$-branes.  We perform our analysis in
  $AdS^{5}\times S^{5}$ and $AdS^{5}\times T^{1,1}$ and find
  qualitatively similar behavior in each case.  We then extend our
  consideration to the approximately conical region of the
  Klebanov-Strassler solution and find that the effect is
  parametrically larger than in the $AdS^{5}\times X^{5}$ examples.
  Additionally, we discuss how these behaviors are modified by the
  presence of other flavors by considering the smeared backreaction of
  such flavor branes in $AdS^{5}\times X^{5}$. Finally, we touch upon
  some of the implications that our results may have for model
  building and argue that the deformation of the worldvolume results
  in $\cO\left(1\right)$ corrections to soft terms in the low-energy
  description.}

\begin{document}

\maketitle

\section{\label{sec:intro}Introduction}

Warped geometries continue to play a prominent role in string theory.
The furthest-reaching application of such constructions is the
gauge/gravity
correspondence~\cite{Maldacena:1997re,Witten:1998qj,Gubser:1998bc,Aharony:1999ti}
which provides for a deeper understanding of strongly coupled gauge
theories and the nature of quantum gravity.  As a consequence of this
duality, warped geometries also provide a fruitful corner of the
landscape in which to search for controlled and phenomenologically
promising vacua.  For example, the large redshift resulting from
strong warping can naturally accommodate an exponentially large
separation of mass
scales~\cite{Randall:1999ee,Verlinde:1999fy,Giddings:2001yu}, just as
dimensional transmutation produces such hierarchies in gauge theories.
Additionally, warping brings to string constructions the powerful
tools of effective field theories in which unknown bulk effects can be
parametrized by Wilson coefficients and systematically incorporated
(see, e.g.,~\cite{Baumann:2010sx,Gandhi:2011id}).

However, in order to reach the goal of phenomenological viability, the
string modes that give rise to visible-sector fields must also be
included.  In the widely explored GKP-like warped constructions of the
type-IIB superstring~\cite{Giddings:2001yu}, such modes are the
open-string fluctuations of either (fractional or whole) $\uD
3$-branes or $\uD 7$-branes\footnote{In more general F-theoretic
  constructions, we could of course consider other types of $7$-branes
  or even in principle other types of fractional $3$-branes.}.
Although $\uD 3$-branes provide a rich framework for model building
and have the extra benefit of being very local, there are many
phenomenological advantages to considering scenarios in which the
visible sector arises from $\uD 7$-branes.  For example, one of the
successes of the so-called ``bulk'' Randall-Sundrum (RS) scenarios is
that a hierarchical structure of Yukawa couplings can be arranged by
$\cO\left(1\right)$ tunings of the 5d masses of fermions propagating
in a warped geometry~\cite{Grossman:1999ra}; such 5d masses control
the internal profiles of the fermions and hence their overlap with
scalar fields.  In a stringy realization of an RS-like scenario, such
modes correspond to the excitations of $\uD 7$-branes that extend some
distance into the warped geometry.  Chiral modes in particular are
realized on the intersections of magnetized $\uD 7$-branes and hence
the string theory ``lift'' of 5d mass is, at least morally if not
literally, the position of $\uD 7$-branes that extend into the warped
region\footnote{The story will be a little more complicated in detail.
  For example, the localization of fermionic modes also depends on the
  magnetic flux as in the D-brane examples
  of~\cite{Cremades:2004wa,Acharya:2006mx}.}.  Hence, it is expected
that $\cO\left(1\right)$ adjustments of the positions of $\uD
7$-branes can produce hierarchical Yukawa textures.  The addition of
such $\uD 7$-branes to a warped geometry has the well-known
interpretation of adding a global flavor group to the dual gauge
theory~\cite{Karch:2002sh} and hence the branes are often called
flavor branes.

A further phenomenological constraint is the absence of supersymmetry,
at least at long wavelengths.  Unfortunately, the construction of
non-supersymmetric configurations in string theory is difficult to
achieve. In additional to the practical difficulties associated with
solving second-order as opposed to first-order differential equations,
the stability of non-supersymmetric geometries is often an
issue\footnote{For some recent examples of non-supersymmetric warped
  geometries, see \cite{Kachru:2009kg,Dymarsky:2011ve}.}.  One method
of breaking supersymmetry in a relatively controlled way is the
addition of anti-branes to a warped geometry.  For example, the branes
and fluxes that build a GKP-like warped region carry positive $\uD
3$-brane charge and so an $\overline{\uD 3}$-brane, which carries
negative $\uD 3$-brane charge and hence breaks the supersymmetry
preserved by the geometry, will be naturally attracted to the point of
strongest warping and largest redshift.  The warping then serves to
restrict the influence of the $\overline{\uD 3}$-brane to the strongly
warped region so that remainder of the geometry is approximately
supersymmetric. Furthermore, in examples such as the
Klebanov-Strassler (KS) solution~\cite{Klebanov:2000hb} in which the
warping is not sourced by integer $\uD 3$-branes, the $\overline{\uD
  3}$-branes will be perturbatively stable so long as the number of
anti-branes is small with respect to the number of flux quanta that
support the geometry~\cite{Kachru:2002gs}.  Due to the localized
nature of the $\overline{\uD 3}$-branes, their backreaction falls off
sufficiently fast with the distance from the branes that the
configuration is dual to a particular non-supersymmetric state in a
supersymmetric theory~\cite{DeWolfe:2008zy}, a fact that has been used
to explore gravitational duals of gauge
mediation~\cite{Benini:2009ff,McGuirk:2009am,Fischler:2011xd,McGuirk:2011yg}
(see
also~\cite{Gabella:2007cp,McGarrie:2010yk,Skenderis:2012bs,Argurio:2012cd,Argurio:2012bi,McGarrie:2012fi})\footnote{Note
  that as a consequence of the non-linear behavior of the supergravity
  equations of motion, the backreaction of $\overline{\uD 3}$-branes
  in a GKP construction results in singular $3$-form flux despite the
  fact that such $3$-branes do not carry any NS-NS or R-R $2$-form
  charge~\cite{McGuirk:2009xx,Bena:2009xk,Bena:2010ze,Bena:2011hz,Bena:2011wh,Massai:2012jn,Bena:2012bk,Dymarsky:2011pm}
  (see
  also~\cite{Blaback:2011nz,Blaback:2011pn,Blaback:2012nf,Bena:2012tx}
  for a T-dual case).  However, as this behavior is directly linked to
  the brane singularities that are expected to be resolved in the full
  string theory, in our opinion the singularities are likely an
  artifact of the supergravity approximation which breaks down near
  the position of the branes.  As the background at large distances is
  essentially determined by the net charge and mass of the solution,
  if it is the case that there is some stringy resolution to the
  singularity, then the effect of such resolution on $\uD 7$ probes is
  probably subleading as long as the $\uD 7$s are sufficiently far
  from the supergravity singularity.}.

Warping, $\uD 7$-branes, and $\overline{\uD 3}$-branes also play an
important role in the KKLT scenario for achieving metastable de Sitter
spaces~\cite{Kachru:2003aw} and the related framework for
inflation~\cite{Kachru:2003sx}.  In each of these cases, gaugino
condensation on $\uD 7$-branes provides a mechanism for the
stabilization of the K\"ahler structure of the internal space while
$\overline{\uD 3}$-branes are central in the construction of the
uplift and inflationary potentials.

Given the many examples in which $\uD 7$-branes and $\overline{\uD
  3}$-branes appear together in warped geometries with $\uD 3$ charge,
it is important to understand how these objects interact with each
other.  As is well known, $\uD 7$s do not experience a net force from
``parallel'' $\uD 3$-branes.  Furthermore, since a $\uD 7$ carries no
$\uD 3$-brane charge (unless such a charge is induced by generalized
worldvolume flux or curvature) and hence cannot distinguish a $\uD
3$-brane from an $\overline{\uD 3}$-brane, $\uD 7$s do not experience
a force from parallel $\overline{\uD 3}$s either.  However, due to the
non-linear nature of supergravity, when both $\uD 3$s and
$\overline{\uD 3}$s are present there is no guarantee that the $\uD
7$-branes will experience no force.  One may further expect that in
the presence of a non-supersymmetric background, parallel $\uD
7$-branes may exert a force on each other, despite the fact that when
only $\uD 3$-branes or $\overline{\uD 3}$-branes were present there
was an exact cancellation between the NS-NS attraction and the R-R
repulsion.

It is precisely such interactions that we explore in this work.  In
section~\ref{sec:prelim} we argue generally from the equations of
motion that $\uD 7$-branes which feel no force in the presence of only
one type of $\uD 3$-brane will bend in the presence of both.  As much
of the paper contains lengthy calculations, section~\ref{sec:prelim}
also contains a summary of our results. In section~\ref{sec:flat} we
perform a more quantitative analysis by considering $\uD 7$-probes of
$AdS^{5}\times S^{5}$ perturbed by a small number of $\overline{\uD
  3}$-branes and find that indeed the $\uD 7$-branes tend to fall a
little toward the non-BPS stack of $3$-branes.  We additionally show
how this effect can be modified by the presence of other $\uD7
$-branes by considering their backreaction as well.  In
section~\ref{sec:conifold}, we perform the analogous analysis for
Kuperstein probes of $AdS^{5}\times T^{1,1}$ and find that, due to
common conical structure of the geometries, the qualitative behavior
is the same.  In section~\ref{sec:DKM} we consider $\uD 7$-probes of
the Klebanov-Tseytlin geometry perturbed by $\overline{\uD 3}$-branes.
In this case, due to the presence of the background $3$-form flux, the
probe $\uD 7$s experience a parametrically stronger attraction to the
non-supersymmetric source of warping.  In section~\ref{sec:model} we
briefly and qualitatively discuss some aspects of the impact the
deformations will have on model building in warped
compacitifcations. Some concluding remarks are made in
section~\ref{sec:conclude}.  Our conventions and some technical
details regarding brane backreaction are relegated to appendices.

\section{\label{sec:prelim}Preliminaries and summary of results}

As is very well known, $\uD 3$-branes and $\uD 7$-branes are, in many
situations, mutually BPS objects and hence there is no net force
between them (see, e.g.~\cite{Polchinski:1998rr}).  As a simple
example, consider a stack of $N$ $\uD 3$-branes in $R^{9,1}$.  A $\uD
7$-brane that sits parallel to the $\uD 3$s and fills four orthogonal
directions will preserve half of the supersymmetry preserved by the
$\uD 3$s.  The absence of a force can be seen by considering the $\uD
7$ as a probe of the $\uD 3$ background.  The backreaction of the $\uD
3$-branes takes the familiar form\footnote{Our conventions are
  presented in appendix~\ref{app:conv}.}
\begin{equation}
  \ud s_{10}^{2}=\ue^{2A}\eta_{\mu\nu}\ud x^{\mu}\ud x^{\nu}
  +\ue^{-2A}\delta_{mn}\ud y^{m}\ud y^{n},
\end{equation}
in which the warp factor $A$ is a function of the coordinates $y^{m}$
that are transverse to the $\uD 3$s.  In addition, $\uD 3$s are
both electric and magnetic sources for $C_{4}$ and hence
\begin{equation}
  F_{5}=\bigl(1+\hat{\ast}\bigr)g_{\us}^{-1}\ud\ue^{4A}\wedge\ud\vol_{R^{3,1}},
\end{equation}
in which $\vol_{R^{3,1}}=\ud x^{0}\wedge\ud x^{1}\wedge \ud
x^{2}\wedge \ud x^{3}$ is the volume element for $R^{3,1}$.  In the
absence of brane flux, the $\uD 7$ does not couple to $C_{4}$ and
hence the potential does not contribute to the $\uD 7$ CS
action~\eqref{eq:Dbrane_action}.  Similarly, in the absence of
fluctuations of the $\uD 7$, the $\uD 3$s do not contribute to the DBI
action for the $\uD 7$; the metric on the $\uD 7$ worldvolume takes
the form
\begin{equation}
  \ud s_{8}^{2}=\hat{g}_{\alpha\beta}\ud x^{\alpha}\ud x^{\beta}
  =\ue^{2A}\eta_{\mu\nu}\ud x^{\mu}\ud x^{\nu}
  +\ue^{-2A}\delta_{ab}\ud y^{a}\ud y^{b},
\end{equation}
and hence the volume element for the $\uD 7$ worldvolume, $\ud
\vol_{\cW^{8}}=\ud^{8}x\sqrt{-\det\left(\hat{g}_{\alpha\beta}\right)}$, is
independent of the warp factor.  Since neither the warp factor nor
$C_{4}$ contribute to the non-derivative parts of the $\uD 7$ action,
the $\uD 7$ feels no force from the $\uD 3$.

The absence of a force can also be seen from the $\uD 3$ perspective.
Instead of $C_{4}$, the $\uD 7$ sources the axiodilaton to which a
$\uD 3$-brane does not couple, while the backreaction on the metric is
``limited'' to a deficit angle in the plane transverse to the $\uD 7$
and hence is not felt by the $\uD 3$.

From either point of view, the Ramond-Ramond potential from the other
brane did not contribute to probe brane action and hence the absence
of a force will result if we replace \textit{all} of the $\uD
3$-branes by $\overline{\uD 3}$s.  Equivalently, of the sixteen
supercharges preserved by the $\uD 7$, eight are also preserved by
$\uD 3$s, while the other eight are preserved by $\overline{\uD 3}$s
and so $\uD 7$s cannot, by themselves, distinguish the sign of the
charge of the $3$-branes.  In a linear theory, this would imply
that a $\uD 7$-brane in the presence of both types of $3$-branes would
feel no force.  However, supergravity is a non-linear theory and we
will find that there is an interaction when both $3$-branes are
present.  Indeed in the presence of both $\uD 3$s and $\overline{\uD
  3}$s, no supercharges are preserved and hence there is no BPS bound
to protect against the development of a force.

To see this more explicitly, we consider a more general warped ansatz
\begin{equation}
  \label{eq:warped_ansatz}
  \ud s_{10}^{2}=\ue^{2A\left(y\right)}\eta_{\mu\nu}\ud x^{\mu}\ud x^{\nu}
  +\ue^{-2A\left(y\right)}g_{mn}\ud y^{m}\ud y^{n},
\end{equation}
in which $\eta_{\mu\nu}$ is the Minkowski metric on $R^{3,1}$ and
$g_{mn}$ is the metric on a space $Y^{6}$.  The geometry is again
supported by various fluxes
\begin{equation}
  F_{5}=\bigl(1+\hat{\ast}\bigr)g_{\us}^{-1}\ud\omega\bigl(y\bigr)
  \wedge\ud\vol_{R^{3,1}},\quad
  \iota_{\partial_{\mu}}G_{3}=0,
\end{equation}
in which $\iota$ is the interior product.  We orient the 3-branes
along the Minkowski directions and take any $\uD 7$-branes to fill the
Minkowski directions and wrap $4$-cycles $\Sig^{4}_{i}$.  It is useful
to define the quantities~\cite{Giddings:2001yu}
\begin{equation}
  \Phi_{\pm}=\ue^{4A}\pm\omega,\quad
  G_{\pm}=\bigl(\ui\pm\ast\bigr)G_{3},\quad
  \Lambda=\Phi_{+}G_{-}+\Phi_{-}G_{+}.
\end{equation}
Then, keeping in mind that $\uD 7$s are magnetic sources for $C_{0}$,
the equations of motion following from the bosonic action and the
self-duality condition for $F_{5}$ are
\begin{subequations}
\label{eq:IIB_eom}
\begin{align}
  0=&\nabla^{2}\Phi_{\pm}-\frac{2}{\Phi_{+}+\Phi_{-}}
  \bigl(\partial\Phi_{\pm}\bigr)^{2}
  -\frac{g_{\us}\left(\Phi_{+}+\Phi_{-}\right)^{2}}
  {16\,\mathrm{Im}\,\tau}\bigl\lvert G_{\pm}\bigr\rvert^{2}\notag\\
  &-\frac{\left(\Phi_{+}+\Phi_{-}\right)^{2}}{2}
  \bigl(2\pi\bigr)^{4}\alpha'^{2}g_{\us}\sum_{\uD 3^{\pm}}
  \frac{\delta^{6}\left(y-y_{a}\right)}{\sqrt{g}},\\
  0=&R_{mp}
  -\frac{2}{\left(\Phi_{+}+\Phi_{-}\right)^{2}}
  \partial_{\left(m\right.}\Phi_{+}\partial_{\left.p\right)}\Phi_{-}
  -\frac{1}{2\left(\mathrm{Im}\,\tau\right)^{2}}
  \partial_{\left(m\right.}\tau\partial_{\left.p\right)}\overline{\tau}\notag\\
  &+\frac{g_{\us}\left(\Phi_{+}+\Phi_{-}\right)}{16\cdot 2!\,\mathrm{Im}\,\tau}
  \bigl[G_{+\left(m\right.}^{\phantom{+\left(m\right.}n_{1}n_{2}}
  \overline{G}_{\left.-p\right)n_{1}n_{2}}
  +G_{-\left(m\right.}^{\phantom{+\left(m\right.}n_{1}n_{2}}
  \overline{G}_{\left.+p\right)n_{1}n_{2}}\bigr]\notag\\
  &-\frac{\ue^{\phi}}{2}
  \sum_{\uD 7}\frac{1}{\sqrt{\Omega_{i}^{2}}}\bigl(\Omega_{i}\bigr)_{mn}
  \bigl(\Omega_{i}\bigr)_{p}^{\phantom{p}n},\\
  0=&\nabla^{2}\tau+\frac{\ui}{\mathrm{Im}\,\tau}
  \bigl(\partial\tau\bigr)^{2}
  +\frac{\ui g_{\us}\left(\Phi_{+}+\Phi_{-}\right)}{8}
  G_{+}\cdot G_{-}
  +\ui \sum_{\uD 7}\sqrt{\Omega_{i}^{2}},\\
  0=&\ud F_{1}+\sum_{\uD 7}\Omega_{i},\\
\intertext{\newpage}
  0=&\ud\Lambda+\frac{\ui}{2\,\mathrm{Im}\,\tau}
  \ud\tau\wedge\bigl(\Lambda+\overline{\Lambda}\bigr),
  \label{eq:Lambda_eom}\\
  0=&\ud \bigl(G_{+}-G_{-}\bigr)+
  \frac{\ui}{2\,\mathrm{Im}\,\tau}
  \ud\tau\wedge\bigl(G_{+}+\overline{G}_{+}
  -G_{-}-\overline{G}_{-}\bigr),
  \label{eq:G_bianchi}
\end{align}
\end{subequations}
in which $\Omega_{i}$ are the Poincar\'e duals in $Y^{6}$ of the
$4$-cycles wrapped by the $\uD 7$s (see, e.g.~\cite{Benini:2006hh}),
the $\uD 3$-branes and $\overline{\uD 3}$-branes (denoted also $\uD
3^{+}$ and $\uD 3^{-}$ respectively) are localized at points in
$Y^{6}$, and $R_{mp}$ and related quantities are built from the
unwarped 6d metric $g_{mp}$. In writing these equations of motion, we
have neglected the presence of worldvolume flux on any of the branes.

We first consider the case in which $G_{3}$ vanishes and the $\uD
7$-brane contribution to the equations of motion can be neglected
(i.e. treating the $\uD 7$s as probe branes).  In such a setup, when
only $\uD 3$ or only $\overline{\uD 3}$s are present (and hence only
one of $\Phi_{+}$ or $\Phi_{-}$ is non-vanishing), the
internal\footnote{Note that since we are not considering explicit
  compactifications, the term ``internal'' is inaccurate here.  We use
  the term to describe directions transverse to the Minkowski
  directions as these are the ones that would eventually become the
  coordinates on a compact space.}  metric is unchanged by the
presence of $3$-branes and correspondingly a probe $\uD 7$ will not
feel their influence.  However, when both types of branes are present,
$\Phi_{+}$ and $\Phi_{-}$ together source $g_{mp}$ so that the
internal metric can no longer be Ricci flat.  Hence, although the warp
factor cancels out of the DBI action for a probe $\uD 7$-brane, the
non-trivial metric will generically cause the $\uD 7$ to bend as it
extremizes its volume.

We can also easily see the non-trivial interactions from the point of
view of probe $3$-branes.  We consider for example a background with
$N$ $\uD 3$-branes and $N_{f}$ $\uD 7$-branes.  The equations of
motion will then imply that $\Phi_{+}^{-1}$ is harmonic.  However,
since the $\uD 7$s backreact on the internal metric forcing it to no
longer be Ricci flat, $\Phi_{+}^{-1}$ will take on a different form
than it would in the absence of $\uD 7$ branes.  Since $\Phi_{+}$ is
precisely the potential felt by an $\overline{\uD 3}$ brane, such a
brane will feel a potential with terms proportional to $NN_{f}$ in
addition to those depending on just $N$.  Similar arguments could be
for $\uD 3$-probes.

In the following sections, we will work out some of the details of
these interactions in particular examples.  However, since our
presentation is somewhat explicit, we will briefly summarize our
results here.  We work in strongly warped geometries so that the
backreaction of the $\uD 7$s is under some degree of control.  This is
easiest to understand in the dual gauge theory as the addition of
$N_{f}$ $\uD 7$-branes to a warped geometry of the type considered
here corresponds to the addition of a $\U{N_{f}}$ flavor
group~\cite{Karch:2002sh}.  Such flavors can be treated as
non-dynamical objects in the 't~Hooft limit in which we take the
number of colors $N\to \infty$ while letting $\frac{N_{f}}{N}\to 0$
(see, for example, the discussion in~\cite{Nunez:2010sf}), and hence
their influence on the gauge theory can be largely neglected for a
wide range of energies.  Correspondingly, many components of the
backreacted geometry scale as $\frac{N_{f}}{N}$ and hence for
$N_{f}\ll N$, the backreaction can be largely neglected, at least for
large ranges of $r$.

The simplest examples of such strongly warped geometries the
near-brane geometries of $N$ $\uD 3$s sitting at the apex of a
Calabi-Yau cone\footnote{Note that in taking the near-brane limit of
  the $\uD 3$ background, we remove the explicit $\uD 3$ sources from
  and dissolve the flux into the geometry. Thus our analysis will not
  completely capture the physics of the interactions between the $\uD
  7$s and the different $\uD 3$-branes (except near the horizon).
  Nevertheless, the analysis should capture the qualitative behavior,
  just as the attraction of probe $\overline{\uD 3}$ brane to the
  origin of $AdS^{5}$ reflects the attraction of $\overline{\uD 3}$
  branes to ${\uD 3}$s}.  The near-brane geometry is then
$AdS^{5}\times X^{5}$
\begin{equation}
  \ud s_{10}^{2}=\frac{r^{2}}{L^{2}}\eta_{\mu\nu}\ud x^{\mu}\ud x^{\nu}
  +\frac{L^{2}}{r^{2}}\ud r^{2}+L^{2}\ud s_{X^{5}}^{2},
\end{equation}
in which $X^{5}$ is the Sasaki-Einstein base of the Calabi-Yau cone
and $L^{4}\sim g_{\us}N\alpha'^{2}$ sets the radius of both the
$AdS^{5}$ and $X^{5}$ factors.  A $\uD 7$ that is holomorphically
embedded into the Calabi-Yau preserves some of the supersymmetry of
the geometry and hence does not experience a force.  On the other
hand, the backreaction of $p\ll N$ $\overline{\uD 3}$s (which was
found in~\cite{DeWolfe:2008zy} and is reviewed in
appendix~\ref{app:DKM_1}) includes a squashing of the metric so that
it is no longer Ricci flat and hence the shape of the
volume-minimizing configuration must change .  By analyzing the probe
$\uD 7$ action, we show that when the Sasaki-Einstein base is $S^{5}$
or $T^{1,1}$ the $\uD 7$ brane deforms so that the minimum radius
reached by the $\uD 7$ decreases (see~\eqref{eq:min_radius_flat}
and~\eqref{eq:min_distance_conifold})
\begin{equation}
  \label{eq:min_distance_change}
  \frac{\Delta r}{r_{\mathrm{min}}}\sim-\frac{p}{N}\frac{L^{8}}{r_{\mathrm{min}}^{8}},
\end{equation}
where $r_{\mathrm{min}}$ is the minimum radius obtained by the $\uD
7$-brane in the supersymmetric case.  The details of the shape depend
on the initial embedding and which Einstein-Sasaki base space is used.
A cartoon of the deformed embedding in the $S^{5}$ case is given in
figure~\ref{fig:comparing_falling_branes}.

If we treat the $\overline{\uD 3}$s as a probe, then the above
discussion suggests that the force experienced by the $\overline{\uD
  3}$ probe will be modified by the backreaction of the $\uD 7$s.  The
backreaction of such $\uD 7$s is difficult to find and to make
progress we use the smeared solutions reviewed in~\cite{Nunez:2010sf}.
As a consequence of such smearing, the geometry at radii inside the
minimum radius of the smeared branes is not modified by the presence
of the branes.  Hence, to obtain a non-vanishing force, we must place
the $\overline{\uD 3}$ brane at large radii and so the configuration
is not the same as the configuration that lead
to~\eqref{eq:min_distance_change} and we should not expect to
reconstruct the precise physics. However the resulting modification of
the potential confirms the non-trivial interaction of the branes.  In
this case, the potential for the canonically normalized field
representing the position of the $\overline{\uD 3}$ brane takes the
same functional form (at least when the $\uD 7$s are treated as a
perturbation) $V\bigl(\sig\bigr)\sim \sig^{4/3}$, but the coefficient
is reduced by $a\frac{g_{\us}N_{f}}{2\pi}$ where
$a\sim\cO\left(1\right)$ (see~\eqref{eq:anti_d3_potential}).  We only
explicitly perform this calculation in the $S^{5}$ case, but the same
qualitative behavior arises from any Einstein-Sasaki base.

The potential for $\uD 3$ probes when the backreaction of the $\uD 7$s
and $\overline{\uD 3}$s (now returned to small $r$) are taken into
account takes the schematic form~\eqref{eq:D3_potentials}
\begin{equation}
  V\bigl(\sig\bigr)\sim -\frac{p}{N}\bigl(1+\frac{g_{\us}N_{f}}{2\pi}\bigr)
  \frac{1}{\sig^{4/3}}-\frac{p^{2}}{N^{2}}\frac{1}{\sig^{4}},
\end{equation}
where we have specialized to the case where the $\overline{\uD 3}$s
and $\uD 7$s are treated as comparable perturbations (see
appendix~\ref{app:backreaction_D7s}).  We again only explicitly
considered the $S^{5}$ case, but other bases have the same scaling.

When $\overline{\uD 3}$s are present and supersymmetry is broken, the
cancellation of NS-NS and R-R forces between two $\uD 7$-branes should
no longer occur.  We can confirm this fact by considering a probe $\uD
7$-brane in the geometry produced by backreacting $\overline{\uD 3}$s
and other $\uD 7$s.  Indeed, we find that at large distances the
backreacting $\uD 7$s repel the probe $\uD 7$, which has again been
placed at a larger radius.  Schematically,
(see~\eqref{eq:min_flat_conifold_backreacting_D7s}
and~\eqref{eq:min_radius_conifold_backreacting_D7s})
\begin{equation}
  \label{eq:flavor_interaction_summary}
  \frac{\Delta r}{r_{\mathrm{min}}}\sim
  -\frac{p}{N}\frac{L^{8}}{r_{\mathrm{min}}^{8}}
  -\frac{p^{2}}{N^{2}}\frac{L^{16}}{r_{\mathrm{min}}^{16}}
  -\frac{p}{N}\frac{g_{\us}N_{f}}{2\pi}\frac{L^{8}}{r_{\mathrm{min}}^{8}}
  \biggl[1+\log\frac{r_{\mathrm{min}}}{r_{\us}}\biggr],
\end{equation}
in which $r_{\us}$ is an integration constant characterizing the
backreaction of the $\uD 7$s.  Again, to find this expression we treat
the $\overline{\uD 3}$s and $\uD 7$s as comparable perturbations to
the geometry.  Because of the logarithmic behavior of the backreaction
of the $\uD 7$s, treating the $\uD 7$s as a small perturbation is a
valid approximation only over a certain range of radii.  However, it
is interesting to note that naive application
of~\eqref{eq:flavor_interaction_summary} for sufficiently small
$r_{\mathrm{min}}$ indicates that the backreacting $\uD 7$s repel the
probe $\uD 7$.

Finally, we turn to the case of $\uD 7$ probes in the warped deformed
conifold perturbed by $\overline{\uD 3}$-branes.  The full linearized
solution provided
by~\cite{Bena:2009xk,Bena:2010ze,Bena:2011hz,Bena:2011wh,Massai:2012jn,Bena:2012bk,Dymarsky:2011pm}
is rather involved and we use the solution provided
by~\cite{DeWolfe:2008zy} in which $\overline{\uD 3}$s are added to the
Klebanov-Tseytlin solution.  Because of the increased complexity of
the geometry even in this case, we do not attempt to backreact any
$\uD 7$s, but for a probe $\uD 7$ we again find a tendency for the
branes to deform (see~\eqref{eq:DKM_min_r}
\begin{equation}
  \label{eq:min_distance_change_KT}
  \frac{\Delta r}{r_{\mathrm{min}}}\sim -
  \cS\frac{\alpha'^{2}}{r_{\mathrm{min}}^{4}},
\end{equation}
in which $\cS$ characterizes the perturbation by the $\overline{\uD
  3}$s.  The different functional form of this behavior is a
consequence of the $3$-form flux present in the geometry.

In all of the above examples we have treated the $\overline{\uD
  3}$-branes as a perturbation to the geometry.  However, since the
solution grows with decreasing $r$, perturbation theory breaks down at
sufficiently small radii.  Hence, our treatment is valid only for
sufficiently large $r$.  Fortunately, the deformation of the
worldvolumes of the $\uD 7$ is sufficiently small that the $\uD 7$s,
even after bending, do not reach the point where the perturbative
treatment breaks down and so~\eqref{eq:min_distance_change}
and~\eqref{eq:min_distance_change_KT} can be trusted.  Similar
statements apply when we include the backreaction of $\uD 7$s, however
there is the additional constraint that the geometry becomes singular
for large $r$ as well (corresponding to the presence of a Landau pole
in a dual theory).  For small $g_{\us}N_{f}$, we can easily arrange
things so that there is a wide range of initial embeddings such that
the perturbative treatment is valid\footnote{Note that although $\uD
  7$s are noncompact and hence extend out to infinite radius, the
  deviation from the supersymmetric embedding is localized in the
  region where a perturbative treatment is valid.}.

\section{\label{sec:flat}Interactions in flat space}

We now return to the flat space case and examine more precisely how a
$\uD 7$-brane will respond to the presence of both types of
$3$-branes.  Keeping in mind the uplift scenario
of~\cite{Kachru:2003aw}, we will consider the addition of $p$ $\uD
3$-$\overline{\uD 3}$ pairs to $N\gg p$ $\uD 3$s.  When the $3$-branes
are coincident, we can take the ansatz for the backreaction of the
$3$-branes to be of the type~\eqref{eq:warped_ansatz} with
\begin{equation}
  g_{mn}\ud y^{m}\ud y^{n}=\ue^{2B}\bigl(\ud r^{2}+r^{2}\ud s^{2}_{S^{5}}\bigr),
\end{equation}
where $\ud s^{2}_{S^{5}}$ is the metric for a unit $S^{5}$, the branes
are located at $r=0$, and all fields other than the metric on $S^{5}$
can be taken as functions of $r$ alone. In the more general case of
coincident $3$-branes stacked at the bottom of a Calabi-Yau cone, the
same ansatz applies where $S^{5}$ is replaced by the appropriate
Sasaki-Einstein manifold.  In this setup, the coincident branes are
perturbatively unstable and so the configuration should be viewed as a
warm-up for the metastable configuration discussed later.

The solution for any values of $p$ and $N$ were presented
in~\cite{Brax:2000cf} (following~\cite{Zhou:1999nm}).  As discussed in
the introduction, our interest is in strongly warped geometries.  Such
a geometry follows from taking the near-brane limit of the $3$-brane
geometry.  The backreaction in this limit and for $p\ll N$ was
discussed in~\cite{DeWolfe:2008zy}, which we review and carry out to
higher order in $\frac{p}{N}$ in appendix~\ref{app:DKM_1}. The result
is
\begin{subequations}
\label{eq:anti_D3s_in_cy_cone}
\begin{align}
  \ue^{-4A}=&\frac{L^{4}}{r^{4}}
  +\frac{4p}{5N}\frac{L^{12}}{r^{12}}
  +\frac{54p^{2}}{125N^{2}}\frac{L^{20}}{r^{20}},\\
  \omega=&\frac{r^{4}}{L^{4}}
  +\frac{6p}{5N}\frac{L^{4}}{r^{4}}
  -\frac{24 p^{2}}{125 N^{2}}\frac{L^{12}}{r^{12}},\\
  \ue^{2B}=&1-\frac{p}{5N}\frac{L^{8}}{r^{8}}
  -\frac{p^{2}}{50N^{2}}\frac{L^{16}}{r^{16}}.
\end{align}
\end{subequations}

$S^{5}$ is an Einstein-Sasaki space and therefore can be written as a
$\U{1}$ fibration over an Einstein-K\"ahler base, a fact that was
exploited in a related context in~\cite{Benini:2006hh}. Indeed writing
\begin{equation}
\label{eq:C3_coords}
  z^{1}=r\cos\frac{\ga}{2}\cos\frac{\theta}{2}
  \ue^{{\ui}\left(\psi+{\eta}/{2}+{\vp}/{2}\right)},\quad
  z^{2}=r\cos\frac{\ga}{2}\sin\frac{\theta}{2}
  \ue^{{\ui}\left(\psi+\eta/2-\vp/2\right)},\quad
  z^{3}=r\sin\frac{\ga}{2}\ue^{\ui\psi},
\end{equation}
in which $r\in\left[0,\infty\right)$,
$\ga,\theta\in\left[0,\pi\right]$, $\vp,\psi\in\left[0,2\pi\right)$,
and $\eta\in\left[0,4\pi\right)$, the metric for $R^{6}=C^{3}$ takes the form
\begin{align}
  \label{eq:flat_space_metric}
  \ud s_{6}^{2}=&\ud z^{i}\ud\bar{z}^{\bar{i}}=\ud r^{2}+r^{2}\bigl(\ud\psi+\cA\bigr)^{2}+r^{2}\ud s_{CP^{2}}^{2},
\end{align}
in which
\begin{equation}
  \cA=\frac{1}{2}\cos^{2}\frac{\ga}{2}
  \bigl(\ud\eta+\cos\theta\ud\vp\bigr).
\end{equation}
The 4d metric
\begin{equation}
  \ud s_{CP^{2}}^{2}=\frac{1}{4}\ud\ga^{2}
  +\frac{1}{4}\cos^{2}\frac{\ga}{2}
  \bigl(\ud\theta^{2}+\sin^{2}\theta\ud\vp^{2}\bigr)
  +\frac{1}{4}\cos^{2}\frac{\ga}{2}\sin^{2}\frac{\ga}{2}
  \bigl(\ud\eta+\cos\theta\ud\vp\bigr)^{2},
\end{equation}
is the Fubini-Study metric on the Einstein-K\"ahler space $CP^{2}$.
This can be confirmed by considering, for example, the inhomogeneous
coordinates $w^{1,2}=\frac{z^{1,2}}{z^{3}}$ on the $z^{3}\neq 0$
patch.  Finally, we note that the K\"ahler form on $CP^{2}$ is given
by
\begin{equation}
  \cJ=-\frac{1}{4}\sin\frac{\ga}{2}\cos\frac{\ga}{2}
  \ud\ga\wedge\bigl(\ud\eta+\cos\theta\ud\vp\bigr)
  -\frac{1}{4}\cos^{2}\frac{\ga}{2}\sin\theta\,\ud\theta\wedge\ud\vp
  =\frac{1}{2}\ud\cA,
\end{equation}
and that the K\"ahler form on $C^{3}$ is
\begin{equation}
  J=r\ud r\wedge\bigl(\ud\psi+\cA\bigr)+r^{2}\cJ.
\end{equation}

When $p=0$, the geometry is $AdS^{5}\times S^{5}$ which is famously
dual to the conformal $\cN=4$ $\SU{N}$ super Yang-Mills
theory~\cite{Maldacena:1997re}.

\subsection{\label{subsec:D7_probe_in_flat_space}Bending of probe D7s}

We now introduce a probe $\uD 7$ into the above geometry.  Note that
since the backreaction of the $\overline{\uD 3}$ grows with decreasing
radius, we must have that the probe $\uD 7$ does not extend to very
small $r$.  When $p=0$, the geometry~\eqref{eq:anti_D3s_in_cy_cone} is
supersymmetric and, since we have taken $G_{3}=0$, a probe $\uD 7$ is
supersymmetric if it is holomorphically embedded into the geometry and
the worldvolume field strength $f_{2}$ vanishes.  In this case, we
will take the $\uD 7$ embedding to satisfy
\begin{equation}
  \label{eq:flat_space_embedding}
  z^{3}=\mu,
\end{equation}
where $\mu$ is a fixed positive real number.  The perturbative
treatment of the $\overline{\uD 3}$s will be valid for our probe
analysis when $\mu\gg \left( \frac{p}{N}\right)^{1/8}L$.  Note that
this embedding preserves an $\SO{4}$ subgroup group of the $\SO{6}$
isometry of $R^{6}$.  When $p\neq 0$, we expect that this embedding
will no longer satisfy the $\uD 7$ equations of motion.  However,
since the $\overline{\uD 3}$s do not break the isometries preserved by
the $\uD 7$, we expect the $\SO{4}$ symmetry to remain even after the
branes are bent.  It is convenient to rewrite the coordinates
as\footnote{There are more convenient coordinate definitions that we
  can make.  However, this set will be more convenient when we include
  the backreaction of other $\uD 7$-branes.}
\begin{equation}
\label{eq:flat_space_foliation_coordinates}
  z^{1}=\mu\, u\cos\frac{\theta}{2}
  \ue^{{\ui}\left(2\psi+\eta+\vp\right)/2},\quad
  z^{2}=\mu\, u\sin\frac{\theta}{2}
  \ue^{{\ui}\left(2\psi+\eta-\vp\right)/2},\quad
  z^{3}=\mu\, \bigl(1+\chi\bigr)\ue^{\ui\psi},
\end{equation}
in which $u\ge 0$ and $\chi\in\left[-1,\infty\right)$.  This corresponds to
\begin{equation}
  \sin\frac{\ga}{2}=\frac{\mu}{r}\left(1+\chi\right),\quad
  \cos\frac{\ga}{2}=\frac{\mu}{r}u,
\end{equation}
and so
\begin{equation}
  r^{2}=\mu^{2}\bigl[\bigl(1+\chi\bigr)^{2}+u^{2}\bigr].
\end{equation}
Then the metric takes the form
\begin{align}
  \ud s_{6}^{2}=\mu^{2}\ue^{2B}\biggl\{&\ud u^{2}
  +\ud\chi^{2}+\bigl(u^{2}+\bigl(1+\chi\bigr)^{2}\bigr)
  \ud\psi^{2}
  +\frac{u^{2}}{4}\bigl[\bigl(h^{1}\bigr)^{2}+\bigl(h^{2}\bigr)^{2}
  +\bigl(h^{3}\bigr)^{2}\bigr]
  +u^{2}\ud\psi\,h^{1}\biggr\},
\end{align}
in which we have defined
\begin{equation}
  h^{1}=\ud\eta+\cos\theta\ud\vp,\quad
  h^{2}=\sin\theta\ud\vp,\quad
  h^{3}=\ud\theta.
\end{equation}
This effectively expresses $C^{3}$ as a foliation of surfaces of the
type $z^{3}=\nu$ for $\nu\in C$.  The
embedding~\eqref{eq:flat_space_embedding} corresponds to $\psi=0$,
$\chi=0$ while more general embeddings respecting the isometries will
have $\chi$ and $\psi$ as functions of $u$.

Our goal is to find the embedding that extremizes the probe brane
action~\eqref{eq:Dbrane_action}.  In the presence of non-trivial
$G_{3}$, such a configuration might be accompanied by non-vanishing
worldvolume flux $f_{2}$.  Since we will consider such a case later,
we will include the $f_{2}$ terms in the $\uD 7$ action, but from
Lorentz invariance we can impose that the field strength has no legs
in the external directions.  Similarly, we can take the ansatz that
the open-string fields depend only on $u$.  With this ansatz, the CS
contribution to the $\uD 7$-brane action automatically vanishes and
hence
\begin{equation}
  S_{\uD 7}=-\tau_{\uD 7}\int\ud^{8}\xi\,\sqrt{\det\left(M_{ab}\right)},
\end{equation}
in which $M_{ab}$ is the internal generalized metric
\begin{equation}
  M_{ab}=\mathrm{P}\bigl[g\bigr]_{ab}+2\pi\alpha'\ue^{2A}f_{ab}.
\end{equation}
Writing the worldvolume gauge field as $a_{1}=a_{h^{i}}h^{i}$, we
find
\begin{subequations}
\begin{align}
  M_{uu}=&\mu^{2}\ue^{2B}\bigl\{1+\chi'^{2}+
  \bigl(u^{2}+\bigl(1+\chi\bigr)^{2}\bigr)\psi'^{2}\bigr\},\\
  M_{h^{i}h^{j}}=&\mu^{2}\ue^{2B}\frac{u^{2}}{4}\delta_{ij},\\
  M_{uh^{1}}=&\mu^{2}\ue^{2B}\frac{u^{2}}{2}\psi'+
  2\pi\alpha'\ue^{2A}a_{h^{1}}',\\
  M_{uh^{2}}=&2\pi\alpha'\ue^{2A}a_{h^{2}}',\\
  M_{uh^{3}}=&2\pi\alpha'\ue^{2A}a_{h^{3}}'.
\end{align}
\end{subequations}
Then,
\begin{equation}
  S_{\uD 7}=-\frac{\tau_{\uD 7}\mu^{4}}{8}
  \int\ud^{4}x\,\ud u\,\ud^{3}h\,\sqrt{W},
\end{equation}
in which $\ud^{3}h:=h^{1}h^{2}h^{3}$ and
\begin{equation}
  \label{eq:flat_space_H}
  W=u^{6}\ue^{8B}\biggl\{1+\chi'^{2}+\bigl(1+\chi\bigr)^{2}\psi'^{2}
    +\frac{4}{u^{2}}\frac{\left(2\pi\alpha'\right)^{2}}{\mu^{4}}
    \ue^{4A-4B}a_{h^{i}}'a_{h^{i}}'\biggr\}.
\end{equation}
We first consider the equations of motion for $\psi$ and
$a_{h^{i}}$.  These are
\begin{equation}
  0=\partial_{u}\biggl[u^{6}
  \ue^{8B}\frac{\left(1+\chi\right)^{2}}{\sqrt{W}}\partial_{u}\psi\biggr],\quad
  0=\partial_{u}\biggl[u^{4}\frac{\ue^{4A+4B}}{\sqrt{W}}\partial_{u}a_{h^{i}}\biggr].
\end{equation}
Hence we see that $\psi=0$ and $a_{h^{i}}=0$ are solutions whether or
not $\overline{\uD 3}$-branes are present.  Inserting these solutions,
the action becomes
\begin{equation}
  \label{eq:flat_space_simplified_action}
  S_{\uD 7}=-\frac{\tau_{\uD 7}\mu^{4}}{8}
  \int\ud^{4}x\,\ud u\,\ud^{3}h\,u^{3}\ue^{4B}
  \bigl(1+\chi'^{2}\bigr)^{1/2}.
\end{equation}
The factor $\ue^{4B}$ is a function of $r$ and therefore of $\chi$ and
$u$. Hence the embedding is determined by
\begin{equation}
  0=\partial_{u}\left[\frac{u^{3}\ue^{4B}}
  {\sqrt{1+\bigl(\partial_{u}\chi\bigr)^{2}}}
  \partial_{u}\chi\right]
  -u^{3}\bigl(\partial_{\chi}\ue^{4B}\bigr)
  \sqrt{1+\bigl(\partial_{u}\chi\bigr)^{2}}.
\end{equation}
Variation of the action also yields the boundary condition
\begin{equation}
  0=\left[\frac{u^{3}\ue^{4B}}
  {\sqrt{1+\bigl(\partial_{u}\chi\bigr)^{2}}}
  \partial_{u}\chi\,\delta\chi\right]_{u=0}^{u=\infty}\!\!\!\!\!\!\!\!\!\!\!\!.
\end{equation}
We will satisfy this by requiring that the brane asymptotes to the
unperturbed solution $\chi=0$ as $u\to\infty$, while at $u=0$ we impose
\begin{equation}
  \label{eq:flat_space_bc}
  0=\frac{u^{3}\ue^{4B}}
  {\sqrt{1+\bigl(\partial_{u}\chi\bigr)^{2}}}
  \partial_{u}\chi.
\end{equation}
When $B=0$, the solution to the equation of motion takes the form
\begin{equation}
  \chi=\tilde{c}_{1}
  +\int\ud u\frac{\tilde{c}_{2}}{\sqrt{u^{6}-\tilde{c}_{2}^{2}}},
\end{equation}
and it is easy to see that the only solution satisfying the boundary
conditions is $\tilde{c}_{1}=\tilde{c}_{2}=0$, reproducing the
unperturbed supersymmetric embedding.

As anticipated, the equation of motion and boundary conditions do not
depend directly on the warping, but only on the internal metric.  The
equation is nonlinear and difficult to solve.  However, we can use the
fact that the background is characterized by a small parameter
$\frac{p}{N}\ll 1$ to solve the above equation perturbatively.  To
this end, we write
\begin{equation}
  \label{eq:chi_expansion}
  \chi=\sum_{n=0}^{\infty}\left(\frac{p}{N}\right)^{n}
  \chi^{\left(n\right)}.
\end{equation}
The zeroth order equation of motion is
\begin{equation}
  \label{eq:zeroth_order_flat_space_eom}
  0=\frac{u^{3}\partial_{u}^{2}\chi^{\left(0\right)}}
  {\left[1+\left(\partial_{u}\chi^{\left(0\right)}\right)^{2}\right]^{3/2}}
  +\frac{3u^{2}\partial_{u}\chi^{\left(0\right)}}
  {\left[1+\left(\partial_{u}\chi^{\left(0\right)}\right)^{2}\right]^{1/2}}.
\end{equation}
This is solved by (among other things) $\chi^{\left(0\right)}=0$,
corresponding to the unperturbed embedding.

\begin{figure}[t]
\begin{center}
  \includegraphics[scale=1]{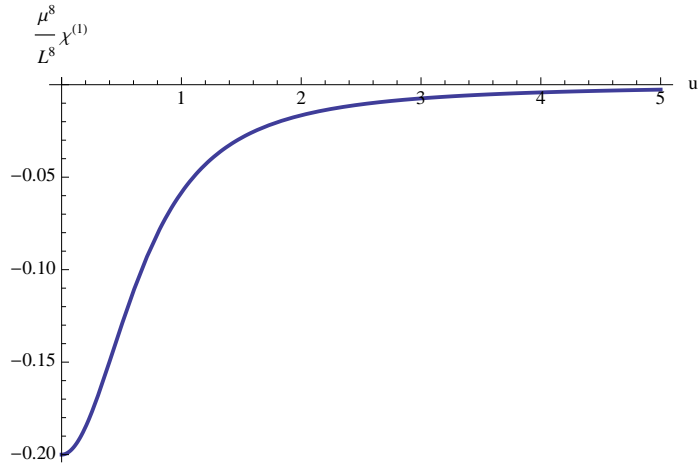}
  \caption{\label{fig:flat_space_embedding}Profile of a $\uD 7$ probe
    of $AdS^{5}\times S^{5}$ perturbed by $p\ll N$ $\uD
    3$-$\overline{\uD 3}$ pairs.  The curve is specified
    by~\eqref{eq:flat_space_embedding_perturbed}.}
\end{center}
\end{figure}

The next order equation is
\begin{equation}
  0=u^{3}\partial_{u}^{2}\chi^{\left(1\right)}
  +3u^{2}\partial_{u}\chi^{\left(1\right)}
  -\frac{16L^{8}}{5\mu^{8}}
  \frac{u^{3}}{\left(1+u^{2}\right)^{5}},
\end{equation}
which has the solution
\begin{equation}
  \chi^{\left(1\right)}
  =c_{1}+\frac{2L^{8}+c_{2}\mu^{8}\left(1+u^{2}\right)^{3}}
  {30u^{2}\left(1+u^{2}\right)^{3}\mu^{8}}.
\end{equation}
Requiring $\chi\to 0$ as $u\to\infty$ sets $c_{1}=0$ while
imposing~\eqref{eq:flat_space_bc} at $u=0$ sets
$c_{2}=-2L^{8}/\mu^{8}$, giving
\begin{equation}
  \label{eq:flat_space_embedding_perturbed}
  \chi^{\left(1\right)}=-\frac{L^{8}}{15\mu^{8}}
  \frac{u^{4}+3u^{2}+3}{\left(1+u^{2}\right)^{3}}.
\end{equation}
A plot of this function is shown in
figure~\ref{fig:flat_space_embedding}.

The above solution naively indicates that at least part of the $\uD 7$
falls toward the $\overline{\uD 3}$s.  However, we must be careful to
disentangle the effect of the motion of the $\uD 7$ branes and the
fact that the backreaction of the $\overline{\uD 3}$ changes the
metric.  That is, the physically relevant quantity is of course the
change in proper distance.  However, because the unperturbed geometry
is anti-de Sitter, the proper distance between the $\uD 3$-brane and
the $\uD 7$s branes is divergent.  Equivalently, the dual theory is a
conformal theory and so there is no natural scale against which to
compare energies.  We might consider measuring the unwarped distance
(that is, using the internal metric $g_{mn}$) between the $\uD
7$-brane and the $\uD 3$-branes, but this quantity will be formally
divergent in the perturbed geometry.  Indeed the proper unwarped
distance between two points on the same point on the $S^{5}$ is
\begin{equation}
  \Delta s=\int_{r_{1}}^{r_{2}}\ud r\,\ue^{B}
  =r_{2}-r_{1}-\frac{\cS L^{8}}{70}\frac{r_{2}^{7}-r_{1}^{7}}
  {r_{1}^{7}r_{2}^{7}},
\end{equation}
which diverges as either $r_{1}$ or $r_{2}$ approaches $0$.  However,
even if this were somehow convergent, the backreaction of the
$\overline{\uD 3}$ becomes large as $r\to 0$ and the solution will
differ significantly from the perturbative one we work with here and
so the result could not be trusted.

To avoid these issues, we consider two probe $\uD 7$-branes, $\uD
7_{1}$ and $\uD 7_{2}$, which asymptote to the embeddings
$z^{3}=\mu_{i=1,2}$.  When $p=0$ the branes are parallel so the
unwarped distance between them is simply $\mu_{2}-\mu_{1}$ (where we
have taken $\mu_{2}>\mu_{1}$ for definiteness).  To characterize the
warped distance between them, we consider the points of the $\uD 7$s
that are closest to the $\uD 3$s.  These are simply where $r=\mu_{i}$
and the geodesic between the two points is a radial path.  Then
\begin{equation}
  \label{eq:SUSY_warped_distance}
  \Delta\hat{s}=\int_{\mu_{1}}^{\mu_{2}}\, \ue^{-A+B}\ud r
  =L\log\frac{\mu_{2}}{\mu_{1}}.
\end{equation}

\begin{figure}[t]
\begin{center}
  \includegraphics[scale=1]{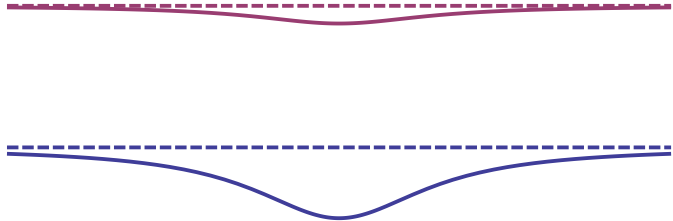}
  \caption{\label{fig:comparing_falling_branes}Sketch of two initially
    parallel branes (represented by the dashed lines) that are deformed
    by the presence of an attractive source.  The lower brane
    experiences a greater attraction than the higher one, and hence
    the distance between the branes increases.}
\end{center}
\end{figure}

In the perturbed geometry, the minimal radii are, to leading order
in $\frac{p}{N}$,
\begin{equation}
\label{eq:min_radius_flat}
  r_{i}=\mu_{i}\biggl(1-\frac{p}{5N}\frac{L^{8}}{\mu_{i}^{8}}\biggr),
\end{equation}
and the unwarped and warped distances between the branes are
\begin{subequations}
\label{eq:interbrane_distance_d7}
\begin{align}
  \Delta s=&\mu_{2}-\mu_{1}
  +\frac{13 p L}{70N}\biggl(\frac{L^{7}}{\mu_{1}^{7}}
  -\frac{L^{7}}{\mu_{2}^{7}}\biggr),\\
  \Delta\hat{s}=&L\log\frac{\mu_{2}}{\mu_{1}}
  +\frac{17 pL}{80N}\biggl(\frac{L^{8}}{\mu_{1}^{8}}
  -\frac{L^{8}}{\mu_{2}^{8}}\biggr).
\end{align}
\end{subequations}
For both of these measures, the proper distance between the branes
increases when we take into account the perturbations.  This is
consistent with the fact that the branes are being attracted to
non-BPS stack of branes; the $\uD 7$ brane that begins closer to the
$\overline{\uD 3}$s is attracted more strongly and so moves a greater
distance than the $\uD 7$ that is further away from them, leading to
an increase in proper distance (see
figure~\ref{fig:comparing_falling_branes}).  We note however one
subtlety in characterizing the movement using the warped distance: if
the geometry is perturbed by just the addition of $\uD 3$-branes, then
the warped distance~\eqref{eq:SUSY_warped_distance} increases even
though the $\uD 3$s do not exert a force on the $\uD 7$s.  The
unwarped proper distance, which does not change, is therefore a more
honest representation of the reaction of the $\uD 7$s.

We can also characterize the influence of the $\overline{\uD 3}$s by
the change in the energy density (that is, energy per unit volume of
Minkowski space) of the $\uD 7$.  Since the configuration that we are
considering is static, this is simply given by the Lagrangian density
defined by $S=\int\ud\vol_{R^{3,1}}\cL$,
\begin{equation}
  \Delta \cE=-\Delta \cL.
\end{equation}
Defining
\begin{equation}
  \cV_{Y^{3}}=\int h^{1}h^{2}h^{3},
\end{equation}
and
\begin{equation}
  T=-\frac{\tau_{\uD 7}\mu^{4}}{8}\cV_{Y^{3}},
\end{equation}
the energy density is
\begin{equation}
  \cE=T\int\ud{u}\, u^{3}\ue^{4B}\sqrt{1+\chi'^{2}}.
\end{equation}
When $p=0$, $\chi=0$ and so the perturbed wavefunction only
contributes to the energy at $\cO\left(\frac{p^{2}}{N^{2}}\right)$.
Since we are working in a non-compact geometry, $\cE$ diverges.
However, the change in the energy defined in this way is convergent
\begin{equation}
  \label{eq:change_in_energy}
  \Delta \cE=\cE_{p\neq 0}-\cE_{p=0}=-T\biggl\{\frac{p}{N}\frac{L^{8}}{30\mu^{8}}
  +\frac{p^{2}}{N^{2}}\frac{L^{16}}{175\mu^{16}}\biggr\}.
\end{equation}
If we artificially compactify the holographic coordinate by hand, this
term would behave like a potential in the 4d theory and indeed we see
that it favors small values of $\mu$, indicating an attraction between
the $3$-branes and the $\uD 7$ that is absent in the $p=0$ case.  Note
that in computing the above quantity we have made use of the
$\cO\left(\frac{p^{2}}{N^{2}}\right)$ corrections to the
geometry~\eqref{eq:anti_D3s_in_cy_cone}.

Finally, we comment briefly on the dual field theory analysis of the
above behavior.  As argued in~\cite{Karch:2002sh}, the open-string
field $\chi$, which corresponds to transverse deformations of the
brane, is dual to a field-theory operator with scaling dimension
$\Delta=3$.  We can confirm this fact quickly by observing that the
linearized solutions of~\eqref{eq:zeroth_order_flat_space_eom} scale
as $u^{-2}$ and $u^{0}$ when $p=0$.  At large $u$, we have $r\approx
\mu u$ and hence the asymptotic solutions act as
\begin{equation}
  \chi\sim \chi_{\mathrm{dom}}+\chi_{\mathrm{sub}}r^{-2}.
\end{equation}
Comparing this to the result for a canonically normalized scalar
\begin{equation}
  \Phi\sim \Phi_{\mathrm{dom}}r^{\Delta-4}+\Phi_{\mathrm{sub}}r^{-\Delta},
\end{equation}
and taking ratios of the terms (which corrects for the fact that
$\chi$ is not canonically normalized) we recover $\Delta=3$.
Evidently the canonically normalized field is proportional to
$\chi/u$, a fact that is easily confirmed by
expanding~\eqref{eq:flat_space_simplified_action}.  A constant shift
in $\chi$ is a solution to~\eqref{eq:zeroth_order_flat_space_eom} (in
the $p=0$ case), which is a reflection of the absence of a force
between the $\uD 3$s and the $\uD 7$.  Such a shift in $\mu$ results
in a shift in the mass of the quarks of the dual theory.  That is, the
addition of $N_{f}$ $\uD 7$-branes at $z^{3}=\mu$ corresponds to the
addition of quarks, fields transforming as bifundamentals of
$\SU{N}\times \U{N_{f}}$. The corresponding term in the superpotential
is
\begin{equation}
  W\ni m_{Q}\tilde{Q}Q,
\end{equation}
where $Q$ and $\tilde{Q}$ transform under conjugate representations,
the mass is set by the embedding parameter, $m_{Q}\propto \mu$, and we
have suppressed interactions between the quarks and the adjoint fields
of the $\cN=4$ sector.  A constant shift of the $\uD 7$s shifts
$m_{Q}$ and hence $\chi$ is dual to the superpotential operator
$\tilde{Q}Q$.  In terms of the chiral superfields $\cQ\sim
Q+\theta\psi_{Q}+\theta^{2}F_{Q}$, $\chi$ corresponds to
$\left.\tilde{\cQ}\cQ\right\rvert_{\theta^{2}}$.  For $p\neq 0$, the
solution~\eqref{eq:flat_space_embedding_perturbed} scales as
$u^{-2}\sim r^{-2}$ at large $u$, agreeing with the expected $r^{-3}$
scaling of the canonically normalized field, and hence the dual theory
has a vacuum expectation value for
$\left.\tilde{\cQ}\cQ\right\rvert_{\theta^{2}}$ that is proportional
to $\frac{p}{N}$.

The analysis of this section was performed for $\uD 7$s with a finite
embedding parameter~\eqref{eq:flat_space_embedding}.  The case for
vanishing $\mu$ cannot be immediately carried over since the
coordinate system~\eqref{eq:flat_space_foliation_coordinates} made use
of a finite $\mu$. Furthermore, the backreaction of the $\overline{\uD
  3}$-brane increases as we approach the brane and hence cannot treat
the addition of the anti-branes as a perturbation to the geometry even
when $\frac{p}{N}\ll 1$.  However, following~\cite{Bigazzi:2009bk} we
can argue that such a $\uD 7$ brane will not be deformed in the sense
that $z^{3}=0$ is still a solution to the $\uD 7$-brane equations of
motion in the presence of the $\overline{\uD 3}$s.  This argument is
reviewed in a more general context in
appendix~\ref{app:backreaction_D7s_in_flat_space}.

\subsection{\label{subsec:d3_probes_flat_flavored}Backreacting D7s}

In the previous section, we considered the backreaction of $N$ $\uD
3$-branes coincident with $p\ll N$ $\uD 3$-$\overline{\uD 3}$ pairs
and argued that a $\uD 7$ probe of such a geometry feels a small
attractive force and hence bends toward the $3$-branes.  We should be
able to observe such an interaction from the perspective of the
$3$-branes as well.  This could in principle be accomplished by
placing probe $3$-branes in the geometry resulting from the
backreaction of such a $\uD 7$. Unfortunately, the backreaction of
codimension-$2$ objects is difficult to compute except in particularly
simple cases and the configuration we will ultimately consider is
different from the one considered in the previous section so a direct
comparison will not be possible.

One strategy to make progress on the problem is to consider the
backreaction of a smeared distribution of $\uD 7$-branes.  That is,
rather than considering the backreaction of a single $\uD 7$
satisfying the embedding condition $z^{3}=\nu$, we consider $N_{f}$
$\uD 7$-branes each of which satisfy an $\SU{3}\times\U{1}$ rotation
of that embedding
\begin{equation}
  a_{i}z^{i}=\nu\ue^{\ui\alpha},
\end{equation}
in which $\alpha\in\left[0,2\pi\right)$ and the $a_{i}$ are complex
parameters satisfying $a_{i}\bar{a}_{\bar{i}}=1$.  We then assume that
$N_{f}$ is large and that the branes are placed in such a way that the
distribution can, to good approximation, be treated as following from
uniformly integrating over the $a_{i}$ and $\alpha$.  The result for
$p=0$ was found for $\nu=0$ in~\cite{Benini:2006hh} and for $\nu\neq
0$ in~\cite{Bigazzi:2009bk}.  For simplicity, we will focus on the
$\nu=0$ case which we review in
appendix~\ref{app:backreaction_D7s_in_flat_space}.  The corresponding
quarks in the dual theory are massless and so we will occasionally
abuse terminology and refer to such $\uD 7$s as massless $\uD
7$s. With the appropriate smearing, the metric for the internal space
can be written in the form
\begin{equation}
  \label{eq:flavored_C3_metric}
  \ud s_{6}^{2}=\ue^{2B}\bigl\{\ud r^{2}+r^{2}\bigl(\ud\psi+\cA\bigr)^{2}\bigr\}
  +\ue^{2C}r^{2}\ud s^{2}_{CP^{2}}.
\end{equation}
The solution describing the backreaction of the smeared branes is
\begin{subequations}
\begin{align}
  F_{1}=&-\frac{N_{f}}{2\pi}
  \bigl(\ud\psi+\cA\bigr),\\
  \ue^{-\phi}=&
  \frac{1}{g_{\us}}\biggl(
  1-\frac{g_{\us}N_{f}}{2\pi}
  \log\frac{r}{r_{\us}}\biggr),\\
  \ue^{2B}=&\left(1-\frac{g_{\us}N_{f}}{2\pi}
    \biggl[\log\frac{r}{r_{\us}}
    -\frac{1}{6}\biggr]\right)^{1/3},\\
  \ue^{2C}=&g_{\us}\ue^{-\phi-4B},\\
  \Phi_{+}^{-1}=&-8\pi g_{\us}N\alpha'^{2}\int_{r_{\mathrm{LP}}}^{r}
  \frac{\ud x}{x^{5}}
  \ue^{-4B\left(x\right)}.
\end{align}
\end{subequations}
Here, $r_{\us}$ is an arbitrarily chosen scale such that
$\phi\left(r_{\us}\right)=\log g_{\us}$.  Note that the dual theory
exhibits a Landau pole at $r_{\mathrm{LP}}=
r_{\us}\ue^{2\pi/g_{\us}N_{f}}$ where $\ue^{-\phi}=0$.

We now consider an $\overline{\uD 3}$ probe of the above geometry.
In the absence of $\uD 7$-branes, the $\overline{\uD 3}$ will feel a
force from the $\uD 3$ charge given by\footnote{Note that this
  expression is slightly different, but not inconsistent with, that
  appearing in~\cite{Kachru:2003sx} since the authors there consider
  the interaction between a stationary $\overline{\uD 3}$ and a
  mobile, distant, $\uD 3$ brane.  In contrast, here we are
  considering a mobile $\overline{\uD 3}$ interacting only with the
  background charge of $AdS^{5}\times S^{5}$.}
\begin{equation}
  F=-\tau_{\uD 3}\frac{\ud\Phi_{+}}{\ud r}
  =-8\tau_{\uD3}\frac{r^{3}}{L^{4}}.
\end{equation}
In order to make a precise comparison of the potential before and
after the addition of $\uD 7$-branes, we must account for the fact
that both $\Phi_{+}$ and the metric factor $g_{rr}$ are modified when
$N_{f}=0$ is finite.  Said differently, from the point of view of the
worldvolume field theory of the $\overline{\uD 3}$, the radial
position is a scalar field that has different normalizations before
and after the introduction of $\uD 7$s and so to make a comparison of
the potentials we ought to give the fields the same normalization.  To
leading order in momenta along the $\overline{\uD 3}$ worldvolume, the
action for the $\overline{\uD 3}$ is
\begin{equation}
  \label{eq:anti_d3_action}
  S_{\overline{\uD 3}}=-\tau_{\uD 3}\int\ud^{4}x\,\biggl\{\frac{1}{2}\ue^{4A+2B}
  \eta^{\mu\nu}\partial_{\mu}r\partial_{\nu}r+\Phi_{+}\bigl(r\bigr)
  \biggr\},
\end{equation}
where we have now considered an $\overline{\uD 3}$-brane that is
moving entirely in the radial direction so that the position of the
brane is described by $r\left(x^{\mu}\right)$.  When $N_{f}=0$, this gives
\begin{equation}
  S_{\overline{\uD 3}}=-\tau_{\uD 3}\int\ud^{4}x\,\biggl\{\frac{1}{2}\frac{r^{4}}{L^{4}}
  \eta^{\mu\nu}\partial_{\mu}r\partial_{\nu}r
  +\frac{2r^{4}}{L^{4}}\biggr\}.
\end{equation}
From the perspective of the $\overline{\uD 3}$-brane, the canonically
normalized field is
\begin{equation}
  \sig=\sqrt{\tau_{\uD 3}}\frac{r^{3}}{3L^{2}},
\end{equation}
in terms of which the potential is
\begin{equation}
  V\bigl(\sig\bigr)=2\cdot 3^{4/3}\frac{\tau_{\uD 3}^{1/3}}{L^{4/3}}
  \sig^{4/3}.
\end{equation}

When $N_{f}>0$, we cannot analytically find such a canonically
normalized field.  However when
$\frac{g_{\us}N_{f}}{2\pi}\abs{\log\frac{r}{r_{\us}}}\ll 1$ the
backreaction of the $\uD 7$ can be treated perturbatively and we have
\begin{equation}
  \ue^{2B}\approx\biggl(1-\frac{g_{\us}N_{f}}{6\pi}
  \biggl[\log\frac{r}{r_{\us}}+\frac{1}{3}\biggr]\biggr),
\end{equation}
where we have taken $\frac{g_{\us}N_{f}}{2\pi}$ to be small and have
neglected terms of higher order in that parameter.  Then,
\begin{equation}
  \Phi_{+}\approx\frac{2r^{4}}{L^{4}}
  \biggl(1-\frac{g_{\us}N_{f}}{3\pi}
  \biggl[\log\frac{r}{r_{\us}}+\frac{1}{12}\biggr]\biggr).
\end{equation}
Hence to this approximation
\begin{equation}
  S_{\overline{\uD 3}}=-\tau_{\uD 3}\int\ud^{4}x\biggl\{\frac{1}{2}
  \frac{r^{4}}{L^{4}}
  \biggl(1-\frac{g_{\us}N_{f}}{2\pi}\log\frac{r}{r_{\us}}\biggr)
  \eta^{\mu\nu}\partial_{\mu}r\partial_{\nu}r
  +\frac{2r^{4}}{L^{4}}
  \biggl(1-\frac{g_{\us}N_{f}}{3\pi}
  \biggl[\log\frac{r}{r_{\us}}+\frac{1}{12}\biggr]\biggr)\biggr\}.
\end{equation}
We can construct a normalized field $\sig$ in terms of which
\begin{equation}
  r=\biggl(\frac{3L^{2}}{\sqrt{\tau_{\uD 3}}}\sig\biggr)^{1/3}
  \biggl(1+\frac{g_{\us}N_{f}}{36\pi}
  \biggl[\log\frac{3L^{2}\sig}{\sqrt{\tau_{\uD 3}}r_{\us}^{3}}-
  \frac{1}{2}\biggr]\biggr),
\end{equation}
and the potential is
\begin{equation}
  \label{eq:anti_d3_potential}
  V\bigl(\sig\bigr)=2\cdot 3^{4/3}\frac{\tau_{\uD 3}^{1/3}}{L^{4/3}}
  \biggl(1-\frac{g_{\us}N_{f}}{12\pi}\biggr)\sig^{4/3}.
\end{equation}
Hence the potential experienced by an $\overline{\uD 3}$ is reduced by
the presence of the $\uD 7$s.  Note that, at least to this order of
approximation, the result is independent of the integration constant
$r_{\us}$.  A useful check of our results would be to confirm
that~\eqref{eq:anti_d3_potential} is consistent with the results of
the probe D7 analysis section~\ref{subsec:D7_probe_in_flat_space}.
However, in the probe D7 analysis, the flavor branes were placed at
larger radii than the $\overline{\uD 3}$-branes while in the analysis
leading to~\eqref{eq:anti_d3_potential}, the $\overline{\uD 3}$-branes
were placed at larger $r$.  Hence~\eqref{eq:anti_d3_potential} cannot
be used to confirm the probe D7 analysis.  Due to the smearing
procedure that we used, the force on an $\overline{\uD 3}$-brane
placed at smaller $r$ would not be modified by the presence of the
flavor branes, and so to check the probe D7 analysis we would need to
move beyond the smeared approximation, an analysis which is beyond the
scope of our current work.

Finally, we can consider a $\uD 3$ probe of the geometry resulting
from the backreaction of $\overline{\uD 3}$s and $\uD 7$s.  The
precise thing to do would be to backreact the perturbed $\uD 7$ branes
which requires solving the $\uD 7$ equations of motion together with
the supergravity equations of motion.  However analogous to what was
argued in~\cite{Bigazzi:2009bk}, $\uD 7$-branes characterized by
$\nu=0$ are not perturbed by the addition of $\overline{\uD 3}$-branes
even once their smeared backreaction is taken into account as well.
Even still, the backreaction of such branes is difficult to compute.
We can make progress by considering the (admittedly non-generic) case
in which the $\overline{\uD 3}$s and the $\uD 7$s can be treated as
comparable perturbations.  Writing $\delta_{p}=\frac{p}{N}$ and
$\delta_{f}=\frac{g_{\us}N_{f}}{2\pi}$, the corrections resulting from
both $\overline{\uD 3}$s and $\uD 7$s comes in at
$\cO\left(\delta_{p}\delta_{f}\right)$ and hence we must use the
solution through $\cO\left(\delta^{2}\right)$ where we assume
$\delta\sim\delta_{p}\sim\delta_{f}$.  The resulting solution should
then be valid when
$\mathrm{min}\left(\delta^{1/8}L,r_{\us}\ue^{-1/\delta}\right)\ll r\ll
r_{\us}\ue^{1/\delta}$.  Consistency of this condition requires us to
chose $r_{\us}$ such that $r_{\us}\gg \delta^{1/8}L\ue^{-1/\delta}$
and hence $L \delta^{1/8}\gg r_{\us}\ue^{-1/\delta}$.  Thus the
solution is valid when $L\delta^{1/8}\ll r\ll
r_{\us}\ue^{1/\delta}$. The result is (see
appendix~\ref{app:backreaction_D7s_in_flat_space})
\begin{subequations}%
\label{eq:D7_backreaction_in_flat_space_r}
\begin{align}
  \ue^{4A}=&\frac{L^{4}}{r^{4}}
  +\delta_{f}\frac{2}{3}\frac{L^{4}}{r^{4}}\biggl(\log\frac{r}{r_{\us}}
  +\frac{1}{12}\biggr)
  +\delta_{p}\frac{4}{5}\frac{L^{12}}{r^{12}}
  +\delta_{f}^{2}\frac{5}{9}
  \frac{L^{4}}{r^{4}}\biggl(\log^{2}\frac{r}{r_{\us}}
  +\frac{1}{6}\log\frac{r}{r_{\us}}+\frac{5}{72}\biggr)\notag\\
  &+\delta_{f}\delta_{p}\frac{8}{5}\frac{L^{12}}{r^{12}}
  \biggl(\log\frac{r}{r_{\us}}+\frac{1}{12}\biggr)
  +\delta_{p}^{2}\frac{54}{125}\frac{L^{20}}{r^{20}},\\
  \omega=&\frac{r^{4}}{L^{4}}
  -\delta_{f}\frac{2}{3}\frac{r^{4}}{L^{4}}\biggl(\log\frac{r}{r_{\us}}
  +\frac{1}{12}\biggr)
  +\delta_{p}\frac{6}{5}\frac{L^{4}}{r^{4}}
  -\frac{1}{9}\delta_{f}^{2}\frac{r^{4}}{L^{4}}
  \biggl(\log^{2}\frac{r}{r_{\us}}+\frac{1}{6}\log\frac{r}{r_{\us}}
  +\frac{23}{72}\biggr)\notag\\
  &+\frac{4}{5}\delta_{f}\delta_{p}\frac{L^{4}}{r^{4}}
  \biggl(\log\frac{r}{r_{\us}}+\frac{1}{12}\biggr)
  -\delta_{p}^{2}\frac{24}{125}\frac{L^{12}}{r^{12}},\\
  F_{1}=&-\frac{1}{g_{\us}}\delta_{f}\bigl(\ud\psi+\cA\bigr),\\
  \ue^{-\phi}=&\frac{1}{g_{\us}}
  \biggl(1-\delta_{f}\log\frac{r}{r_{\us}}
  +\delta_{f}\delta_{p}\frac{1}{10}\frac{L^{8}}{r^{8}}\biggr),\\
  \ue^{2B}=&1-\delta_{f}\frac{1}{3}\biggl(\log\frac{r}{r_{\us}}
  +\frac{1}{3}\biggr)-\delta_{p}\frac{1}{5}\frac{L^{8}}{r^{8}}
  -\delta_{f}^{2}\frac{1}{9}\biggl(\log^{2}\frac{r}{r_{\us}}
  +\frac{2}{3}\log\frac{r}{r_{\us}}-\frac{5}{36}\biggr)\notag\\
  &-\delta_{f}\delta_{p}\frac{1}{5}\frac{L^{8}}{r^{8}}
  \log\frac{r}{r_{\us}}
  -\delta_{p}^{2}\frac{1}{50}\frac{L^{16}}{r^{16}},\\
\intertext{\newpage}
  \ue^{2C}=&1-\delta_{f}\frac{1}{3}\biggl(\log\frac{r}{r_{\us}}
  -\frac{1}{6}\biggr)-\delta_{p}\frac{1}{5}\frac{L^{8}}{r^{8}}
  -\delta_{f}^{2}\frac{1}{9}\biggl(\log^{2}\frac{r}{r_{\us}}
  -\frac{1}{3}\log\frac{r}{r_{\us}}+\frac{1}{36}\biggr)\notag\\
  &-\delta_{f}\delta_{p}\frac{1}{5}\frac{L^{8}}{r^{8}}
  \biggl(\log\frac{r}{r_{\us}}+\frac{1}{6}\biggr)
  -\delta_{p}^{2}\frac{1}{50}\frac{L^{16}}{r^{16}}.
\end{align}
\end{subequations}
Here, we have suppressed the additive constants to $\ue^{4A}$ and
$\omega$ which are present to ensure that $\ue^{-4A}$ and
$\frac{1}{\omega}$ vanish at the Landau pole.

The relevant action for calculating the $\uD 3$ potential
is~\eqref{eq:anti_d3_action} with the replacement
$\Phi_{+}\to\Phi_{-}$.  Then, in terms of a canonically normalized
field $\sig$, the potential is
\begin{equation}
  \label{eq:D3_potentials}
  V\bigl(\sig)=-\delta_{p}\frac{2}{3^{4/3}}
  \frac{\tau_{\uD 3}^{5/3}L^{4/3}}{\sig^{4/3}}
  -\delta_{p}\delta_{f}\frac{1}{3^{7/3}}
  \frac{\tau_{\uD 3}^{5/3}L^{4/3}}{\sig^{4/3}}
  -\delta_{p}^{2}\frac{2}{405}\frac{L^{4}\tau_{\uD 3}^{3}}{\sig^{4}}.
\end{equation}
Hence the presence of the $\uD 7$s results in a steeper potential for
a probe $\uD 3$.

We again note that the above configurations of $\uD 7$s and $3$-branes
are not quite what we would want to consider to compliment the
discussion in section~\ref{subsec:D7_probe_in_flat_space}.  In the
earlier analysis, the $3$-branes were located at smaller radii than
the $\uD 7$s while in this section the bottom of the $\uD 7$s is
localized at smaller $r$ than the $3$-branes.  Unfortunately, the
reverse situation is more difficult to describe.  The backreaction of
smeared $\uD 7$-branes characterized by $\nu>0$ is non-trivial only
for $r>\nu$ and hence the influence of such massive branes on
$3$-branes at $r=0$ cannot be captured in the smeared approximation.

\subsection{\label{subsec:flavor_forces}Forces between flavors}

In the absence of $\overline{\uD 3}$-branes, parallel $\uD 7$s in the
above background will not feel a force between them.  However, this
will not be the case once supersymmetry is broken by the presence of
the anti-branes.  To be more precise, we consider two $\uD
7$s that asymptote to $z^{3}=\mu_{1,2}$ and take
$\abs{\mu_{2}}>\abs{\mu_{1}}$.  When $p=0$, the embeddings $z^{3}=\mu$
will solve the equations of motion precisely, even when the
backreaction of the $\uD 7$s are taken into account due to the
supersymmetry of the configuration.  However, when the $\overline{\uD
  3}$s backreact, this cancellation no longer occurs.

We can be more quantitative by considering a $\uD 7$ probe of the
geometry~\eqref{eq:D7_backreaction_in_flat_space_r} describing the
backreaction of smeared massless $\uD 7$s and $\overline{\uD 3}$s in
$AdS^{5}\times S^{5}$.  In order for the perturbative treatment of the
$\uD 7$ backreaction to be valid, we must again consider the situation
in which $\left\lvert\log\frac{\mu}{r_{\us}}\right\rvert\ll
\frac{1}{\delta_{f}}$.  That is, the embedding parameter of our probe
$\uD 7$ is neither too close to the origin nor too close to the Landau
pole.  Of course, since the $\uD 7$ extends to the UV in the radial
direction, there will be part of the probe worldvolume that reaches
the Landau pole; however, at that point the influence of the
$\overline{\uD 3}$s is small so that the configuration is
approximately supersymmetric and the deviation from the $z^{3}=\mu$
embedding is expected to be negligible, a fact that we will confirm.

The process that we follow is largely the same as without the
backreacting $\uD 7$s.  We again adopt the
coordinates~\eqref{eq:flat_space_foliation_coordinates} in terms of
which the metric~\eqref{eq:flavored_C3_metric} takes the form
\begin{align}
  \ud s_{6}^{2}=&
  \frac{\ue^{2B}-\ue^{2C}}{r^{2}}
  \bar{z}^{\bar{A}}\ud z^{A}\,z^{B}\ud \bar{z}^{\bar{B}}
  +\ue^{2C}\ud z^{A}\ud\bar{z}^{\bar{A}},
\end{align}
or
\begin{align}
  \ud s_{6}^{2}=&
  \mu^{2}\biggl\{
  \frac{u^{2}\ue^{2B}+\left(1+\chi\right)^{2}\ue^{2C}}
  {u^{2}+\left(1+\chi\right)^{2}}\ud u^{2}
  +\frac{u^{2}\ue^{2C}+\left(1+\chi\right)^{2}\ue^{2B}}
  {u^{2}+\left(1+\chi\right)^{2}}\ud\chi^{2}
  +\ue^{2B}\bigl[u^{2}+\bigl(1+\chi\bigr)^{2}\bigr]
  \ud\psi^{2}\notag\\
  &\phantom{\mu^{2}\biggl\{}+\frac{u^{2}}{4}
  \biggl(\frac{u^{2}\ue^{2B}+\left(1+\chi\right)^{2}\ue^{2C}}
  {u^{2}+\left(1+\chi\right)^{2}}\bigl(h^{1}\bigr)^{2}
  +\ue^{2C}\bigl(h^{2}\bigr)^{2}+\ue^{2C}
  \bigl(h^{3}\bigr)^{2}\biggr)
  +\ue^{2B}u^{2}\ud\psi\,h^{1}\notag\\
  &\phantom{\mu^{2}\biggl\{}
  +\frac{2u\left(1+\chi\right)\left(\ue^{2B}-\ue^{2C}\right)}
  {u^{2}+\left(1+\chi\right)^{2}}
  \ud u\,\ud\chi\biggr\}.
\end{align}
Then the internal generalized metric on the $\uD 7$ probe is
\begin{subequations}
\begin{align}
  M_{uu}=&\mu^{2}\biggl\{
  \frac{u^{2}\ue^{2B}+\bigl(1+\chi\bigr)^{2}\ue^{2C}}
  {u^{2}+\left(1+\chi\right)^{2}}+
  \frac{2u\left(1+\chi\right)\left(\ue^{2B}-\ue^{2C}\right)}
  {u^{2}+\left(1+\chi\right)^{2}}\chi'\notag\\
  &\phantom{\mu^{2}\biggl\{}+
  \frac{u^{2}\ue^{2C}+\bigl(1+\chi\bigr)^{2}\ue^{2B}}
  {u^{2}+\left(1+\chi\right)^{2}}\chi'^{2}
  +\ue^{2B}\bigl(u^{2}+\bigl(1+\chi\bigr)^{2}\bigr)\psi'^{2}\biggr\},\\
  M_{h^{1}h^{1}}=&
  \mu^{2}\frac{u^{2}\ue^{2B}+\left(1+\chi\right)^{2}\ue^{2C}}
  {u^{2}+\left(1+\chi\right)^{2}}\frac{u^{2}}{4},\\
  M_{h^{2}h^{2}}=M_{h^{3}h^{3}}=&\mu^{2}\ue^{2C}\frac{u^{4}}{4},\\
  M_{uh^{1}}=&\mu^{2}\ue^{2B}\frac{u^{2}}{2}\psi'
  +2\pi\alpha'g_{\us}^{1/2}\ue^{2A-\phi/2}a'_{h^{1}},\\
  M_{uh^{2}}=&2\pi\alpha'g_{\us}^{1/2}\ue^{2A-\phi/2}a'_{h^{2}},\\
  M_{uh^{3}}=&2\pi\alpha'g_{\us}^{1/2}\ue^{2A-\phi/2}a'_{h^{3}}.
\end{align}
\end{subequations}
This gives the DBI action
\begin{equation}
  S_{\uD 7}^{\mathrm{DBI}}
  =-\frac{\tau_{\uD 7}\mu^{4}}{8}\int\ud^{4}x\,\ud u\,\ud^{3}h
  \sqrt{W},
\end{equation}
where now
\begin{equation}
  W=W_{0}+W_{1}\chi'+W_{2}\chi'^{2}+Y\psi'^{2}
  +H_{ij}a'_{h^{i}}a'_{h^{j}},
\end{equation}
with
\begin{subequations}
\begin{align}
  W_{0}=&g_{\us}^{-2}\ue^{4C+2\phi}
  \frac{u^{6}\left(u^{2}\ue^{2B}+\left(1+\chi\right)^{2}\ue^{2C}\right)^{2}}
  {\left(u^{2}+\left(1+\chi\right)^{2}\right)^{2}}
  ,\\
  W_{1}=&g_{\us}^{-2}\ue^{4C+2\phi}
  \frac{2u^{7}\left(1+\chi\right)\left(\ue^{2B}-\ue^{2C}\right)
    \left(u^{2}\ue^{2B}+\left(1+\chi\right)^{2}\ue^{2C}\right)}
  {\left(u^{2}+\left(1+\chi\right)^{2}\right)},\\
  W_{2}=&g_{\us}^{-2}\ue^{4C+2\phi}
  \frac{u^{6}\left(u^{2}\ue^{2C}+\left(1+\chi\right)^{2}\ue^{2B}\right)
    \left(u^{2}\ue^{2B}+\left(1+\chi\right)^{2}\ue^{2C}\right)}
  {\left(u^{2}+\left(1+\chi\right)^{2}\right)^{2}},\\
  Y=&g_{\us}^{-2}\ue^{2B+6C+2\phi}u^{6}\left(1+\chi\right)^{2},\\
  H_{11}=&
  \frac{\left(2\pi\alpha'\right)^{2}}{\mu^{4}}4u^{4}
  g_{\us}^{-1}\ue^{4A+4C+\phi},\\
  H_{22}=H_{33}=&
  \frac{\left(2\pi\alpha'\right)^{2}}{\mu^{4}}4u^{4}
  g_{\us}^{-1}\ue^{4A+4C+\phi}\frac{u^{2}\ue^{2B-2C}+\left(1+\chi\right)^{2}}
  {u^{2}+\left(1+\chi\right)^{2}},
\end{align}
\end{subequations}
and other components of $H_{ij}$ vanishing.

In addition to the DBI action, there is a contribution from the
Chern-Simons action due to the nontrivial $C_{0}$ sourced by the
backreacting $\uD 7$s.  $\uD 7$s couple magnetically to $C_{0}$ and so
we construct
\begin{equation}
  F_{9}=-g_{\us}^{-2}\ue^{2\phi}\hat{\ast}\ud F_{1}
  =-Q_{f}\frac{r^{3}}{2}g_{\us}^{-2}\ue^{2\phi}\ue^{4C}
  \ud\vol_{R^{3,1}}\wedge\ud r\wedge\cJ\wedge\cJ,
\end{equation}
where we have written $F_{1}=-Q_{f}\left(\ud\psi+\cA\right)$ and made
use of the fact that the flat space volume element is
\begin{equation}
  \ud\vol_{C^{3}}=\frac{1}{3!}J\wedge J\wedge J=\frac{1}{2}r^{5}\ud
  r\wedge\left(\ud\psi+\cA\right)\wedge\cJ\wedge\cJ,
\end{equation}
where the factor of $\frac{1}{2}$ comes from the fact that
$\ud\vol_{CP^{2}}=\frac{1}{2}\cJ\wedge\cJ$. In the absence of $3$-form
flux, we have $F_{9}=\ud C_{8}$ and so we can take
\begin{equation}
  \label{eq:8form_pot}
  C_{8}=Q_{f}\frac{r^{3}}{4}g_{\us}^{-2}\ue^{2\phi+4C}
  \ud\vol_{R^{3,1}}\wedge\ud r\wedge\cA\wedge\cJ.
\end{equation}
In terms of the
coordinates~\eqref{eq:flat_space_foliation_coordinates}, this is
\begin{equation}
  C_{8}=-Q_{f}\frac{\mu^{4}u^{4}}{32\left(u^{2}+\left(1+\chi\right)^{2}\right)}
  g_{\us}^{-2}\ue^{2\phi+4C}\ud\vol_{R^{3,1}}\wedge
  \bigl[u\ud u+\bigl(1+\chi\bigr)\ud\chi\bigr]\wedge
  \ud\eta\wedge\sin\theta\ud\theta\wedge\ud\vp,
\end{equation}
from which we easily calculate the pullback to the probe $\uD 7$.
\begin{equation}
  \mathrm{P}\bigl[C_{8}\bigr]=
  -Q_{f}\frac{\mu^{4}u^{4}\left(u+\left(1+\chi\right)\chi'\right)}
  {32\left(u^{2}+\left(1+\chi\right)^{2}\right)}
  g_{\us}^{-2}\ue^{2\phi+4C}\ud\vol_{R^{3,1}}\wedge
  \ud u\wedge
  \ud\eta\wedge\ud\theta\wedge\ud\vp.
\end{equation}
When $p=0$ the $\uD 7$-brane is supersymmetric implying that the
volume of the 4-cycle transverse to $R^{3,1}$ is calibrated in the sense that
\begin{equation}
  \ud\vol_{\Sig^{4}}=-\frac{1}{2}\mathrm{P}\bigl[J\wedge J\bigr]
  \sim+\ud u\wedge\ud\eta\wedge\ud\theta\wedge\ud\vp.
\end{equation}
Although the volume form will not be the same for $p\neq 0$, this
defines the orientation for a $\uD 7$ (an $\overline{\uD 7}$ has the
opposite orientation).  With this orientation, the CS contribution to
the $\uD 7$ action is
\begin{equation}
  S_{\uD 7}^{\mathrm{CS}}=-\frac{\mu^{4}\tau_{\uD 7}g_{\us}}{8}\int\ud^{4}x\,\ud u
  \,\ud^{3}h
  \frac{u^{4}\left(u+\left(1+\chi\right)\chi'\right)}
  {4\left(u^{2}+\left(1+\chi\right)^{2}\right)}
  Q_{f}
  g_{\us}^{-2}\ue^{2\phi+4C}.
\end{equation}
The total action is then
\begin{equation}
  S_{\uD 7}=-\frac{\mu^{4}\tau_{\uD 7}}{8}
  \int\ud^{4}x\,\ud u
  \,\ud^{3}h
  \biggl\{\sqrt{W}
  +
  \frac{u^{4}\left(u+\left(1+\chi\right)\chi'\right)}
  {4\left(u^{2}+\left(1+\chi\right)^{2}\right)}
  Q_{f}
  g_{\us}^{-1}\ue^{2\phi+4C}\biggr\}.
\end{equation}

It is easy to see that $\psi=a_{h^{i}}=0$ is again a solution to the
equations of motion.  Setting these fields on shell, the action takes
the form
\begin{equation}
  S_{\uD 7}=-\frac{\mu^{4}\tau_{\uD 7}}{8} \int\ud^{4}x\,\ud u
  \,\ud^{3}h\,\biggl\{
  \sqrt{W_{0}+W_{1}\chi'+W_{2}\chi'^{2}}+W_{3}+W_{4}\chi'\biggr\},
\end{equation}
Then the equation of motion determining the perturbed embedding is
\begin{align}
  0=&\partial_{u}\biggl[\frac{W_{2}}{\sqrt{W}}\partial_{u}\chi\biggr]
  -\frac{1}{2}\frac{\partial_{\chi}W_{2}}{\sqrt{W}}\chi'^{2}
  -\frac{1}{2}\frac{\partial_{\chi}W_{1}}{\sqrt{W}}\chi'
  -\partial_{\chi}W_{4}\chi'\notag\\
  &+\partial_{u}\biggl[\frac{W_{1}}{2\sqrt{W}}\biggr]
  +\partial_{u}W_{4}
  -\frac{\partial_{\chi}W_{0}}{2\sqrt{W}}
  -\partial_{\chi}W_{3}.
  \label{eq:chieom}
\end{align}
The variation of the action also leads to the boundary
terms\footnote{Note that the $u=\infty$ boundary condition is imposed
  where there is perturbative control which fails at around the Landau
  pole at $r_{\mathrm{LP}}=r_{\us}\ue^{2\pi/g_{\us}N_{f}}$.  Due to
  the corresponding UV singularity, a more appropriate boundary
  condition to impose might be that $\chi$ vanishes at some finite
  $r\lesssim r_{\mathrm{LP}}$ as in~\cite{Bigazzi:2009gu}.  However,
  since for $\delta_{f}\ll 1$ this occurs at exponentially large
  radii, the difference between formally imposing a boundary condition
  at $u=\infty$ and this finite part will be exponentially small.  A
  similar problem does not occur for the $u=0$ boundary condition
  since $u=0$ corresponds to $r\approx\mu$ which we are choosing to be
  within the region of control.}
\begin{equation}
  \label{eq:chibc}
  0=\biggl[\biggl(\frac{W_{2}}{\sqrt{W}}\chi'
  +\frac{W_{1}}{2\sqrt{W}}+W_{4}\biggr)\delta\chi\biggr]_{u=0}^{u=\infty}
  \!\!\!\!\!\!\!\!\!\!\!.
\end{equation}
We again satisfy this by imposing that $\chi$ vanishes as
$u\to\infty$ and imposing the Neumann condition at $u=0$.

Performing an expansion in $\delta\sim\delta_{f}\sim\delta_{p}$ in
both $W_{i}$ and $\chi$ schematically as
\begin{equation}
  \chi=\sum_{n=0}^{\infty}\delta^{n}\chi^{\left(n\right)},
\end{equation}
the $\cO\left(\delta^{0}\right)$ and $\cO\left(\delta^{1}\right)$
corrections are the same as they were in the $N_{f}=0$ case of
section~\ref{subsec:D7_probe_in_flat_space}.  This is a manifestation
of the fact that the $\uD 7$ probe and the backreacting $\uD 7$s are
mutually supersymmetric when $p=0$.  Hence through
$\cO\left(\delta^{1}\right)$ we again
have~\eqref{eq:flat_space_embedding_perturbed}
\begin{equation}
  \chi=-\delta_{p}\frac{L^{8}}{15\mu^{8}}
  \frac{u^{4}+3u^{2}+3}{\left(1+u^{2}\right)^{3}}.
\end{equation}

The $\cO\left(\delta^{2}\right)$ corrections to the equations of
motion then take the form
\begin{align}
  0=&\delta^{2}\bigl[u^{3}\bigl(\chi^{\left(2\right)}\bigr)''+3u^{2}
  \bigl(\chi^{\left(2\right)}\bigr)'\bigr]
  +\delta_{p}^{2}\frac{16L^{16}}{75\mu^{16}}
  \frac{u^{3}}{\left(1+u^{2}\right)^{9}}
  \bigl(3u^{6}+2u^{4}-12u^{2}-33\bigr)\notag\\
  &-\delta_{p}\delta_{f}\frac{64L^{8}}{15\mu^{8}}
  \frac{u^{3}}{\left(1+u^{2}\right)^{5}}
  \biggl[\log\frac{\sqrt{1+u^{2}}\mu}{r_{\us}}
  -\frac{u^{8}+6u^{6}+14u^{4}-2u^{2}-26}{96\left(1+u^{2}\right)}\biggr].
\end{align}
With the above boundary conditions the solution is
\begin{align}
  \chi=&-\delta_{p}\frac{L^{8}}{15\mu^{8}}
  \frac{u^{4}+3u^{2}+3}{\left(1+u^{2}\right)^{3}}\notag\\
  &-\delta_{p}^{2}\frac{L^{16}}{7875\mu^{2}}
  \frac{270u^{12}+1890u^{10}+5670u^{8}+9450u^{6}+9513u^{4}+5761u^{2}+1813}
  {\left(1+u^{2}\right)^{7}}\notag\\
  &-\delta_{p}\delta_{f}
  \frac{L^{8}}{1080\mu^{8}}
  \frac{1}{\left(1+u^{2}\right)^{4}}\biggl\{
  37u^{6}+160 u^{4}+262u^{2}+148
  +96\bigl(u^{6}+4u^{4}+6u^{2}+3\bigr)\log\frac{\mu}{r_{\us}}\notag\\
  &\phantom{-\delta_{p}\delta_{f}
  \frac{L^{8}}{1080\mu^{8}}
  \frac{1}{\left(1+u^{2}\right)^{4}}\biggl\{}
  -12\biggl(u^{6}+4u^{4}+6u^{2}+8+\frac{5}{u^{2}}\biggr)\log\bigl(1+u^{2}\bigr)
  \biggr\}.
\end{align}
From this we see that the minimal coordinate value reached by the $\uD
7$ is
\begin{equation}
\label{eq:min_flat_conifold_backreacting_D7s}
  r_{\mathrm{min}}=\mu\biggl\{1-\delta_{p}\frac{L^{8}}{5\mu^{8}}
  \biggl[1+\delta_{p}\frac{259L^{8}}{225\mu^{8}}
  +\delta_{f}\frac{4}{3}
  \biggl(\log\frac{\mu}{r_{\us}}+\frac{11}{36}\biggr)
  \biggr]\biggr\}.
\end{equation}
Hence whether or not the backreacting $\uD 7$s attract or repel the
probe depends on the ratio $\frac{\mu}{r_{\us}}$.  The backreacting
$\uD 7$s attract $\uD 7$ probes so long as $\mu\ge
r_{\us}\ue^{-11/36}$ but will repel branes that are closer.  However,
the repulsion will not cancel out the attraction from the $3$-branes
until $\mu$ is exponentially small,
$\log\frac{\mu}{r_{\us}}\sim-\frac{1}{\delta_{f}}$, at which point the
$\uD 7$s can no longer be treated as perturbation.  Our interpretation
is that the presence of the $\overline{\uD 3}$s causes the NS-NS
forces and R-R forces between $\uD 7$s to no longer balance and at
sufficiently small distances, the repulsive R-R forces dominate.  This
repulsion may indicate a possible instability of $\uD 7$-stacks in
this background; if one of the $\uD 7$s in the stacks fluctuates, then
the repulsion between the $\uD 7$s will reinforce the fluctuation and
push the $\uD 7$s apart.  However, in order to confirm this effect, we
would need to extend our analysis beyond perturbative treatment of the
$\uD 7$ backreaction and to include the backreaction of branes at
finite embedding parameter.  Moreover, we have imposed by hand that
the $\uD 7$s asymptote to the unperturbed embedding and hence any
fluctuation respecting this boundary condition will eventually relax
to the initial configuration.  In a more complete treatment, the
asymptotic boundary condition would be replaced by fluxes in a UV
completion of the geometry.  Such an effect must also be taken into
account to quantify the conditions for stability.

To be more precise about the response of the $\uD 7$, we should again
consider the proper distances. Due to the increased complexity of the
geometry we will not report the result here, but they do not
qualitatively change the above picture.

\section{\label{sec:conifold}Interactions on the conifold}

The previous section explored the behavior of $\uD 7$s in the presence
of both $\uD 3$-branes and $\overline{\uD 3}$-branes.  However, the
$\uD 3$-$\overline{\uD 3}$ pairs will annihilate into closed strings
and so a significant drawback of the construction is the absence of
stability.  A more realistic scenario is the one suggested
in~\cite{Kachru:2002gs} in which $\overline{\uD 3}$s are placed at the
bottom of a warped geometry sourced by fluxes rather than by integer
$\uD 3$-branes.  The best known example, and the one explicitly
considered in~\cite{Kachru:2002gs}, is the Klebanov-Strassler solution
which results from the backreaction of fractional $\uD 3$s in the
deformed conifold.  However, the KS geometry is a comparatively
complicated solution and even more so when the backreaction of
$\overline{\uD 3}$s are taken into
account~\cite{DeWolfe:2008zy,McGuirk:2009xx,Bena:2009xk,Bena:2010ze,Bena:2011hz,Bena:2011wh,Massai:2012jn,Bena:2012bk,Dymarsky:2011pm}.
Hence we will first consider the case of $\uD 3$s and $\overline{\uD
  3}$s located at a conifold singularity, deferring the discussion of
the metastable case to the next section.

The conifold\footnote{See~\cite{Candelas:1989js} for a review of the
  conifold and related geometries.} can be embedded into $C^{4}$ by
\begin{equation}
  z^{A}z^{A}=0,\quad z^{A=1,2,3,4}\in C.
\end{equation}
It admits a Calabi-Yau metric
\begin{align}
  \ud s_{6}^{2}=-\frac{3}{4r^{4}}z^{A}\ud\bar{z}^{\bar{A}}\,
  \bar{z}^{\bar{B}}\ud z^{B}
  +\frac{\sqrt{3}}{\sqrt{2}r}\ud z^{A}\ud\bar{z}^{\bar{A}},
\end{align}
in which
\begin{equation}
  z^{A}\bar{z}^{\bar{A}}=\left(\frac{2}{3}\right)^{3/2}r^{3}.
\end{equation}
The metric takes the form of a cone over the Einstein-Sasaki space
$T^{1,1}$
\begin{equation}
  \ud s_{6}^{2}=\ud r^{2}+r^{2}\ud s_{T^{1,1}}^{2}.
\end{equation}
$T^{1,1}$ can in turn be written as an $S^{1}$ fibration over the
Einstein-K\"ahler space $S^{2}\times S^{2}$.  Indeed, writing
\begin{subequations}
\label{eq:conifold_coordinates}
\begin{align}
  z^{1}=&\left(\frac{2}{27}\right)^{1/4}r^{3/2}\ue^{\ui\psi/2}
  \biggl[
  \cos\biggl(\frac{\theta^{1}+\theta^{2}}{2}\biggr)
  \cos\biggl(\frac{\phi^{1}+\phi^{2}}{2}\biggr)
  +\ui
  \cos\biggl(\frac{\theta^{1}-\theta^{2}}{2}\biggr)
  \sin\biggl(\frac{\phi^{1}+\phi^{2}}{2}\biggr)\biggr],\\
  z^{2}=&\left(\frac{2}{27}\right)^{1/4}r^{3/2}\ue^{\ui\psi/2}
  \biggl[
  -\cos\biggl(\frac{\theta^{1}+\theta^{2}}{2}\biggr)
  \sin\biggl(\frac{\phi^{1}+\phi^{2}}{2}\biggr)
  +\ui
  \cos\biggl(\frac{\theta^{1}-\theta^{2}}{2}\biggr)
  \cos\biggl(\frac{\phi^{1}+\phi^{2}}{2}\biggr)\biggr],\\
  z^{3}=&\left(\frac{2}{27}\right)^{1/4}r^{3/2}\ue^{\ui\psi/2}
  \biggl[
  -\sin\biggl(\frac{\theta^{1}+\theta^{2}}{2}\biggr)
  \cos\biggl(\frac{\phi^{1}-\phi^{2}}{2}\biggr)
  +\ui
  \sin\biggl(\frac{\theta^{1}-\theta^{2}}{2}\biggr)
  \sin\biggl(\frac{\phi^{1}-\phi^{2}}{2}\biggr)\biggr],\\
  z^{4}=&\left(\frac{2}{27}\right)^{1/4}r^{3/2}\ue^{\ui\psi/2}
  \biggl[
  -\sin\biggl(\frac{\theta^{1}+\theta^{2}}{2}\biggr)
  \sin\biggl(\frac{\phi^{1}-\phi^{2}}{2}\biggr)
  -\ui
  \sin\biggl(\frac{\theta^{1}-\theta^{2}}{2}\biggr)
  \cos\biggl(\frac{\phi^{1}-\phi^{2}}{2}\biggr)\biggr],
\end{align}
\end{subequations}
in which $\theta^{i}\in\left[0,\pi\right]$,
$\phi^{i}\in\left[0,2\pi\right)$, and
$\psi\in\left[0,4\pi\right)$, the metric takes the form
\begin{equation}
  \ud s_{6}^{2}=\ud r^{2}
  +r^{2}\biggl(\frac{1}{3}\ud\psi+\cA\biggr)^{2}
  +r^{2}\ud s^{2}_{S^{2}\times S^{2}},
\end{equation}
where
\begin{equation}
  \cA=\frac{1}{3}\sum_{i=1}^{2}\cos\theta^{i}\ud\phi^{i},
\end{equation}
and
\begin{equation}
  \ud s^{2}_{S^{2}\times S^{2}}=\frac{1}{6}\sum_{i=1}^{2}
  \bigl[\bigl(\ud\theta^{i}\bigr)^{2}+\sin^{2}\theta^{i}
  \bigl(\ud\phi^{i}\bigr)^{2}\bigr].
\end{equation}
The K\"ahler form for the conifold is
\begin{equation}
  J=r\,\ud r\wedge\biggl(\frac{1}{3}\ud\psi+\cA\biggr)
  +r^{2}\cJ,
\end{equation}
where
\begin{align}
  \cJ=
  \frac{\sqrt{3}\ui}{2\sqrt{2}r}
  z^{A}\wedge\bar{z}^{\bar{A}}
  -\frac{9\ui}{8r^{4}}
  \bar{z}^{\bar{A}}\ud z^{A}\wedge z^{B}\ud\bar{z}^{\bar{B}}
  =-\frac{1}{6}\sum_{i=1}^{2}\sin\theta^{i}\ud\theta^{i}\wedge\ud\phi^{i}
  =\frac{1}{2}\ud\cA.
\end{align}
is the K\"ahler form on $S^{2}\times S^{2}$.

\begin{figure}[t]
\begin{center}
  \includegraphics[scale=.5]{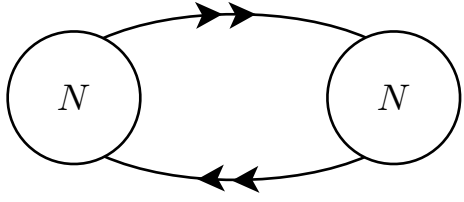}
  \caption{\label{fig:KW_quiver}Quiver for the Klebanov-Witten gauge
    theory.}
\end{center}
\end{figure}

Due to the common conical structure, the backreaction of $N+p$ $\uD 3$
branes with $p$ $\overline{\uD 3}$-branes is captured by
\begin{equation}
\label{eq:conifold_antibranes}
  \ud s_{6}^{2}=\ue^{2B}\bigl(\ud r^{2}+r^{2}\ud s_{T^{1,1}}^{2}\bigr),
\end{equation}
where $B$, the warp factor, and the $C_{4}$ potential are again given
by~\eqref{eq:anti_D3s_in_cy_cone}.  The $p=0$ case is the
Klebanov-Witten solution~\cite{Klebanov:1998hh} while the finite $p$
case was first considered in~\cite{DeWolfe:2008zy}.  The
Klebanov-Witten geometry is dual to an $\cN=1$ $\SU{N}\times \SU{N}$
conformal gauge theory summarized by the quiver in
figure~\ref{fig:KW_quiver}.

\subsection{\label{subsec:k_probes}Kuperstein probes}

\begin{figure}[t]
\begin{center}
  \includegraphics[scale=.15]{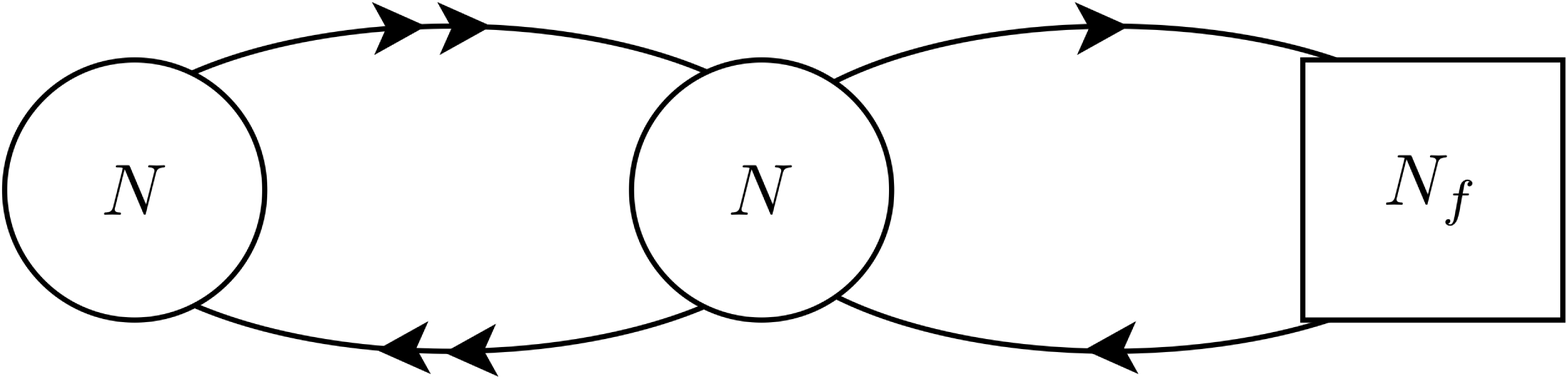}
  \caption{\label{fig:flavoredKW_quiver}Quiver for the Klebanov-Witten
    gauge theory flavored by Kuperstein quarks.}
\end{center}
\end{figure}

Just as in the $AdS^{5}\times S^{5}$ case, a probe $\uD 7$ brane will
be supersymmetric if it is holomorphically embedded into the
Calabi-Yau.  Here we focus on the Kuperstein
embedding~\cite{Kuperstein:2004hy}
\begin{equation}
  \label{eq:Kuperstein_embedding}
  z^{4}=\left(\frac{2}{27}\right)^{1/4}\mu.
\end{equation}
Although our analysis could be extended to Ouyang
embeddings~\cite{Ouyang:2003df}
\begin{equation}
  z^{3}+\ui z^{4}=\mu,
\end{equation}
such embeddings are not supersymmetric in the KS geometry without the
presence of worldvolume flux~\cite{Benini:2007kg,Chen:2008jj} and so
we will not consider them here.  Upon adding $N_{f}$ such branes, the
dual gauge theory exhibits a $\U{N_{f}}$ flavor group such that
resulting quiver is shown in figure~\ref{fig:flavoredKW_quiver}.

It is
convenient to write the conifold as a foliation of Kuperstein
embeddings, similar to what was written in~\cite{Benini:2009ff}
\begin{subequations}
\label{eq:Kuperstein_embedding_foliation_coordinates}
\begin{align}
  z^{1}=&\ui\left(\frac{2}{27}\right)^{1/4}
  \mu\bigl(1+\chi\bigr)\ue^{\ui\zeta}
  \biggl[\cos\phi\cosh\biggl(\frac{t+\ui\ga}{2}\biggr)
  \cos\theta-\ui\sin\phi\sinh\biggl(\frac{t+\ui\ga}{2}\biggr)\biggr],\\
  z^{2}=&\ui\left(\frac{2}{27}\right)^{1/4}
  \mu\bigl(1+\chi\bigr)\ue^{\ui\zeta}
  \biggl[\sin\phi\cosh\biggl(\frac{t+\ui\ga}{2}\biggr)
  \cos\theta+\ui\cos\phi\sinh\biggl(\frac{t+\ui\ga}{2}\biggr)\biggr],\\
  z^{3}=&\ui\left(\frac{2}{27}\right)^{1/4}
  \mu\bigl(1+\chi\bigr)\ue^{\ui\zeta}
  \cosh\biggl(\frac{t+\ui\ga}{2}\biggr)\sin\theta,\\
  z^{4}=&\left(\frac{2}{27}\right)^{1/4}\mu\left(1+\chi\right)\ue^{\ui\zeta},
\end{align}
\end{subequations}
in which $\chi\in\left[-1,\infty\right)$,
$\zeta\in\left[0,2\pi\right)$, $t\in\left[0,\infty\right)$,
$\theta\in\left[0,\pi\right]$, $\phi\in\left[0,2\pi\right)$, and
$\ga\in\left[0,4\pi\right)$.  Our Kuperstein
embedding~\eqref{eq:Kuperstein_embedding} is then specified by
$\chi=\zeta=0$.  The embedding retains an $\SO{3}$ invariance under
which $z^{A=1,2,3}$ are rotated among each other. $t$ is an
$\SO{3}$-invariant radial-like coordinate satisfying\footnote{The
  awkward factors of $\left(\frac{2}{27}\right)^{1/4}$ appearing
  in~\eqref{eq:Kuperstein_embedding}
  and~\eqref{eq:Kuperstein_embedding_foliation_coordinates} were
  chosen so that this expression has no numerical prefactor.}
\begin{equation}
  r=\mu^{2/3}\left(1+\chi\right)^{2/3}\cosh^{2/3}\frac{t}{2}.
\end{equation}
It is then useful to define
\begin{equation}
  u=\cosh^{2/3}\frac{t}{2},
\end{equation}
so that $u\in\left[1,\infty\right)$.  We also use the $\SO{3}$
left-invariant $1$-forms~\cite{Benini:2009ff}
\begin{subequations}
\begin{align}
  h^{1}=&2\biggl(\cos\frac{\ga}{2}\ud\theta
  -\sin\frac{\ga}{2}\sin\theta\ud\phi\biggr),\\
  h^{2}=&2\biggl(\sin\frac{\ga}{2}\ud\theta
  +\cos\frac{\ga}{2}\sin\theta\ud\phi\biggr),\\
  h^{3}=&\ud\ga-2\cos\theta\ud\phi.
\end{align}
\end{subequations}

In terms of the foliation coordinates, the
metric~\eqref{eq:conifold_antibranes} takes the form
\begin{align}
  \ud s_{6}^{2}=\ue^{2B}\mu^{4/3}\bigl(1+\chi\bigr)^{4/3}\biggl\{&
  \frac{4u^{3}-1}{4u^{3}-4}\ud u^{2}
  +\frac{u^{2}}{9}\biggl(1-\frac{1}{4u^{3}}\biggr)
  \bigl(h^{3}\bigr)^{2}+\frac{u^{2}}{12}\bigl(h^{1}\bigr)^{2}
  +\frac{u^{2}}{12}\biggl(1-\frac{1}{u^{3}}\biggr)
  \bigl(h^{2}\bigr)^{2}\notag\\
  &
  +\frac{4u}{3\left(1+\chi\right)}\ud u\,\ud\chi
  +\frac{4u^{2}}{9}\sqrt{1-\frac{1}{u^{3}}}
  h^{3}\ud\zeta
  +\frac{4u^{2}}{9}
  \biggl(\frac{\ud\chi^{2}}{\left(1+\chi\right)^{2}}
  +\ud\zeta^{2}\biggr)\biggr\}.
\end{align}
The resulting DBI action is then
\begin{equation}
  S_{\uD 7}^{\mathrm{DBI}}=-\frac{\tau_{\uD 7}\mu^{8/3}}{144}
  \int\ud^{4}x\,\ud u\,\ud^{3}h\,
  \sqrt{W_{0}+W_{1}\chi'+W_{2}\chi'^{2}+Y\zeta'^{2}+H_{ij}a'_{h^{i}}
    a'_{h^{j}}},
\end{equation}
in which
\begin{subequations}
\begin{align}
  W_{0}=&\left(4u^{3}-1\right)^{2}\ue^{8B}
  \left(1+\chi\right)^{16/3},\\
  W_{1}=&\frac{16u\left(4u^{3}-1\right)\left(u^{3}-1\right)}{3}\ue^{8B}
  \left(1+\chi\right)^{13/3},\\
  W_{2}=&\frac{16u^{2}\left(4u^{3}-1\right)\left(u^{3}-1\right)}{9}\ue^{8B}
  \left(1+\chi\right)^{10/3},\\
  Y=&\frac{16}{3}u^{2}\left(u^{3}-1\right)\ue^{8B}\left(1+\chi\right)^{16/3},\\
  H_{11}=&\frac{\left(2\pi\alpha'\right)^{2}}{\mu^{8/3}}
  \frac{48\left(4u^{3}-1\right)\left(u^{3}-1\right)}{u^{2}}
  \ue^{4A+4B}
  \left(1+\chi\right)^{8/3},\\
  H_{22}=&\frac{\left(2\pi\alpha'\right)^{2}}{\mu^{8/3}}
  {48u^{3}\left(4u^{3}-1\right)}\ue^{4A+4B}
  \left(1+\chi\right)^{8/3},\\
  H_{33}=&\frac{\left(2\pi\alpha'\right)^{2}}{\mu^{8/3}}{144u\left(u^{3}-1\right)}
  \ue^{4A+4B}
  \left(1+\chi\right)^{8/3},
\end{align}
\end{subequations}
with other components of $H_{ij}$ vanishing.

\begin{figure}[t]
\begin{center}
  \includegraphics[scale=1]{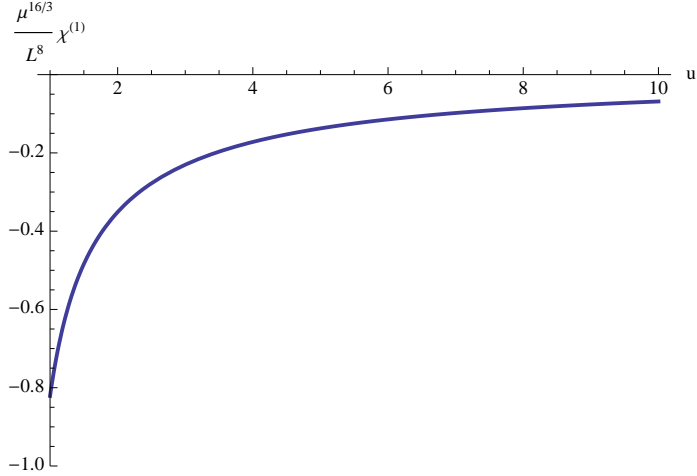}
  \caption{\label{fig:kuperstein_embedding}Profile of a would-be
    Kuperstein $\uD 7$ probe of $AdS^{5}\times T^{1,1}$ perturbed by
    $p\ll N$ $\uD 3$-$\overline{\uD 3}$ pairs.  The curve is specified
    by~\eqref{eq:kuperstein_embedding_perturbed}. Note that the kink
    at $u=1$ is an artifact of the coordinate system used.}
\end{center}
\end{figure}

The equations of motion for $\zeta$ and $a_{h^{i}}$ are satisfied by
setting $a_{h^{i}}=\zeta=0$.  Setting those fields on shell, the
action simplifies to
\begin{equation}
  S_{\uD 7}=-\frac{\tau_{\uD 7}\mu^{8/3}}{144} \int\ud^{4}x\,\ud u\,
  \ud^{3}h
  \sqrt{W_{0}+W_{1}\chi'+W_{2}\chi'^{2}}.
\end{equation}
The resulting equation of motion and boundary conditions again take
the forms~\eqref{eq:chieom} and~\eqref{eq:chibc} after setting
$W_{3}=W_{4}=0$ and replacing the lower bound of the boundary
condition with $u=1$.  Using that $\chi=0$ is a solution at leading
order in $\frac{p}{N}$, the equation of motion at
$\cO\left(\frac{p}{N}\right)$ is
\begin{equation}
  0=\partial_{u}\biggl[\frac{u^{2}\left(u^{3}-1\right)}{4u^{3}-1}
  \partial_{u}\chi^{\left(1\right)}\biggr]
  -\frac{6L^{8}}{5\mu^{16/3}}\frac{1}{u^{8}},
\end{equation}
where we have again written $\chi=\sum_{n}\left(\frac{p}{N}\right)^{n}
\chi^{\left(n\right)}$.  The particular solution satisfying the
boundary conditions is
\begin{align}
  \label{eq:kuperstein_embedding_perturbed}
  \chi^{\left(1\right)}=-\frac{3L^{8}}{700\mu^{16/3}}
  \biggl\{\frac{40u^{7}+60u^{6}+24u^{3}-5}{u^{8}}
  +\frac{240}{\sqrt{3}}\arctan\frac{\sqrt{3}}{1+2u}\biggr\}.
\end{align}
This is plotted in figure~\ref{fig:kuperstein_embedding}.
  
The corresponding minimal radius occurs at $u=1$ and is given by
\begin{equation}
  \label{eq:min_distance_conifold}
  r_{\mathrm{min}}=\mu^{2/3}\biggl[1-\frac{p}{N}
  \frac{357+40\sqrt{3}\pi}{150}
  \frac{L^{8}}{\mu^{16/3}}\biggr].
\end{equation}
Given that $\mu$ carries different dimensions than the same symbol did
in the flat space embedding~\eqref{eq:flat_space_embedding}, this is
qualitatively the same behavior as in the flat space case.
Correspondingly, the distance between two such branes and change in
energy takes the same qualitative forms as
\eqref{eq:interbrane_distance_d7} and~\eqref{eq:change_in_energy}.

By considering the large $u$ behavior when $p=0$, we can show that
$\chi$ is dual to an operator with mass-dimension $\frac{5}{2}$ and
again corresponds to a mass term for the dual quarks.  $\chi$ is
related to a canonically normalized field $\Phi$ such that at large
$u$, $\Phi\sim \chi\, u^{-3/2}$, consistent with the $u^{-1}$ falloff
of the perturbed embedding.

\subsection{Backreacting Kupersteins}

As reviewed in appendix~\ref{subsec:backreacting_Kuperstein}, the
backreaction of smeared massless Kupersteins in $AdS^{5}\times
T^{1,1}$ takes the same form as the smeared branes in $AdS^{5}\times
S^{5}$ does, except that for the case of Kuperstein embeddings,
$\delta_{f}=\frac{3N_{f}}{4\pi}$.  Therefore the forces experienced by
probe $3$-branes in the presence of backreacting will be the same as
in section~\ref{subsec:d3_probes_flat_flavored}.  In this section, we
will therefore focus on the interaction between Kuperstein branes.

Expressed in terms of the
coordinates~\eqref{eq:Kuperstein_embedding_foliation_coordinates}, the
metric resulting from the backreaction of $\uD 7$s takes the form
\begin{align}
  \ud s_{6}^{2}=
  &\ue^{2B}\bigl\{\ud r^{2}+r^{2}\bigl(\frac{1}{3}\ud\psi+\cA\bigr)^{2}\bigr\}
  +\ue^{2C}r^{2}\ud s_{S^{2}\times S^{2}}^{2}\notag\\
  =&-\frac{3}{4r^{4}}\left(3\ue^{2C}-2\ue^{2B}\right)
  \bar{z}^{\bar{A}}\ud z^{A}
  z^{B}\ud\bar{z}^{\bar{B}}
  +\frac{\sqrt{3}}{\sqrt{2}r}\ue^{2C}
  \ud z^{A}\ud\bar{z}^{\bar{A}}\notag\\
  =&
  \mu^{4/3}\bigl(1+\chi\bigr)^{4/3}
  \biggl\{\biggl(\ue^{2B}+\frac{3}{4\left(u^{3}-1\right)}\ue^{2C}\biggr)
  \biggl(\ud u^{2}+\frac{u^{3}-1}{9u}\bigl(h^{3}\bigr)^{2}\biggr)
  \notag\\
  &\phantom{\mu^{4/3}\bigl(1+\chi\bigr)^{4/3}
  \biggl\{}+\frac{u^{2}}{12}\ue^{2C}\bigl(h^{1}\bigr)^{2}
  +\frac{u^{2}}{12}\biggl(1-\frac{1}{u^{3}}\biggr)\ue^{2C}
  \bigl(h^{2}\bigr)^{2}\notag\\
  &\phantom{\mu^{4/3}\bigl(1+\chi\bigr)^{4/3}
  \biggl\{}+\frac{4u}{3\left(1+\chi\right)}\ue^{2B}\ud u\,\ud\chi
  +\frac{4u^{2}}{9}\sqrt{1-\frac{1}{u^{3}}}
  \ue^{2B}h^{3}\ud\zeta\notag\\
  &\phantom{\mu^{4/3}\bigl(1+\chi\bigr)^{4/3}
  \biggl\{}+\frac{4u^{2}}{9}\ue^{2B}
  \biggl(\frac{\ud\chi^{2}}{\left(1+\chi\right)^{2}}+
  \ud\zeta^{2}\biggr)\biggr\}.\label{eq:squashed_conifold_metric}
\end{align}
Again imposing that the open-string fields depend only on $u$, the
induced metric on the $4$-cycle is found by making the replacements
$\ud\chi\to \chi'\ud u$ and $\ud\zeta\to\zeta'\ud u$.  The DBI
Lagrangian is then
\begin{equation}
  S_{\uD 7}^{\mathrm{DBI}}=-\frac{\tau_{\uD 7}\mu^{8/3}}{144}\int\,\ud^{4}x\,
  \ud u\,\ud^{3}h\,
  \sqrt{W_{0}+W_{1}\chi'+W_{2}\chi'^{2}+Y\zeta'^{2}
    +H_{ij}a'_{h^{i}}a'_{h^{j}}},
\end{equation}
in which
\begin{subequations}
\begin{align}
  W_{0}=&
  g_{\us}^{-2}\ue^{4C+2\phi}\left(3\ue^{2C}+4\left(u^{3}-1\right)\ue^{2B}\right)^{2}
  \left(1+\chi\right)^{16/3}
  ,\\
  W_{1}=&\frac{16}{3}u\bigl(u^{3}-1\bigr)g_{\us}^{-2}\ue^{4C+2B+2\phi}
  \bigl(3\ue^{2C}+4\bigl(u^{3}-1\bigr)\ue^{2B}\bigr)\left(1+\chi\right)^{13/3},\\
  W_{2}=&\frac{16}{9}\bigl(u^{3}-1\bigr)u^{2}
  g_{\us}^{-2}\ue^{4C+2B+2\phi}
  \bigl(3\ue^{2C}+4\bigl(u^{3}-1\bigr)\ue^{2B}\bigr)\left(1+\chi\right)^{10/3},\\
  Y=&\frac{16}{3}\bigl(u^{3}-1\bigr)u^{2}g_{\us}^{-2}\ue^{6C+2B+2\phi}
  \bigl(1+\chi\bigr)^{16/3},\\
  H_{11}=&\frac{\left(2\pi\alpha'\right)^{2}}{\mu^{8/3}}
  \frac{48\left(u^{3}-1\right)}{u^{2}}g_{\us}^{-1}\ue^{4A+2C+\phi}
  \bigl(3\ue^{2C}+4\bigl(u^{3}-1\bigr)\ue^{2B}\bigr)
  \bigl(1+\chi\bigr)^{8/3},\\
  H_{22}=&\frac{\left(2\pi\alpha'\right)^{2}}{\mu^{8/3}}
  48u\,g^{-1}_{\us}\ue^{4A+2C+\phi}
  \bigl(3\ue^{2C}+4\bigl(u^{3}-1\bigr)\ue^{2B}\bigr)\bigl(1+\chi\bigr)^{8/3}
  ,\\
  H_{33}=&\frac{\left(2\pi\alpha'\right)^{2}}{\mu^{8/3}}
  144\bigl(u^{3}-1\bigr)u\,g^{-1}_{\us}\ue^{4A+4C+\phi}\bigl(1+\chi\bigr)^{8/3},
\end{align}
\end{subequations}
with other components of $H_{ij}$ vanishing.

The $8$-form potential again takes the form~\eqref{eq:8form_pot}.
Since $\ud\psi$ is closed we can write this as
\begin{equation}
  C_{8}=Q_{f}\frac{r^{3}}{4}g_{\us}^{-2}\ue^{2\phi+4C}
  \ud\vol_{R^{3,1}}\wedge\ud r\wedge\bigl(\frac{1}{3}\ud\psi+\cA\bigr)\wedge\cJ.
\end{equation}
Using
\begin{equation}
  \ud r\wedge\bigl(\frac{1}{3}\ud\psi+\cA\bigr)
  =\frac{3\ui}{4r^{5}}\bar{z}^{\bar{A}}\ud z^{A}\wedge z^{B}
  \ud\bar{z}^{\bar{B}},\qquad
  \cJ=\frac{\ui\sqrt{3}}{2\sqrt{2}r^{3}}
  \ud z^{A}\wedge\ud\bar{z}^{\bar{A}}
  -\frac{9}{8r^{6}}
  \bar{z}^{\bar{A}}\ud z^{A}\wedge z^{B}
  \ud\bar{z}^{\bar{B}},
\end{equation}
we have
\begin{align}
  C_{8}=&Q_{f}\frac{\mu^{8/3}\left(1+\chi\right)^{2}u^{3}}
  {4}
  g_{\us}^{-2}\ue^{2\phi+4C}\ud\vol_{R^{3,1}}\notag\\
  &\wedge
  \biggl\{\frac{\left(1+\chi\right)^{2/3}}{36}
  \biggl[\biggl(1-\frac{1}{u^{3}}\biggr)\ud u
  +\frac{2}{3\left(1+\chi\right)}
  \frac{u^{3}-1}{u^{2}}\ud\chi\biggr]\wedge h^{1}\wedge h^{2}\wedge h^{3}\notag\\
  &\phantom{\wedge
  \biggl\{}-\frac{1}{9\left(1+\chi\right)^{1/3}u^{3/2}\sqrt{u^{3}-1}}
  \ud u\wedge\ud\zeta\wedge\ud\chi\wedge h^{3}\notag\\
  &\phantom{\wedge
  \biggl\{}
  +\frac{\left(1+\chi\right)^{2/3}}{18}
  \biggl[\sqrt{1-\frac{1}{u^{3}}}\ud u+\frac{2}{3\left(1+\chi\right)}
  \sqrt{\frac{u^{3}-1}{u}}\ud\chi\biggr]\wedge\ud\zeta\wedge h^{1}
  \wedge h^{2}\biggr\}.
\end{align}
When pulling back to the worldvolume, we make the replacement
$\ud\chi\to \chi'\ud u$ and $\ud\zeta\to\zeta'\ud u$ and hence
\begin{multline}
  \mathrm{P}\bigl[C_{8}\bigr]=
  Q_{f}\frac{\mu^{8/3}\left(1+\chi\right)^{8/3}}{144}
  g_{\us}^{-2}\ue^{2\phi+4C}\bigl(u^{3}-1\bigr) \biggl[
  1+\frac{2u\chi'}{3\left(1+\chi\right)}
  \biggr]\,\ud\vol_{R^{3,1}}\wedge \ud u\wedge h^{1}\wedge h^{2}\wedge
  h^{3}.
\end{multline}
We can again use the calibration condition in the supersymmetric case
to determine the orientation.  When $p=0$,
\begin{equation}
  \label{eq:Kuperstein_orientation}
  \ud\vol_{\Sig^{4}}=-\frac{1}{2}\mathrm{P}\bigl[J\wedge J\bigr]
  \sim-\ud u\wedge h^{1}\wedge h^{2}\wedge h^{3},
\end{equation}
and so writing
\begin{equation}
  S_{\uD 7}^{\mathrm{CS}}=-\frac{\tau_{\uD 7}\mu^{8/3}}{144}
  \int\ud^{4}x\,\ud u\,\ud^{3}h\,\bigl\{W_{3}+W_{4}\chi'\bigr\},
\end{equation}
we have
\begin{subequations}
\begin{align}
  W_{3}=&Q_{f}\bigl(u^{3}-1\bigr)g_{\us}^{-1}\ue^{2\phi+4C}\left(1+\chi\right)^{8/3}
  ,\\
  W_{4}=&\frac{2}{3}Q_{f}u\bigl(u^{3}-1\bigr)g_{\us}^{-1}\ue^{2\phi+4C}
  \left(1+\chi\right)^{5/3}.
\end{align}
\end{subequations}

The equations of motion for the gauge field and $\zeta$ are solved by
taking $a_{h^{i}}=0$ and $\zeta=0$ while for $\chi$ we again
have~\eqref{eq:chieom} and~\eqref{eq:chibc}. The resulting solution is
a little cumbersome, but for large $u$,
\begin{align}
  \chi\sim-\delta_{p}\frac{24L^{8}}{35\mu^{16/3}u}
  -\delta_{p}^{2}\frac{96\left(88+15\sqrt{3}\pi-45\log 3\right)L^{16}}
  {6125\mu^{32/3}u}
  +\delta_{p}\delta_{f}\frac{8L^{8}\log u}{35\mu^{16/3} u}.
\end{align}
This is again qualitatively the same analytic behavior that we
encountered in the $N_{f}=0$ case, aside from the logarithmic running
which is a result of the non-trivial RG flow induced by the
backreacting flavors.  The minimum radius obtained is
\begin{equation}
  \label{eq:min_radius_conifold_backreacting_D7s}
  r_{\mathrm{min}}=\mu^{2/3}\biggl\{1-\delta_{p}\frac{aL^{8}}{\mu^{16/3}}
  -\delta_{p}^{2}\frac{bL^{16}}{\mu^{32/3}}
  -\delta_{p}\delta_{f}\frac{L^{8}}{\mu^{16/3}}
  \biggl[c\log\frac{\mu^{2/3}}{r_{\us}}+d\biggr]\biggr\},
\end{equation}
where the coefficients can be expressed precisely but have the
approximate values
\begin{equation}
  a\approx 0.547,\quad
  b\approx 1.90,\quad
  c\approx 0.730,\quad
  d\approx 0.961.
\end{equation}
The qualitative behavior is similar to the flat space case: for
$\mu\ge r_{\us}\ue^{-d/c}$ the backreacting flavors add to the
attraction, while for smaller $\mu$, the attraction is reduced.

\section{\label{sec:DKM}D7s and anti-D3s in Klebanov-Tseytlin}

As discussed previously, the above examples are unstable in that the
spectrum of open strings stretching from a $\uD 3$-brane to an
$\overline{\uD 3}$-brane contains a tachyon indicating that the branes
will annihilate into closed strings.  An alternative and metastable
construction is that of~\cite{Kachru:2002gs} in which $\overline{\uD
  3}$s are added to a geometry warped by the presence of $3$-form flux
rather than $\uD 3$-branes

The Klebanov-Tseytlin (KT) solution~\cite{Klebanov:2000nc}
accomplishes this by considering the presence of fractional $\uD
3$-branes resulting from $M$ $\uD 5$s wrapping a collapsing $2$-cycle
of the conifold.  The resulting dual theory is best described in terms
of a cascade of self-similar Seiberg dualities where at each stage the
quiver takes a form similar to that in figure~\ref{fig:KW_quiver},
except at each stage in the cascade the difference in rank between the
two nodes is $M$.  In terms of the angles appearing
in~\eqref{eq:conifold_coordinates}, it is useful to define the
$1$-forms
\begin{align}
  e^{1}=&-\sin\theta^{1}\ud\vp^{1},&
  e^{2}=&\ud\theta^{1},\notag\\
  e^{3}=&\cos\psi\sin\theta^{2}\ud\vp^{2}-\sin\psi\ud\theta^{2},&
  e^{4}=&\sin\psi\sin\theta^{2}\ud\vp^{2}+\cos\psi\ud\theta^{2},\\
  e^{5}=&\ud\psi+\cos\theta^{1}\ud\vp^{1}+\cos\theta^{2}\ud\vp^{2}.\notag
\end{align}
and then~\cite{Minasian:1999tt}
\begin{equation}
  g^{1,3}=\frac{1}{\sqrt{2}}\bigl(e^{1}\mp e^{3}\bigr),\quad
  g^{2,4}=\frac{1}{\sqrt{2}}\bigl(e^{2}\mp e^{4}\bigr),\quad
  g^{5}=e^{5}.
\end{equation}
In terms of these forms, the $3$-form flux resulting from the
fractional $\uD 3$s is
\begin{equation}
  F_{3}=\frac{\alpha' M}{4}g^{5}\wedge\bigl(g^{1}\wedge g^{2}
  +g^{3}\wedge g^{4}\bigr),\quad
  H_{3}=\frac{3\alpha' g_{\us} M}{4r}
  \ud r\wedge\bigl(g^{1}\wedge g^{2}
  +g^{3}\wedge g^{4}\bigr).
\end{equation}
Such flux is imaginary self dual (ISD) in that $\ast G_{3}=\ui G_{3}$
where $G_{3}=F_{3}-\ui\ue^{-\phi}H_{3}$ and $\ast$ is the unwarped
Hodge-$\ast$ built from the 6d metric
\begin{equation}
  \ud s_{6}^{2}=\ud r^{2}+\frac{r^{2}}{9}\bigl(g^{5}\bigr)^{2}
  +\frac{r^{2}}{6}\sum_{i=1}^{4}\bigl(g^{i}\bigr)^{2}.
\end{equation}
Since the background is ISD, we have $\Phi_{-}=0$ and
\begin{equation}
  \ue^{-4A}=\frac{27\pi\alpha' g_{\us}}{4r^{4}}
  \biggl\{N+\frac{3g_{\us}M^{2}}{2\pi
  }\biggl[\log\frac{r}{r_{0}}+\frac{1}{4}\biggr]\biggr\}.
\end{equation}
The naked singularity exhibited by this solution is resolved by the
deformation of the conifold, the effects of which become important at
$r\approx r_{0}$.  The resulting supergravity solution is the
Klebanov-Strassler geometry~\cite{Klebanov:2000hb}.

The backreaction of $\overline{\uD 3}$s on this geometry was first
considered in~\cite{DeWolfe:2008zy} (and is known as the DKM
solution).  The result is a non-ISD flux and a squashed metric
\begin{subequations}
\label{eq:DKM_soln}
\begin{align}
  \ud s_{6}^{2}=&\ue^{2B}\biggl\{
  \ud r^{2}+\frac{r^{2}}{9}\bigl(g^{5}\bigr)^{2}\biggr\}
  +\frac{r^{2}}{6}\ue^{2C}\sum_{i=1}^{4}\bigl(g^{i}\bigr)^{2},\\
  \ue^{2B}=&1+\frac{3\alpha'^{2}\cS}{4r^{4}},\\
  \ue^{2C}=&1-\frac{\alpha'^{2}\cS}{4r^{4}},\\
  F_{3}=&\frac{\alpha' M}{4}g^{5}\wedge\bigl(g^{1}\wedge g^{2}
  +g^{3}\wedge g^{4}\bigr),\\
  H_{3}=&\alpha'\beta'
  \ud r\wedge\bigl(g^{1}\wedge g^{2}
  +g^{3}\wedge g^{4}\bigr),\\
\intertext{\newpage}
  \beta\bigl(r\bigr)=&\frac{3g_{\us}M}{4}
  \log\frac{r}{r_{0}}
  +\frac{3\pi\alpha'^{2}\cS}{8Mr^{4}}
  \biggl\{N+\frac{3g_{\us}M^{2}}{\pi}\biggl[\log\frac{r}{r_{0}}
  +\frac{3}{8}\biggr]\biggr\},\\
  \omega^{-1}=\ue^{-4A}=&\frac{27\pi g_{s}\alpha'^{2}}{4r^{4}}
  \biggl\{N+\frac{3g_{\us}M^{2}}{2\pi}
  \biggl[\log\frac{r}{r_{0}}+\frac{1}{4}\biggr]\biggr\}\notag\\
  &+\frac{135\pi g_{\us}\alpha'^{4}\cS}{32r^{8}}
  \biggl\{N+\frac{12g_{\us}M^{2}}{5\pi}
  \biggl[\log\frac{r}{r_{0}}+\frac{13}{12}\biggr]\biggr\},\\
  \ue^{-\phi}=&\frac{1}{g_{\us}}+\frac{3\alpha'^{2}\cS}{g_{\us}r^{4}}
  \log\frac{r}{r_{0}},
\end{align}
\end{subequations}
in which $\cS\sim\frac{p}{N}\ue^{-8\pi
  N/3g_{\us}M^{2}}\frac{r_{0}^{4}}{\alpha'^{2}}$ and we have again
written the metric as~\eqref{eq:squashed_conifold_metric}.  Note that
the form of our solution differs from the original DKM solution
in~\cite{DeWolfe:2008zy} by an $\cO\left(1\right)$ redefinition of
$\cS$ and a redefinition of the radial coordinate\footnote{It was
  pointed out in~\cite{Benini:2009ff} that the DKM solution, as
  originally expressed in~\cite{DeWolfe:2008zy}, exhibits a
  non-Hermitian metric when expressed in terms of the unperturbed
  complex structure (i.e. the solution of~\cite{DeWolfe:2008zy} has
  $g_{zz}\neq 0$).  This feature is not exhibited
  by~\eqref{eq:squashed_conifold_metric} due to the different ansatz
  used.  This should not be seen as contrary to the expectation that
  non-supersymmetric sources will generically induce such
  non-Hermitian elements (see, e.g., the discussion
  in~\cite{McGuirk:2012sb}) as the coordinate redefinition that brings
  the solution~\cite{DeWolfe:2008zy} to the
  form~\eqref{eq:squashed_conifold_metric} is non-holomorphic.}. As
emphasized in~\cite{DeWolfe:2008zy, Bena:2009xk, Bena:2011wh}, the DKM
solution does not entirely capture the backreaction of an
$\overline{\uD 3}$-brane.  The most obvious deficiency is that
$\Phi_{-}=0$, while it is precisely in the $\Phi_{-}$ equation of
motion~\eqref{eq:IIB_eom} that explicit $\overline{\uD 3}$-branes
appear.  However, the consequences of a non-zero $\Phi_{-}$ fall off
faster than the DKM solution for large $r$
(see~\eqref{eq:anti_D3s_in_cy_cone}).  Less obvious is the fact that
the backreaction of $\overline{\uD 3}$s in the full KS solution have
modes that drop off more slowly than those appearing in
DKM~\cite{Bena:2009xk,Bena:2010ze,Bena:2011hz,Bena:2011wh,Massai:2012jn,Bena:2012bk,Dymarsky:2011pm}
but disappear as we take the deformation of the conifold to vanish.
For these reasons we will consider the analysis of a Kuperstein brane
in DKM as a warm-up to a more complete analysis in future work,
anticipating that much of the behavior will be qualitatively captured
by our analysis in DKM.  Note that in order to trust the DKM solution,
we must ensure that $r\gg r_{0}$ and $r\gg \cS^{1/4}\sqrt{\alpha'}$.

Consider a Kuperstein probe of the geometry.  We can write the NS-NS
potential as (see, e.g.~\cite{Herzog:2001xk})
\begin{equation}
  B_{2}=\alpha'\beta\bigl(g^{1}\wedge g^{2}+g^{3}\wedge g^{4}\bigr)
  =\frac{27\ui\alpha'\beta}{4r^{6}}\ep_{ABCD}
  z^{A}\bar{z}^{\bar{B}}\ud z^{C}\wedge\ud\bar{z}^{\bar{D}},
\end{equation}
where $\ep_{1234}=+1$.  In the $\cS=0$ case, the $\uD 7$ is
holomorphically embedded into the conifold and so
$\cF_{2}=\mathrm{P}\left[B_{2}\right]+2\pi\alpha' f_{2}$ is
$\left(1,1\right)$ when $f_{2}=0$.  Furthermore one can
show~\cite{Kuperstein:2004hy} that $\cF_{2}$ is primitive in the sense
that $\mathrm{P}\left[J\right]\wedge\cF_{2}=0$ where $J$ is the
K\"ahler form of the conifold.  Hence, a Kuperstein embedding is
supersymmetric in KT\footnote{In contrast, while
  $\mathrm{P}\left[B_{2}\right]$ is course still $\left(1,1\right)$
  for an Ouyang embedding, it is not primitive, and hence, as
  mentioned earlier, an Ouyang embedding must be magnetized in order
  to be supersymmetric~\cite{Benini:2007kg,Chen:2008jj}.}.

Consider a would-be Kuperstein probe of the DKM geometry.  Note that
again we must take $\mu$ to be sufficiently large to trust the
perturbative treatment of the $\overline{\uD 3}$s as well treat the
deformation of the conifold singularity as a negligible modification
of the geometry.  The presence of $3$-form flux modifies the DBI and
CS actions from what was considered in previous sections.  In terms of
the coordinates~\eqref{eq:Kuperstein_embedding_foliation_coordinates},
the NS-NS $2$-form takes the form
\begin{equation}
  B_{2}=-\frac{\alpha'\beta}{2u^{5/2}}\bigl\{3u\,\ud u\wedge h^{2}
  +h^{1}\wedge h^{3}\bigr\}.
\end{equation}
The DBI action then takes the form
\begin{equation}
  \label{eq:kuperstein_in_DKM_DBI}
  S_{\uD 7}^{\mathrm{DBI}}=-\frac{\tau_{\uD 7}\mu^{8/3}}{144}
  \int\ud^{4}x\,\ud u\,\ud^{3}h
  \sqrt{W_{0}+W_{1}\chi'+W_{2}\chi'^{2}
    +Y\zeta'^{2}
    +Y_{i}\zeta'a_{h^{i}}'
    +H_{i}a_{h^{i}}'
    +H_{ij}a'_{h^{i}}a'_{h^{j}}},
\end{equation}
in which
\begin{subequations}
\begin{align}
  W_{0}=&g_{\us}^{-2}\ue^{2\phi}
  \biggl[\bigl(1+\chi\bigr)^{8/3}
  \ue^{2C}\bigl(3\ue^{2C}+4\bigl(u^{3}-1\bigr)\ue^{2B}\bigr)
  +\frac{27\alpha'^{2}}{\mu^{8/3}}\frac{\beta^{2}}{u^{4}}
  g_{\us}\ue^{4A-\phi}\biggr]^{2},\\
  W_{1}=&\frac{16}{3}g_{\us}^{-2}\ue^{2\phi}\left(1+\chi\right)^{13/3}
  \ue^{4C+2B}u\bigl(u^{3}-1\bigr)\bigl(3\ue^{2C}+4\bigl(u^{3}-1\bigr)\ue^{2B}\bigr)
  \notag\\
  &+\frac{144\alpha'^{2}}{\mu^{8/3}}
  \left(1+\chi\right)^{5/3}\ue^{2B+2C}\beta^{2}\left(1-\frac{1}{u^{3}}\right)
  g_{\us}^{-1}\ue^{4A+\phi},\\
  W_{2}=&\frac{16}{9}g_{\us}^{-2}\ue^{2\phi}\left(1+\chi\right)^{10/3}
  \ue^{4C+2B}u^{2}
  \bigl(u^{3}-1\bigr)\bigl(3\ue^{2C}+4\bigl(u^{3}-1\bigr)\ue^{2B}\bigr)
  \notag\\
  &+\frac{48\alpha'^{2}}{\mu^{8/3}}\left(1+\chi\right)^{2/3}
  \left(u-\frac{1}{u^{2}}\right)
  \ue^{2B+2C}\beta^{2}g_{\us}^{-1}\ue^{4A+\phi},\\
  Y=&\frac{16}{3}g_{\us}^{-2}\ue^{2\phi}
  \bigl(1+\chi\bigr)^{16/3}\ue^{6C+2B}\bigl(u^{3}-1\bigr)u^{2}\notag\\
  &+\frac{48\alpha'^{2}}{\mu^{8/3}}\left(1+\chi\right)^{8/3}
  \left(u-\frac{1}{u^{2}}\right)
  \ue^{2B+2C}\beta^{2}g_{\us}^{-1}\ue^{4A+\phi},\\
  Y_{1}=&-\frac{192\left(2\pi\alpha'\right)\alpha'}{\mu^{8/3}}
  \left(1+\chi\right)^{8/3}\frac{\left(u^{3}-1\right)^{3/2}}{u^{2}}
  \ue^{2B+2C}\beta g_{\us}^{-1}\ue^{4A+\phi},\\
  H_{2}=&-\frac{72\left(2\pi\alpha'\right)\alpha'}{\mu^{8/3}u^{3/2}}
  \ue^{4A}\beta\biggl[\bigl(1+\chi\bigr)^{8/3}
  \bigl(3\ue^{2C}+4\bigl(u^{3}-1\bigr)\ue^{2B}\bigr)
  g_{\us}^{-1}\ue^{\phi}
  +\frac{27\alpha'^{2}}{\mu^{8/3}u^{4}}\beta^{2}\ue^{4A}\biggr],\\
  H_{11}=&\frac{48\left(2\pi\alpha'\right)^{2}}{\mu^{8/3}}
  \bigl(1+\chi\bigr)^{8/3}\frac{\left(u^{3}-1\right)}{u^{2}}
  \bigl(3\ue^{2C}+4\bigl(u^{3}-1\bigr)\ue^{2B}\bigr)
  g_{\us}^{-1}\ue^{4A+\phi},\\
  H_{22}=&\frac{48\left(2\pi\alpha'\right)^{2}}{\mu^{8/3}}
  \biggl[
  \bigl(1+\chi\bigr)^{8/3}u\,\ue^{2C}
  \bigl(3\ue^{2C}+4\bigl(u^{3}-1\bigr)\ue^{2B}\bigr)
  g^{-1}_{\us}\ue^{4A+\phi}
  +\frac{27\alpha'^{2}}{\mu^{8/3}}\frac{\beta^{2}}{u^{3}}\ue^{8A}\biggr],\\
  H_{33}=&\frac{144\left(2\pi\alpha'\right)^{2}}{\mu^{8/3}}
  \bigl(1+\chi\bigr)^{8/3}
  \bigl(u^{3}-1\bigr)u\,g^{-1}_{\us}\ue^{4A+4C+\phi},
\end{align}
\end{subequations}
with the remaining $H_{ij}$, $H_{i}$, and $Y_{i}$ vanishing.

The Chern-Simons action receives a contribution from the $4$-form potential
\begin{equation}
  S_{\uD 7}^{\mathrm{CS}\left(4\right)}=
  \frac{1}{2}\tau_{\uD 7}g_{\us}\int
  \mathrm{P}\bigl[C_{4}\bigr]\wedge{\cF_{2}}\wedge\cF_{2}.
\end{equation}
Using that the pullback of $B_{2}$ is trivial and taking the
orientation defined by~\eqref{eq:Kuperstein_orientation}, this
contribution becomes
\begin{align}
  S_{\uD 7}^{\mathrm{CS}\left(4\right)}=&
  \frac{\tau_{\uD 7}}{8}\int\ud^{8}\xi\,\omega\,\ep^{abcd}\cF_{ab}\cF_{cd}
  \notag\\
  =&\frac{\tau_{\uD 7}\mu^{8/3}}{144}
  \int\ud^{8}\xi\,\omega
  \biggl\{\frac{27\alpha'^{2}}{\mu^{8/3}}\frac{\beta^{2}}{u^{4}}
  -\frac{36\left(2\pi\alpha'\right)\alpha'}{\mu^{8/3}}
  \frac{\beta}{u^{3/2}}a'_{h^{2}}\biggr\}.
\end{align}
in which $\ep^{u123}=-1$.

In addition, there is a contribution from $C_{6}$, the magnetic dual
of $C_{2}$,
\begin{equation}
  S_{\uD 7}^{\mathrm{CS}\left(6\right)}=
  \tau_{\uD 7}g_{\us}\int
  \mathrm{P}\bigl[C_{6}\bigr]\wedge{\cF_{2}}.
\end{equation}
The potential $C_{6}$ is such that
\begin{equation}
  F_{7}=\ud C_{6}-H_{3}\wedge C_{4}=-g_{\us}^{-1}\ue^{\phi}\hat{\ast}F_{3},
\end{equation}
where $\hat{\ast}$ is the Hodge-$\ast$ built from the 10d
Einstein-frame metric.  Since $\iota_{\partial_{\mu}}F_{3}=0$, we have
\begin{equation}
  \hat{\ast}F_{3}=\ue^{4A}\ud\vol_{R^{3,1}}\wedge \ast F_{3},
\end{equation}
in which $\ast$ is the Hodge-$\ast$ built from the unwarped 6d metric
$g_{mp}$.  Hence,
\begin{equation}
  \ud C_{6}=g_{\us}^{-1}\ue^{\phi}\ud\vol_{R^{3,1}}\wedge
  \bigl(\omega\,\ue^{-\phi}H_{3}-\ue^{4A}\ast F_{3}\bigr).
\end{equation}
It is easy to check that in an ISD background, $\ud C_{6}$ vanishes
and so we can take $C_{6}$ to vanish.  In DKM this becomes
\begin{equation}
  \ud C_{6}=\alpha'g_{\us}^{-1}\ue^{4A}
  \biggl(\frac{\ud\beta}{\ud r}-\ue^{\phi}
  \frac{3 M}{4r}\biggr)\ud r\wedge\bigl(g^{1}\wedge g^{2}
  +g^{3}\wedge g^{4}\bigr).
\end{equation}
Writing
\begin{equation}
  C_{6}=g_{\us}^{-1}\ud\vol_{R^{3,1}}\wedge\tilde{C}_{2},\qquad
  \tilde{C}_{2}=\alpha'\ga\, \bigl(g^{1}\wedge g^{2}
  +g^{3}\wedge g^{4}\bigr),
\end{equation}
we have
\begin{equation}
  \ga=-\frac{2\cS}{9g_{\us}M}\log\frac{r}{r_{0}}.
\end{equation}
Note that $\tilde{C}_{2}$ is parallel to $B_{2}$.
\begin{align}
  S_{\uD 7}^{\mathrm{CS}\left(6\right)}
  =&\frac{\tau_{\uD 7}g_{\us}}{4}\int\ud^{8}\xi\,\ep^{abcd}
  \tilde{C}_{ab}\cF_{cd}\notag\\
  =&\frac{\tau_{\uD 7}\mu^{8/3}}{144}\int\ud^{8}x\,
  \biggl\{\frac{54\alpha'^{2}}{\mu^{8/3}}\frac{\beta\ga}{u^{4}}
  -\frac{36\left(2\pi\alpha'\right)\alpha'}{\mu^{8/3}}
  \frac{\ga}{u^{3/2}}a'_{h^{2}}\biggr\}.
\end{align}
The DKM solution has $C_{0}=0$ and hence there is no $C_{0}$ or
$C_{8}$ contribution to the Chern-Simons action.  Finally, with the
assumption that $f_{\mu\nu}=0$, neither $C_{2}$ nor the magnetic part
of $C_{4}$ contribute.  Hence we can write the total CS action as
\begin{equation}
  \label{eq:kuperstein_in_DKM_CS}
  S_{\uD 7}^{\mathrm{CS}}=-\frac{\tau_{\uD 7}\mu^{8/3}}{144}
  \int\ud^{8}\xi\biggl\{W_{3}+K_{i}a'_{h^{i}}\biggr\},
\end{equation}
with
\begin{subequations}
\begin{align}
  W_{3}=&-\frac{27\alpha'^{2}}{\mu^{8/3}}\frac{\beta}{u^{4}}
  \bigl(\omega\beta+2\ga\bigr),\\
  K_{2}=&\frac{36\left(2\pi\alpha'\right)\alpha'}{\mu^{8/3}}
  \frac{\omega\beta+\ga}{u^{3/2}}.
\end{align}
\end{subequations}

The functions $W_{i}$, $Y$, $Y_{i}$, $H_{i}$, $H_{ij}$ and $K_{i}$
depend only on $u$ and $\chi$.  Hence $\zeta$, $a_{h^{1}}$ and
$a_{h^{3}}$ only appear in the action through the derivatives and
furthermore only in the combinations $\zeta'a'_{h^{1}}$, $\zeta'^{2}$,
$a'^{2}_{h^{1}}$ and $a'^{2}_{h^{3}}$.  Therefore the equations of
motion for those fields are solved by taking
$\zeta=a_{h^{1}}=a_{h^{3}}=0$.  However, due to the non-trivial NS-NS
potential, we cannot also take $a_{h^{2}}=0$.  The remaining equations
of motion are
\begin{subequations}
\begin{align}
  0=&\partial_{u}\biggl[\frac{W_{2}}{\sqrt{W}}\partial_{u}\chi\biggr]
  -\frac{1}{2}\frac{\partial_{\chi}W_{2}}{\sqrt{W}}\chi'^{2}
  -\frac{1}{2}\frac{\partial_{\chi}H_{22}}{\sqrt{W}}a'^{2}_{h^{2}}
  -\frac{1}{2}\frac{\partial_{\chi}W_{1}}{\sqrt{W}}\chi'\notag\\
  &-\frac{1}{2}\frac{\partial_{\chi}H_{2}}{\sqrt{W}}a'_{h^{2}}
  -\partial_{\chi}K_{2}a'_{h^{2}}
  +\frac{1}{2}\partial_{u}\biggl[\frac{W_{1}}{\sqrt{W}}\biggr]
  -\frac{1}{2}\frac{\partial_{\chi}W_{0}}{\sqrt{W}}
  -\partial_{\chi}W_{3},\\
  0=&\partial_{u}\biggl[\frac{H_{22}}{\sqrt{W}}\partial_{u}a_{h^{2}}\biggr]
  +\frac{1}{2}\partial_{u}\biggl[\frac{H_{2}}{\sqrt{W}}\biggr]
  +\partial_{u}K_{2},
\end{align}
\end{subequations}
in which
\begin{equation}
  W=W_{0}+W_{1}\chi'+W_{2}\chi'^{2}
  +H_{2}a_{h^{i}}'
  +H_{22}a'^{2}_{h^{2}}.
\end{equation}
The boundary terms vanish by imposing Dirichlet boundary conditions
$\chi\to 0$ and $a_{h^{2}}\to 0$ as $u\to\infty$, while at $u=1$ we
impose the Neumann conditions
\begin{subequations}
\begin{align}
  0=&\frac{W_{2}}{\sqrt{W}}\chi'+\frac{W_{1}}{2\sqrt{W}},\\
  0=&\frac{H_{22}}{\sqrt{W}}a'_{h^{2}}+\frac{H_{2}}{2\sqrt{W}}
  +K_{2}.
\end{align}
\end{subequations}

Performing the expansions
\begin{equation}
  \chi=\sum_{n=0}^{\infty}\cS^{n}\chi^{\left(n\right)},\quad
  a_{h^{2}}=\sum_{n=0}^{\infty}\cS^{n}a^{\left(n\right)},
\end{equation}
the $\cO\left(\cS^{0}\right)$ contributions to the equations are
satisfied by taking $\chi^{\left(0\right)}=0$ and
$a^{\left(0\right)}=0$.  The $\cO{\left(\cS\right)}$ contributions to
the equations take the form
\begin{subequations}
\begin{align}
  0=&\partial_{u}\bigl[\cK_{\chi\chi}\partial_{u}\chi^{\left(1\right)}\bigr]
  +\partial_{u}\bigl[\cK_{a\chi}\partial_{u}a^{\left(1\right)}\bigr]
  +t_{\chi},\\
  0=&\partial_{u}\bigl[\cK_{aa}\partial_{u}a^{\left(1\right)}\bigr]
  +\partial_{u}\bigl[\cK_{a\chi}\partial_{u}\chi^{\left(1\right)}\bigr]
  +t_{a},
\end{align}
\end{subequations}
in which
\begin{subequations}
\begin{align}
  \cK_{\chi\chi}=&\frac{16 u^{2}\left(u^{3}-1\right)
    \left(8\pi N+3 g_{\us}M^{2}
      + 12 g_{\us} M^{2}\log\frac{u\mu^{2/3}}{r_{0}}
      +6 g_{\us} M^{2}\log^{2}\frac{u\mu^{2/3}}{r_{0}}\right)}
  {54 g_{\us}M^{2}\log^{2}\frac{u\mu^{2/3}}{r_{0}}
    +3\left(4u^{3}-1\right)
    \left(8\pi N+3 g_{\us}M^{2}
      + 12 g_{\us} M^{2}\log\frac{u\mu^{2/3}}{r_{0}}\right)},\\
  \cK_{aa}=&\frac{2048\pi^{2}u^{5}\left(4u^{3}-1\right)}
  {3g_{\us}\left[54 g_{\us}M^{2}\log^{2}\frac{u\mu^{2/3}}{r_{0}}
      +3\left(4u^{3}-1\right)
      \left(8\pi N+3 g_{\us}M^{2}
        + 12 g_{\us} M^{2}\log\frac{u\mu^{2/3}}{r_{0}}\right)\right]},\\
  \cK_{a\chi}=&\frac{512M\pi u^{7/2}\left(u^{3}-1\right)
    \log\frac{u\mu^{2/3}}{r_{0}}}
  {54 g_{\us}M^{2}\log^{2}\frac{u\mu^{2/3}}{r_{0}}
    +3\left(4u^{3}-1\right)
    \left(8\pi N+3 g_{\us}M^{2}
      + 12 g_{\us} M^{2}\log\frac{u\mu^{2/3}}{r_{0}}\right)},
\end{align}
\end{subequations}
and
\begin{subequations}
\begin{align}
  t_{\chi}=&-\frac{2\alpha'^{2}}{\mu^{8/3}}
  \biggl(18\log\frac{u\mu^{2/3}}{r_{0}}-5\biggr),\\
  t_{a}=&
  \frac{8\pi\alpha'^{2}}{g_{\us}Mu^{5/2}\mu^{8/3}}
  \biggl(3\log\frac{u\mu^{2/3}}{r_{0}}-2\biggr).
\end{align}
\end{subequations}
The solutions to these equations are
\begin{subequations}
\begin{align}
  \chi^{\left(1\right)}=&c_{1}-\int\ud u
  \frac{1}{\cK_{\chi\chi}\cK_{aa}-\cK_{a\chi}^{2}}
  \biggl\{\cK_{aa}\biggl(\int \ud u\, t_{\chi}+c_{3}\biggr)
  -\cK_{a\chi}\biggl(\int \ud u\,t_{a}+c_{4}\biggr)
  \biggr\},\\
  a^{\left(1\right)}=&c_{2}+\int\ud u
  \frac{1}{\cK_{\chi\chi}\cK_{aa}-\cK_{a\chi}^{2}}
  \biggl\{\cK_{a\chi}\biggl(\int \ud u\, t_{\chi}+c_{3}\biggr)
  -\cK_{\chi\chi}\biggl(\int \ud u\,t_{a}+c_{4}\biggr)
  \biggr\}.
\end{align}
\end{subequations}

\begin{figure}[t]
\begin{center}
  \includegraphics[scale=1]{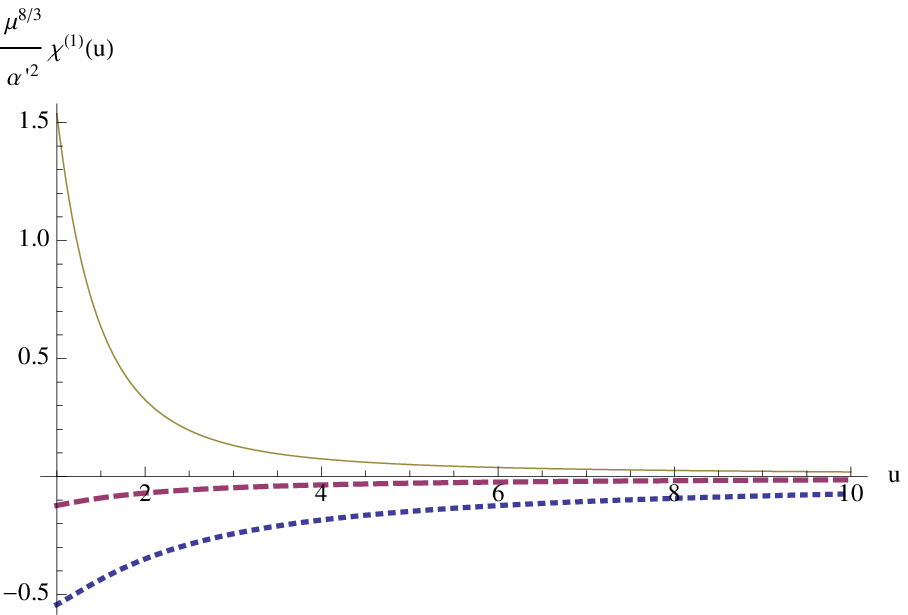}
  \includegraphics[scale=1]{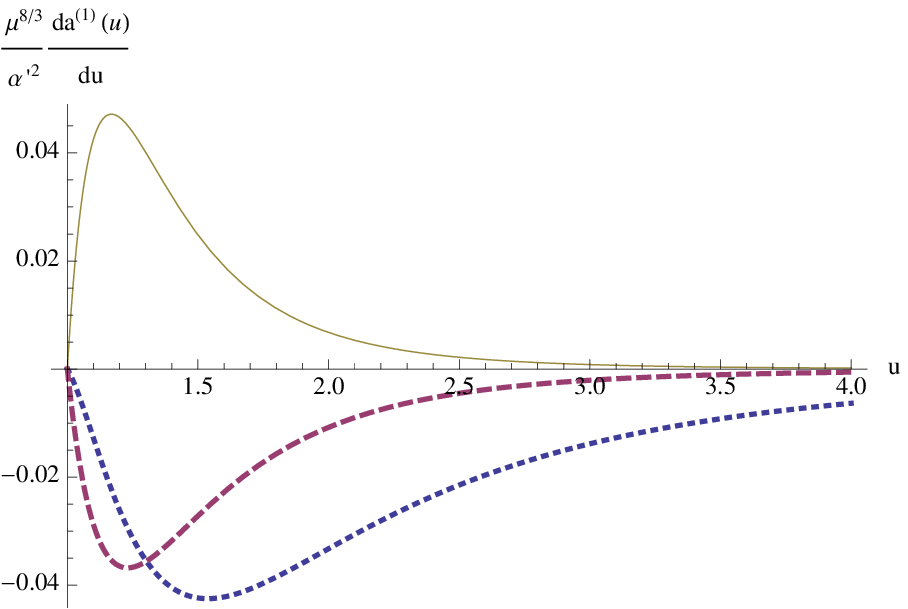}
  \caption{\label{fig:kuperstein_in_DKM}The behavior of a would-be
    Kuperstein brane probing the DKM geometry~\eqref{eq:DKM_soln}.
    \textbf{Top.}  Deformation of the worldvolume given
    by~\eqref{eq:Kuperstein_deformation}.  The dotted blue line is
    $\mu=2r_{0}^{3/2}$, the dashed purple line is $\mu=5r_{0}^{3/2}$
    and the solid mustard line is
    $\mu=.9r_{0}^{3/2}$. \textbf{Bottom.} The magnetization
    $F_{uh^{2}}$, plotted with the same values of $\mu$ and setting
    $g_{\us}=10^{-6}$, $M=10$, $N=1000$.  Note that our analysis is
    only valid for $\mu\gg r_{0}^{3/2}$ and the plotted solutions do
    not satisfy this condition.  However, solutions with large $\mu$
    do not qualitatively change with respect to the $\mu=2r_{0}^{3/2}$
    and $\mu=5 r_{0}^{3/2}$ solutions.}
\end{center}
\end{figure}

\begin{figure}[t]
\begin{center}
  \includegraphics[scale=1]{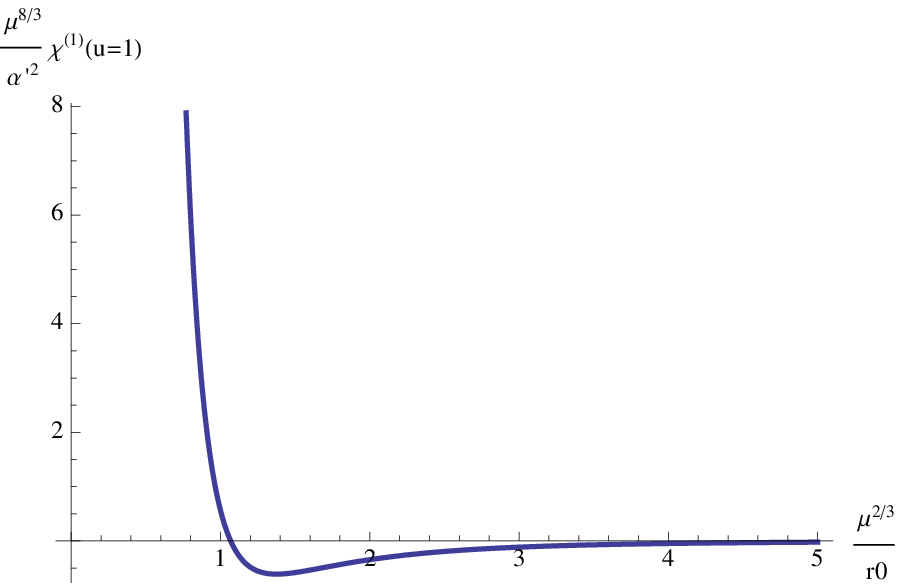}
  \caption{\label{fig:kuperstein_in_DKM_min_chi}The minimum value of $\chi$
    reached by a probe Kuperstein in the DKM solution.  Negative
    values of $\chi$ indicate that the brane is pulled into the center
    of warping.}
\end{center}
\end{figure}

Imposing the boundary conditions, we have
\begin{subequations}
\begin{align}
  \chi^{\left(1\right)}
  =&-\frac{9\alpha'^{2}}{4\mu^{8/3}}
  \int_{\infty}^{u}\frac{\ud x}{x^{5}}
  \biggl\{\log^{2}\frac{x\mu^{2/3}}{r_{0}}
  -x^{3/2}\log\frac{x\mu^{2/3}}{r_{0}}
  \log\frac{\mu^{2/3}}{r_{0}}\notag\\
  &\phantom{-\frac{9\alpha'^{2}}{4\mu^{8/3}}
  \int_{\infty}^{u}\ud x \frac{1}{x^{5}}
  \biggl\{}
  +\frac{4x^{3}-1}{x^{3}-1}
  \biggl[\log\frac{x\mu^{2/3}}{r_{0}}
  -x^{3}\log\frac{\mu^{2/3}}{r_{0}}\biggr]
  -\frac{4x^{3}-1}{18}\biggr\},
  \label{eq:Kuperstein_deformation}\\
  a^{\left(1\right)}=&\frac{27\alpha'^{2}g_{\us}M}{64\pi\mu^{8/3}}
  \int_{\infty}^{u}\frac{\ud x}{x^{13/2}}
  \biggl\{
  \log^{3}\frac{x\mu^{2/3}}{r_{0}}
  -x^{3/2}\log^{2}\frac{x\mu^{2/3}}{r_{0}}
  \log\frac{\mu^{2/3}}{r_{0}}\notag\\
  &\phantom{\frac{27\alpha'^{2}}{64\pi M\mu^{8/3}}
  \int_{\infty}^{u}\frac{\ud x}{x^{13/2}}
  \biggl\{}
  +6\log^{2}\frac{x\mu^{2/3}}{r_{0}}
  -2x^{3/2}\left(1+2x^{3/2}\right)
  \log\frac{x\mu^{2/3}}{r_{0}}\log\frac{\mu^{2/3}}{r_{0}}\notag\\
  &\phantom{\frac{27\alpha'^{2}}{64\pi M\mu^{8/3}}
    \int_{\infty}^{u}\frac{\ud x}{x^{13/2}}
  \biggl\{}
  -\frac{2}{9}x^{3}\log\frac{x\mu^{2/3}}{r_{0}}
  +\frac{1}{18}\left(\frac{24\pi N}{g_{\us}M^{2}}+13\right)
  \log\frac{x\mu^{2/3}}{r_{0}}\notag\\
  &\phantom{\frac{27\alpha'^{2}}{64\pi M\mu^{8/3}}
    \int_{\infty}^{u}\frac{\ud x}{x^{13/2}}
  \biggl\{}
  -\frac{1}{6}\biggl(\frac{8\pi N}{g_{\us}M^{2}}+3\biggr)
  x^{3/2}\log\frac{\mu^{2/3}}{r_{0}}\biggr\}.
\end{align}
\end{subequations}
The resulting profile and magnetization are plotted for a few
different values of $\mu/r_{0}^{3/2}$ in
figure~\ref{fig:kuperstein_in_DKM}.  There is a qualitative change in
the behavior, marked by an apparent repulsion of the probe $\uD 7$, as
$\mu$ crosses $r_{0}$.  Indeed the value of $\chi^{\left(1\right)}$ at
$u=1$, which is the point of closest approach when $\cS=0$, is
\begin{equation}
\label{eq:min_Kuperstein}
  \chi^{\mathrm{m}}:=
  \chi^{\left(1\right)}\bigl(u=1\bigr)=-\frac{3\alpha'^{2}}{3200\mu^{8/3}}
  \biggl\{7475-800\,\psi_{1}\biggl(\frac{1}{3}\biggr)+9084
  \log\frac{\mu^{2/3}}{r_{0}}
  +360\log^{2}\frac{\mu^{2/3}}{r_{0}}\biggr\}
\end{equation}
in which $\psi_{1}$ is the trigamma function.  This is plotted in
figure~\ref{fig:kuperstein_in_DKM_min_chi} and indeed it changes sign
at around $\mu^{2/3}\approx r_{0}$, reflecting the fact that
$\log\frac{\mu^{2/3}}{r_{0}}$ changes sign there.  However, as $r_{0}$
is the position at which the effects of a finite deformation of the
conifold are expected to become important, the solution is not
applicable in that region.  We also note that the deformation of the
worldvolume falls off more slowly this case than it did in the KW case
discussed in section~\ref{subsec:k_probes} where $\delta\chi\sim
\mu^{-16/3}$.  This is a consequence of the $3$-form flux which even
in supersymmetric configurations gives rise to potentials for
$7$-brane deformation moduli and induces positive $\uD 3$-charge on
the $\uD 7$s.  Moreover, again as a consequence of the flux, the
influence of the $\overline{\uD 3}$-branes falls off more slowly than
it did in $AdS^{5}\times X^{5}$.

We now consider, as we did in
section~\ref{subsec:D7_probe_in_flat_space}, two Kuperstein
embeddings.  The minimum radii reached by the branes are (assuming
$\mu_{i}$ is sufficiently large)
\begin{equation}
  \label{eq:DKM_min_r}
  r_{i}=\mu_{i}^{2/3}\biggl(1+\frac{2\cS}{3}\chi^{\mathrm{m}}_{i}\biggr).
\end{equation}
As argued earlier, the unwarped distance is more natural
characterization of the effects of the addition of $\overline{\uD 3}$s
and we have
\begin{equation}
  \Delta s=\mu_{2}^{2/3}-\mu_{1}^{2/3}
  +\frac{\cS}{8}\frac{\alpha'^{2}\left(\mu_{2}^{2}-\mu_{1}^{2}\right)}
  {\mu_{2}^{4}\mu_{1}^{4}}
  +\frac{2}{3}\biggl(\chi^{\mathrm{m}}_{2}\mu_{2}^{2/3}-
  \chi^{\mathrm{m}}_{1}\mu_{1}^{2/3}\biggr).
\end{equation}
For $\mu_{1}\gtrsim 2 r_{0}^{3/2}$, this is positive.  Thus, where the
solution can be trusted, the separation between the probe $\uD 7$
increases, as is consistent with the interpretation that the flavor
branes fall toward the $\overline{\uD 3}$s.  Note that unlike the
conformal $AdS^{5}\times X^{5}$ cases, the Klebanov-Strassler theory
has a mass scale at which the theory becomes confined.  In principle,
we could compare against the motion of the branes against this scale.
However, to reliably do so would require studying the motion of $\uD
7$-branes in the backreaction of $\overline{\uD 3}$s in the full KS
geometry.  Although this solution is
known~\cite{Bena:2009xk,Bena:2010ze,Bena:2011hz,Bena:2011wh,Massai:2012jn,Bena:2012bk,Dymarsky:2011pm},
the analysis is considerably more involved and we will not consider it
here.

Finally, we consider the energy of a Kuperstein brane.  As discussed
previously, the DKM solution only contains some information about the
backreaction of the $\overline{\uD 3}$s.  The
$\cO\left(\cS^{2}\right)$ terms may be sub-dominant compared to these
omitted corrections and so we will not include them in our
consideration.  Therefore, we can only reliably find the energy of the
configuration to linear order in $\cS$ which does not make use of the
perturbed profile.  We find however, that even to this order the
change in energy is divergent.  In particular if the worldvolume is
cutoff at $u_{\infty}$,
\begin{equation}
  \Delta \cE=\cE_{p\neq 0}-\cE_{p=0}
  \sim-\frac{\alpha'^{2}\cS}{\mu^{8/3}}\log^{2}u_{\infty}.
\end{equation}

\section{\label{sec:model}Model building with falling flavors}

In the previous section we quantified the extent to which $\uD
7$-branes probing supersymmetric warped geometries are deformed by the
inclusion of $\overline{\uD 3}$-branes.  The general result was that
the worldvolumes of such $\uD 7$-branes bend toward the $\overline{\uD
  3}$-branes.  In this section we discuss what implications this might
have on model building with such $7$-branes.  We will limit ourselves
to a qualitative discussion, leaving a more quantitative treatment for
future work.  Much of what is presented here is well known in the
literature and so this section serves primarily to place the effects
that we have discussed here into a wider context by pointing out where
the deformation of the $\uD 7$s may have substantial impact.

One immediate consequence is the correction to the soft terms
resulting from dimensional reduction of the $\uD 7$ worldvolume
theory\footnote{Due to the localized nature of the source of
  supersymmetry breaking and the RG filtering of UV effects, we
  anticipate that this discussion should be largely insensitive to the
  details of the UV completion of the warped geometry that is
  necessary to perform the reduction.}.  For example, the 4d gaugino
resulting from such a dimensional reduction receives a mass from
non-supersymmetric fluxes that is given by an integral over the
4-cycle~\cite{Camara:2004jj,Lust:2004fi,Lust:2004dn,Lust:2008zd}
wrapped by the $\uD 7$ that schematically takes the form
\begin{equation}
\label{eq:gaugino_mass}
  m_{1/2} \sim\int_{\Sig^{4}}\ud\vol_{\Sig^{4}}\,
  \eta^{\dagger}\mathrm{P}\bigl[\ga^{mnp}G_{mnp}\bigr]\eta\sim
  \int_{\Sig^{4}}\ud\vol_{\Sig^{4}}\,\overline{\Omega}\cdot G_{3},
\end{equation}
in which $\eta$ is the internal part of the gaugino wavefunction
(which to leading order in the non-supersymmetric perturbation is
proportional to the Killing spinor of the underlying Calabi-Yau),
$\Omega$ is the fundamental $3$-form of the Calabi-Yau, and we have
omitted contributions from the worldvolume flux.  This term describes
a non-vanishing gaugino mass resulting from a $\left(3,0\right)$ flux.
However, the calculation of the pullback in~\eqref{eq:gaugino_mass}
assumes that the $\uD 7$ is holomorphically embedded into the
Calabi-Yau and so holomorphic $\ga$-matrices in the bulk are
pulled-back to holomorphic $\ga$-matrices.  On the other hand, the
deformation of the worldvolume that we found here is non-holomorphic
when expressed in terms of the unperturbed complex structure.  For
example, for a Kuperstein in DKM, the deformed embedding can be
written as $z^{4}=f\bigl(u\bigr)$ where $u$ is a real, as opposed to
holomorphic, function of the other $z^{A}$.  This implies that the
pullback of a holomorphic $\ga$-matrix to the worldvolume will contain
both holomorphic and anti-holomorphic parts, inducing a coupling to
$\left(2,1\right)$ components of the flux.  This is similar to the
effect discussed in~\cite{Benini:2009ff} where the squashing of the
metric can lead to contributions to the soft terms from other
Hodge-types.  When the deformation of the $\uD 7$ is not taken into
account, a probe Kuperstein in DKM does not receive a contribution to
the gaugino mass as a result of the dimensional
reduction~\cite{Benini:2009ff}.  The interpretation provided
by~\cite{Benini:2009ff} was the $Z_{2M}$ R-symmetry of the KT geometry
(a remnant of the $\U{1}$ R-symmetry of KW under which $z^{i}\to
\ue^{\ui\alpha}z^{i}$) which forbids a non-vanishing gaugino mass when
$M>1$. At least morally, the terms appearing in the $\uD 7$-geometry
are essentially pullbacks of R-symmetric terms and so cannot induce a
gaugino mass.  However, the $\uD 7$-brane explicitly breaks the
R-symmetry entirely and hence one should not expect this protection to
be perfect.  Although we again defer a more precise treatment to
future work, the effect of the non-holomorphic pullback is
particularly sensitive to the R-breaking embedding of the $\uD 7$ and
hence we expect that effect to give a non-vanishing contribution.
Furthermore, the deformation of the worldvolume, as characterized
by~\eqref{eq:min_Kuperstein}, has the same scaling with $\cS$ and
$\mu$ (using~\eqref{eq:DKM_min_r}) as the non-supersymmetric
deformations of the geometry~\eqref{eq:DKM_soln}. Therefore we expect
this contribution to be comparable to the naive parametric estimates
for the gaugino mass appearing in~\cite{Benini:2009ff}.  Similar
arguments can be made for other soft terms resulting from the
dimensional reduction.  Hence, the deformation of the worldvolume of
the $\uD 7$s generically introduces $\cO\left(1\right)$ corrections to
the soft Lagrangian of the low-energy effective theory.

In the context of AdS/CFT, the inclusion of $\uD 7$-branes corresponds
to the addition of a global flavor symmetry and quarks that transform
as fundamentals under this flavor group and the strongly coupled gauge
group.  The spectrum of mesons and mesini in the dual theory can be
determined by considering the open-string fluctuations of the $\uD 7$.
This analysis was performed in~\cite{Kruczenski:2003be} for a $\uD 7$
probe $AdS^{5}\times S^{5}$ and in~\cite{Kuperstein:2004hy} for a
subset of the mesons for a Kuperstein $\uD 7$ in Klebanov-Strassler.
The analyses were extended to the non-supersymmetric cases
in~\cite{McGuirk:2011yg} and~\cite{Benini:2009ff} respectively.
However in these cases the deformation of the $\uD 7$ was taken not
into account.  One would again expect that in each of these setups the
deformation would alter the spectrum by $\cO\left(1\right)$ factors
and subsequently the visible-sector soft terms found in models of
holographic gauge mediation models (though we anticipate that the
effect would not change the parametric scaling of such terms).

\begin{figure}[t]
\begin{center}
  \includegraphics[scale=.55]{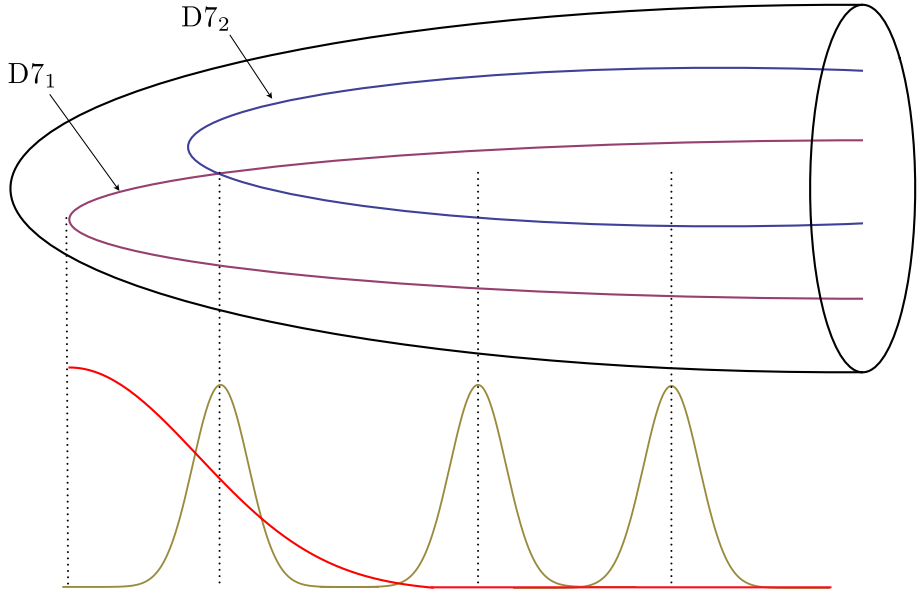}
  \caption{\label{fig:intersecting_d7s}Cartoon of chiral modes
    supported by intersecting $\uD 7$ branes in a warped geometry.
    $\uD 7_{1}$ supports a scalar whose IR-localized profile is in
    red, while the magnetized intersection $\uD 7_{1}$ and $\uD 7_{2}$
    supports chiral modes localized at various positions depending on
    the magnetic flux.}
\end{center}
\end{figure}

As discussed in the introduction, $\uD 7$-branes are expected to
appear in string theory realization of bulk Randall-Sundrum
scenarios~\cite{Randall:1999ee} (see, for
example,~\cite{Acharya:2006mx,Marchesano:2008rg}).  Although gauge
multiplets appear on a single stack of $\uD 7$-branes, bifundamental
chiral matter appears on the intersection of magnetized $\uD
7$-branes.  One possibility (see figure~\ref{fig:intersecting_d7s}) is
to model the visible sector gauge group as $\U{N_{f}}$ realized on
$\uD 7$s characterized by an embedding parameter $\mu$ that intersects
with another stack of $\uD 7$s with a different embedding
parameter\footnote{Note that in order for the branes to intersect,
  different coordinates for the embedding must be used.  For
  Kuperstein embeddings for example, we might use $z^{4}\sim\mu$ for
  one stack and $z^{3}\sim\nu$ for the other.}.  If the intersection
carries a chiral index (i.e. if the integral of $\cF_{2}$ over the
intersection is non-vanishing), then chiral modes will be supported on
the intersection.  In the unwarped case, the wavefunctions for the
chiral modes are Gaussians peaked at different locations along the
intersection with the spacing of the modes set by the chiral index
(see e.g.~\cite{Cremades:2004wa}). Although the warping is difficult
to take into account~\cite{Marchesano:2010bs}, it is reasonable to
assume that this property persists to the warped case, though perhaps
with some modulation due to the warping.  As in RS models and other
string constructions, operators in the 4d effective field theory
description depend on the overlap of such wavefunctions.  As these
wavefunctions will necessarily depend on the shape of the worldvolumes
that support them, the deformations that we considered here must be
taken into account.  However, unlike some of the previous effects we
discussed where the correction is not anticipated to generically lead
to qualitatively different results, in this case there may be some
qualitative differences.  For example, in some supersymmetric string
models only one generation of fermions receive masses from electroweak
symmetry breaking effects (see, for
example,~\cite{Cecotti:2009zf,Marchesano:2009rz}).  Although
non-perturbative effects can resolve this rank-one problem, it would
be important to understand to what extent the breaking of
supersymmetry, and the corresponding deformation of $\uD 7$
worldvolumes, alters the story.

Finally, we consider the effect of the inter-flavor forces induced by
the breaking of supersymmetry that were discussed in
section~\ref{subsec:flavor_forces}.  In that analysis we found some
evidence that the force between different flavor branes may be
repulsive at small distances.  This would imply that stacks of branes
are, under certain circumstances, unstable in these geometries.  In
addition to making the construction of visible-sector non-Abelian
gauge symmetries difficult, it could also interfere with the onset of
gaugino condensation which is an essential ingredient in the moduli
stabilization scheme of~\cite{Kachru:2003aw} and related
constructions.  We again emphasize that our analysis is in a very
particular construction and we treated the $\uD 7$ as a small
perturbation which is inapplicable at small distances.  Furthermore,
in a complete treatment the $\uD 7$ branes are stabilized partially by
fluxes in the bulk and their effect must be considered as well.
Finally, such an effect would be rather surprising from the dual field
theory point of view.  Hence, we refrain from making any definitive
statement of the existence of such an instability, but merely point
out that it is a possible concern worth exploring.

\section{\label{sec:conclude}Concluding remarks}

In this paper we have examined the behavior of $\uD 7$-branes in
non-supersymmetric geometries.  In particular, we focused on GKP-like
warped geometries where the backreaction of $\uD 7$-branes can be
controlled.  In such cases, $\overline{\uD 3}$-branes naturally sit at
origin of the geometry and break supersymmetry in a relatively
well-controlled (and possibly metastable) manner. We argued that in
such a setup, $\uD 7$s experience a small attractive force to the
$3$-branes.  Since we impose that the $\uD 7$s asymptote to their
unperturbed configurations, this results in a small bending of the
flavor branes.  In the $AdS^{5}\times X^{5}$ case, the bending falls
off as $r_{\mathrm{min}}^{-8}$ where $r_{\mathrm{min}}$ is the
distance of closest approach before the $\overline{\uD 3}$s are added.
In the case of $\overline{\uD 3}$s in the KS geometry (which we
modeled as the DKM perturbation of KT), the effect falls off as
$r_{\mathrm{min}}^{-4}$.  In either case, the influence on the $\uD 7$
physics (for example soft terms) from this bending is comparable to
the effects produced when the deformation of the worldvolume is
neglected.

Beyond a more quantitative analysis of the effects on model building,
there are a number of ways in which our analysis can be improved.  For
example, as discussed in section~\ref{sec:DKM}, the DKM
solution~\cite{DeWolfe:2008zy} only captures some of the behavior of
the backreaction of $\overline{\uD 3}$s in the Klebanov-Strassler
geometry.  In addition to terms that scale as $r^{-8}$ that are
present even in KT, there are terms that scale as
$r^{-3}$~\cite{Bena:2011wh} but vanish as we take the deformation
parameter of the deformed conifold is taken to zero.  Since the size
of this deformation is set by the amount of flux in the geometry, we
cannot literally take this limit, and hence it may be worth while to
understand how the effects we compute are modified when we include
this deformation.

We have also examined how the effects of backreacting flavors will
modify the above picture and found some hints that at long distances
the flavor branes attract each other while at smaller distances they
repel.  To make progress in addressing the backreaction of the $\uD
7$s, we considered the smeared approximation and furthermore
considered only the backreaction of ``massless'' $\uD 7$s, meaning
those whose quarks in the dual theory are massless.  Furthermore, due
to the Landau pole in the dual theory, the geometry resulting from
such backreaction becomes singular in the UV.  It would be worth
understanding how to get around both of these issues\footnote{It is
  worth noting that some progress has been made in understanding the
  non-smeared solutions (see, for
  example,~\cite{Grana:2001xn,Burrington:2004id}).}.  In particular,
smearing the $\uD 7$s is a rather dramatic approximation to make,
especially in light of the fact the $\uD 7$s will exert forces on each
other. Although smearing may, to linear order, correctly capture the
radial components of the inter-brane forces, in general there will be
angular components that would probably be important. It would also be
interesting to understand how more general $7$-branes
behave\footnote{$\overline{\uD 7}$s may also be interesting to
  consider as in, for example,~\cite{Ihl:2012bm}.}  However, such an
analysis would require an F-theoretical approach and little has been
done in non-supersymmetric F-theory constructions.  One of the
consequences of a repulsion between $\uD 7$s in this sort of setup is
the difficulty in constructing non-Abelian gauge groups in the
low-energy theory.  Although our analysis is not sufficient to make a
strong statement regarding the existence of this repulsion, it may be
important to understand the interaction at small distances.

Finally, our analysis was entirely in the context of a supergravity
description of the branes.  An interesting and complimentary approach
would be a worldsheet analysis by considering the exchange of closed
strings between a $\uD 3$-$\overline{\uD 3}$-pair and a ``parallel''
$\uD 7$-brane.  Such a calculation can confirm, for example, the
absence of a force between $\uD 3$s and $\uD 7$s.  However, such a
computation may be somewhat involved due to the non-BPS stack of
$3$-branes.

\acknowledgments

It is a pleasure to thank Ben Heidenreich, Cody Long, and Liam
McAllister for helpful discussions.  I also thank Gary Shiu and Yoske
Sumitomo for past conversations on closely related topics which piqued
my initial interest in this question. This work was supported by the
NSF under grant PHY-0757868.

\appendix

\section{\label{app:conv}Type-IIB supergravity}

The bosonic degrees of freedom of type-IIB supergravity consist of the
10d Einstein-frame metric
$\hat{g}_{MN}={g}_{\us}^{1/2}\ue^{-\phi/2}\hat{g}^{\mathrm{string}}_{MN}$,
the axiodilaton $\tau=C_{0}+\ui\ue^{-\phi}$ (normalized such that the
IIB coupling is $g_{\mathrm{IIB}}=\left({\mathrm{Im}\,\tau}\right)^{-1}$), the
$2$-form potentials $B_{2}$ and $C_{2}$, and the $4$-form potential
$C_{4}$.  In terms of the gauge-invariant field strengths
\begin{equation}
  H_{3}=\ud B_{2},\quad F_{1}=\ud C_{0},\quad
  F_{3}=\ud C_{2}-H_{3}C_{0},\quad
  G_{3}=F_{3}-\ui\ue^{-\phi}H_{3},\quad
  F_{5}=\ud C_{4}-H_{3}\wedge C_{2},
\end{equation}
the corresponding pseudo-action is
\begin{align}
  S_{\IIB}=\frac{1}{2\ka_{10}^{2}}
  \int\biggl\{&\hat{\ast}\hat{R}
  +\frac{1}{2\left(\mathrm{Im}\,\tau\right)^{2}}
  \ud\tau\wedge\hat{\ast}\ud \bar{\tau}
  +\frac{g_{\us}}{2\,\mathrm{Im}\,\tau}
  G_{3}\wedge\hat{\ast}\overline{G}_{3}\notag\\
  &+\frac{g_{\us}^{2}}{4}F_{5}\wedge
  \hat{\ast}F_{5}
  +\frac{\ui g_{\us}^{2}}{4\mathrm{Im}\,\tau}
  C_{4}\wedge G_{3}\wedge\overline{G}_{3}\biggr\},
\end{align}
in which $2\ka_{10}^{2}=\left(2\pi\right)^{7}\alpha'^{4}g_{\us}^{2}$,
$\hat{R}$ is the Ricci scalar built from $\hat{g}_{MP}$ and we have
defined the Hodge-$\ast$ in $D$ dimensions by
\begin{equation}
  \bigl(\hat{\ast}\Omega_{p}\bigr)_{M_{1}\cdots M_{D-p}}
  =\frac{1}{p!}\hat{\ep}_{M_{1}\cdots M_{D-p}}
  ^{\phantom{M_{1}\cdots M_{D-p}}N_{1}\cdots N_{p}}
  \Omega_{N_{1}\cdots N_{p}},
\end{equation}
where we make use of the 10d volume form
\begin{equation}
  \hat{\ep}_{0\cdots 9}=+\sqrt{-\det\left(\hat{g}\right)}.
\end{equation}
The resulting equations of motion must be supplemented by the
self-duality constraint
\begin{equation}
  F_{5}=\hat{\ast}F_{5}.
\end{equation}

We will couple the supergravity modes to $\uD 3$-branes,
$\overline{\uD 3}$-branes, and $\uD 7$-branes\footnote{We will not
  work with explicit compactifications so that we need not consider
  orientifold planes.}.  The bosonic action for a $\uD p$-brane is a
sum of the Dirac-Born-Infeld and Chern-Simons actions
\begin{subequations}
\label{eq:Dbrane_action}
\begin{align}
  S_{\uD p^{\pm}}=&S_{\uD p}^{\mathrm{DBI}}+S_{\uD p^{\pm}}^{\mathrm{CS}},\\
  S_{\uD p}^{\mathrm{DBI}}=&-\tau_{\uD p}\int\ud^{p+1}\xi\,
  \bigl(g_{\us}^{-1}\ue^{\phi}\bigr)^{\left(p-3\right)/4}
  \sqrt{-\det\left(\mathrm{P}\left[\hat{g}+g_{\us}^{1/2}\ue^{-\phi/2}
        B\right]_{\alpha\beta}
        +2\pi\alpha' g_{\us}^{1/2}\ue^{-\phi/2}f_{\alpha\beta}\right)},\\
  S_{\uD p^{\pm}}^{\mathrm{CS}}=&
  \pm\tau_{\uD p}\int\mathrm{P}\biggl[\mathcal{C}\wedge\ue^{B_{2}}\biggr]
  \wedge\ue^{2\pi\alpha' f_{2}},
\end{align}
\end{subequations}
in which $\tau_{\uD
  p}^{-1}=\left(2\pi\right)^{p}\alpha'^{\left(p+1\right)/2}g_{\us}$
gives the Einstein-frame brane tension, $\mathrm{P}$ denotes the
pullback to the worldvolume of the brane, and $f_{2}=\ud a_{1}$ is the
field strength of the $\U{1}$ gauge field supported by the brane. In
the Chern-Simons action, the upper sign applies to $\uD p$-branes and
the lower sign to $\overline{\uD p}$-branes. $\mathcal{C}$ is the
formal sum of Ramond-Ramond potentials and their magnetic duals
defined by
\begin{equation}
  F_{7}=\ud C_{6}-H_{3}\wedge C_{4}=
  -g_{\us}^{-1}\ue^{\phi}\,\hat{\ast}F_{3},\qquad
  F_{9}=\ud C_{8}-H_{3}\wedge C_{6}
  =g_{\us}^{-2}\ue^{2\phi}\,\hat{\ast}\ud F_{1},
\end{equation}
in which the factors of the string coupling appear because we are
working in the 10d Einstein frame while these forms are more naturally
defined in the string frame.

\section{\label{app:DKM_1}Backreacting 3-branes on Calabi-Yau cones}

In this appendix we consider the backreaction of $N$ $\uD 3$-branes
and $p\ll N$ $\uD 3$-$\overline{\uD 3}$ pairs at the tip of a
Calabi-Yau cone.  The ansatz and relevant field definitions are
reviewed in section~\ref{sec:prelim}.  For the internal metric we take
the ansatz
\begin{equation}
  \ud s_{6}^{2}=\ue^{2B}\bigl(\ud r^{2}+r^{2}\ud s^{2}_{X^{5}}\bigr),
\end{equation}
in which $\ud s^{2}_{X^{5}}=\breve{g}_{\theta\phi}\ud y^{\theta}\ud
y^{\phi}$ is the metric on a Sasaki-Einstein space normalized such
that its Ricci tensor is
$\breve{R}_{\theta\phi}=4\breve{g}_{\theta\phi}$.  One way to proceed
in the flat space case would be to take a small $\frac{p}{N}$ limit of
the exact supergravity solutions presented in~\cite{Brax:2000cf}.
However, our interest is in the near-brane limit in which the space
asymptotes to $AdS^{5}\times X^{5}$ rather than flat space as assumed
in~\cite{Brax:2000cf}.  The near-brane limit of the solution
of~\cite{Brax:2000cf} was performed and presented to leading order in
$\frac{p}{N}$ in~\cite{DeWolfe:2008zy}.

In this section we review and extend the result
of~\cite{DeWolfe:2008zy}.  We will exploit the observation
of~\cite{Gandhi:2011id} that the supergravity equations of motion
describing a perturbation from a background with $\Phi_{-}=0$,
$G_{-}=0$, and constant dilaton take a block-triangular form.
Moreover, this structure persists order-by-order in perturbation
theory.  Before taking the near-brane limit, the backreaction of $N$
$\uD 3$s at the tip of the cone takes the form
\begin{equation}
  \Phi_{+}^{-1}=\frac{1}{2}+\frac{L^{4}}{2r^{4}},
  \quad\tau=\frac{\ui}{g_{\us}},
  \quad\ue^{2B}=1,
  \quad
  L^{4}=\frac{4\pi^{4}g_{\us}N\alpha'^{2}}{\cV_{X^{5}}},
\end{equation}
in which $\cV_{X^{5}}$ is the volume of the compact Sasaki-Einstein
space and all other fields vanish.  We then perturb this geometry by
the addition of $\uD 3$-$\overline{\uD 3}$ pairs.  In doing so, we can
take $G_{\pm}$ and $\tau$ to remain trivial since the $\overline{\uD
  3}$ is not charged under any of the relevant fields and they are
trivial before the addition of the $\overline{\uD 3}$s.  The equations
of motion~\eqref{eq:IIB_eom} take the form
\begin{subequations}
\label{eq:D7_backreaction_in_flat_space}
\begin{align}
  0=&\Phi_{-}''+\bigl(4B'+\frac{5}{r}\bigr)\Phi_{-}'
  -\frac{2\Theta_{+}}{1+\Phi_{-}\Theta_{+}}
  \Phi'^{2}_{-}-\frac{\left(1+\Phi_{-}\Theta_{+}\right)^{2}}{2\Theta_{+}^{2}}
  \bigl(4\pi^{2}\alpha'\bigr)^{2}p g_{\us}
  \frac{\delta^{6}\left(y\right)}{r^{5}\ue^{4B}},
  \label{eq:Phi_minus_eq_1}\\
  0=&-5B''-\frac{5}{r}B'+
  \frac{2}{\left(1+\Phi_{-}\Theta_{+}\right)^{2}}
  \Theta_{+}'\Phi_{-}',
  \label{eq:Einstein_1}\\
  0=&B''+4B'^{2}+\frac{9}{r}B',
  \label{eq:Einstein_2}\\
  0=&\Theta_{+}''+\bigl(4B'+\frac{5}{r}\bigr)\Theta_{+}'
  -\frac{2\Phi_{-}}{1+\Phi_{-}\Theta_{+}}
  \Theta_{+}'^{2}+\frac{\left(1+\Phi_{-}\Theta_{+}\right)^{2}}{2}
  \bigl(4\pi^{2}\alpha'\bigr)^{2}\left(N+p\right) g_{\us}
  \frac{\delta^{6}\left(y\right)}{r^{5}\ue^{4B}},
  \label{eq:Phi_plus_eq_1}
\end{align}
\end{subequations}
in which we have defined $\Theta_{+}=\Phi_{+}^{-1}$ and have taken all
of the fields to depend only on $r$.  The Ricci tensor in the
perturbed geometry is
\begin{equation}
  R_{rr}=-5B''-\frac{5}{r}B',\quad
  R_{\theta\phi}=-r^{2}\bigl(B''+4B'^{2}+\frac{9}{r}B'\bigr)
  \breve{g}_{\theta\phi}.
\end{equation}
Since we are treating $\frac{p}{N}$ as a small parameter, it useful to
expand the fields as a power series in $\frac{p}{N}$.  To that end, we
write
\begin{equation}
  \Phi_{-}=\sum_{n=0}^{\infty}\left(\frac{p}{N}\right)^{n}
  \Phi_{-}^{\left(n\right)},
\end{equation}
and similarly for other fields.  The equations of motion are then
straightforwardly solved order-by-order in $\frac{p}{N}$.  At any
given order, we first first solve~\eqref{eq:Phi_minus_eq_1}, followed
by~\eqref{eq:Einstein_1} and~\eqref{eq:Einstein_2} and
finally~\eqref{eq:Phi_plus_eq_1}. The result is
\begin{subequations}
\begin{align}
  \Phi_{-}=&-\frac{p}{N}\frac{2L^{4}}{r^{4}}
  +\left(\frac{p}{N}\right)^{2}\biggl(\frac{2L^{8}}{r^{8}}
  +\frac{2L^{12}}{5r^{12}}\biggr),\\
  B=&-\frac{p}{N}\frac{L^{8}}{10r^{8}}
  +\left(\frac{p}{N}\right)^{2}
  \biggl(-\frac{L^{8}}{10r^{8}}-\frac{L^{16}}{50 r^{16}}\biggr),\\
  \Theta_{+}=&\frac{1}{2}+\frac{L^{4}}{2r^{4}}
  +\frac{p}{N}\biggl(\frac{L^{4}}{2r^{4}}-\frac{L^{12}}{10 r^{12}}\biggr)
  +\left(\frac{p}{N}\right)^{2}
  \biggl(-\frac{L^{12}}{5r^{12}}+\frac{2L^{20}}{125 r^{20}}\biggr).
\end{align}
\end{subequations}
The integration constants have been fixed by
integrating~\eqref{eq:Phi_minus_eq_1} and~\eqref{eq:Phi_plus_eq_1} and
by requiring that the space asymptotes to the $R^{9,1}$.  The warp
factor for this solution is
\begin{equation}
  \ue^{-4A}=1+\frac{L^{4}}{r^{4}}
  +\frac{p}{N}\biggl(\frac{2L^{4}}{r^{4}}
  +\frac{2L^{8}}{r^{8}}+\frac{4L^{12}}{5r^{12}}\biggr)+\cdots
\end{equation}
To obtain the near-horizon limit, we follow~\cite{DeWolfe:2008zy} and
perform the rescaling
\begin{equation}
  x^{\mu}= \tilde{x}^{\mu}Z^{-1/4},\quad
  r= \tilde{r}Z^{-1/4},\quad
  L^{4}= \tilde{L}^{4}Z^{-2},\quad
  N=\tilde{N} Z^{-2},\quad
  p=\tilde{p},
\end{equation}
and then consider the limit in which $Z\to 0$ while holding fixed
$\tilde{L}$ and other quantities with tildes.  Then, $\ue^{2A}\ud
x_{4}^{2}=Z^{-1/2}\ue^{2A}\ud x_{4}^{2}$ and so the warp factor in the
near-brane limit is $\ue^{2\tilde{A}}=Z^{-1/2}\ue^{2A}$.  Taking $Z\to
0$ and dropping the tildes for notational simplicity, we find
\begin{subequations}
\begin{align}
  \ue^{-4A}=&\frac{L^{4}}{r^{4}}
  +\frac{4p}{5N}\frac{L^{12}}{r^{12}}
  +\frac{54p^{2}}{125N^{2}}\frac{L^{20}}{r^{20}},\\
  \omega=&\frac{r^{4}}{L^{4}}
  +\frac{6p}{5N}\frac{L^{4}}{r^{4}}
  -\frac{24 p^{2}}{125 N^{2}}\frac{L^{12}}{r^{12}},\\
  \ue^{2B}=&1-\frac{p}{5N}\frac{L^{8}}{r^{8}}
  -\frac{p^{2}}{50N^{2}}\frac{L^{16}}{r^{16}}.
\end{align}
\end{subequations}

\section{\label{app:backreaction_D7s}Backreaction of smeared D7s}

\subsection{\label{app:backreaction_D7s_in_flat_space}D7s and 3-branes
  in flat space}

Here we construct the supergravity solution corresponding to the
backreaction of $\uD 7$s smeared over the near-brane background
created by $\uD 3$s and $\overline{\uD 3}$s in flat space.  The
strategy will be to first review the backreaction of $\uD 7$s in
$AdS^{5}\times S^{5}$ and then perturb the resulting geometry by
$\overline{\uD 3}$-branes.  Our starting point is the warped
ansatz~\eqref{eq:warped_ansatz} where $g_{mn}$ is a metric on
$R^{6}\simeq C^{3}$ which we express
as~\eqref{eq:flat_space_metric}. In terms of the radial coordinate
defined by~\eqref{eq:C3_coords}, it useful to define a new radial
coordinate $\rho$ by
\begin{equation}
  \label{eq:flat_radial_coord_2}
  r=\alpha'^{1/2}\ue^{\rho}.
\end{equation}
Then the metric~\eqref{eq:flat_space_metric} on $C^{3}$ takes the form
\begin{equation}
  \ud s_{6}^{2}=\alpha'\ue^{2\rho}\bigl\{\ud\rho^{2}
  +\bigl(\ud\psi+\cA\bigr)^{2}+\ud s_{CP^{2}}^{2}\bigr\}.
\end{equation}
We then consider a $\uD 7$-brane embedded into the geometry according
to~\eqref{eq:flat_space_embedding}
\begin{equation}
  \label{eq:flat_space_embedding_2}
  z^{3}=\alpha'^{1/2}\ue^{\rho_{\nu}},
\end{equation}
in which $\rho_{\nu}$ is a constant.  The backreaction of such
codimension two objects is quite difficult to compute and so we
consider a smeared approximation in which we take large number of $\uD
7$s distributed in such a way that they can be treated as uniformly
spread along the internal space.  Here we follow the smearing
suggested by~\cite{Benini:2006hh} and implemented for this
configuration in $AdS^{5}\times S^{5}$ in~\cite{Bigazzi:2009bk} which
we quickly review.  An $\SU{3}\times\U{1}$ rotation of the
embedding~\eqref{eq:flat_space_embedding_2} gives
\begin{equation}
  \label{eq:flat_space_embedding_smeared}
  a_{i}z^{i}=\alpha'^{1/2}\ue^{\rho_{\nu}+\ui\alpha},
\end{equation}
where $\alpha\in\left[0,2\pi\right)$ while the $a_{i}$ are complex
numbers satisfying
\begin{equation}
  a_{i}\bar{a}_{i}=1.
\end{equation}
The $a_{i}$ thus parametrize a unit $S^{5}$, which we denote
$\tilde{S}^{5}$, and so we can write
\begin{subequations}
\begin{align}
  a_{1}=&\cos\frac{\tilde{\ga}}{2}\cos\frac{\tilde{\theta}}{2}
  \ue^{{\ui}\left(\tilde{\psi}+\tilde{\eta}/{2}+\tilde{\vp}/{2}\right)},\\
  a_{2}=&\cos\frac{\tilde{\ga}}{2}\sin\frac{\tilde{\theta}}{2}
  \ue^{{\ui}\left(\tilde{\psi}+\tilde{\eta}/2-\tilde{\vp}/2\right)},\\
  a_{3}=&\sin\frac{\tilde{\ga}}{2}\ue^{\ui\tilde{\psi}},
\end{align}
\end{subequations}
in which, as before,
$\tilde{\ga},\tilde{\theta}\in\left[0,\pi\right]$,
$\tilde{\vp},\tilde{\psi}\in\left[0,2\pi\right)$, and
$\tilde{\eta}\in\left[0,4\pi\right)$.  The distribution of $\uD 7$s is
specified by a density function on
$\tilde{S}^{5}\times\left[0,2\pi\right)$.  The density function
$\rho_{\uD 7}$ is normalized such that
\begin{equation}
  N_{f}=\int\ud\vol_{\tilde{S}^{5}}\ud\alpha\,\rho_{\uD 7}
\end{equation}
in which $N_{f}$ is the total number of $\uD 7$s.  A uniform
distribution corresponds to a constant $\rho_{\uD 7}$
\begin{equation}
  \rho_{\uD 7}=\frac{N_{f}}{2\pi^{4}}.
\end{equation}
With this configuration of magnetic charges for $C_{0}$, $F_{1}$ must
take the form\footnote{Note that the relative sign with respect to the
  solutions reviewed in~\cite{Nunez:2010sf} is conventional.}
\begin{equation}
  F_{1}=-Q\left(\rho\right)\bigl(\ud\psi+\cA\bigr).
\end{equation}
Writing the CS action for the $\uD 7$s as
\begin{equation}
  \tau_{\uD 7}g_{\us}\sum_{\uD 7}\int_{\cW^{i}} C_{8}
  =\tau_{\uD 7}g_{\us}\sum_{\uD 7}\int\Omega_{i}\wedge C_{8}
  =:\tau_{\uD 7}g_{\us}\int\Omega\wedge C_{8},
\end{equation}
we have
\begin{equation}
  \ud F_{1}=-\Omega.
\end{equation}

We can write the embedding~\eqref{eq:flat_space_embedding_smeared} as
$f_{1}=f_{2}=0$ in which
\begin{subequations}
\begin{align}
  f_{1}=&2\bigl(\psi+\tilde{\psi}\bigr)+\eta+\tilde{\eta}
  +\vp+\tilde{\vp}+2\arg\Ga-2\alpha+4\pi n,\\
  f_{2}=&\bigl\lvert\Ga\bigr\rvert^{2}-\ue^{2\left(\rho_{\nu}-\rho\right)},
\end{align}
\end{subequations}
with $n\in Z$ and
\begin{equation}
  \Ga=\cos\frac{\ga}{2}\cos\frac{\tilde{\ga}}{2}
  \cos\frac{\theta}{2}\cos\frac{\tilde{\theta}}{2}
  +\cos\frac{\ga}{2}\cos\frac{\tilde{\ga}}{2}
  \sin\frac{\theta}{2}\sin\frac{\tilde{\theta}}{2}
  \ue^{-\ui\left(\vp+\tilde{\vp}\right)}
  +\sin\frac{\ga}{2}\sin\frac{\tilde{\ga}}{2}
  \ue^{-\frac{\ui}{2}\left(\eta+\tilde{\eta}+\vp+\tilde{\vp}\right)}.
\end{equation}
Then
\begin{equation}
  \Omega=\int
  \ud\vol_{\tilde{S}^{5}}\ud\alpha\,\rho^{\uD 7}
  \delta\bigl(f_{1}\bigr)\delta\bigl(f_{2}\bigr)\ud f_{1}\wedge \ud f_{2},
\end{equation}
where we are again integrating over the space
$\tilde{S}^{5}\times\left[0,2\pi\right)$ and the $\ud f_{i}$, which
are $1$-forms in the physical space $C^{3}$, are not integrated over.
Due to the isometries we can evaluate the integral at particular point
in $S^{5}$ (for example, $\ga=\pi$ is a particularly simple choice)
and we find a distribution consistent with the ansatz for $F_{1}$
with~\cite{Bigazzi:2009bk}
\begin{equation}
  Q=\begin{cases}
    0 & \rho<\rho_{\nu} \\
    \frac{N_{f}}{2\pi}\left(1-\ue^{2\left(\rho_{\nu}-\rho\right)}\right)^{2}
    & \rho\ge \rho_{\nu}
  \end{cases}.
\end{equation}
We then take the limit in which $\rho_{\nu}\to\-\infty$, corresponding
to the massless limit of the quarks in the dual theory.

The $\uD 7$s backreact on the metric and with this smearing procedure
the most general metric consistent with the isometries that we can
write is
\begin{equation}
  \label{eq:squashed_flat_space}
  \ud s_{6}^{2}=\alpha'\bigl\{\ue^{2f}
  \bigl[\ud\rho^{2}+\bigl(\ud\psi+\cA\bigr)^{2}\bigr]
  +\ue^{2g}\ud s_{CP^{2}}^{2}\bigr\}.
\end{equation}
The metric functions and other fields can be solved by an analysis of
the Killing spinor equations.  The result is~\cite{Benini:2006hh}
\begin{subequations}
\begin{align}
  \ue^{-\phi}=&\frac{1}{g_{\us}}\biggl\{1-\frac{g_{\us}N_{f}}{2\pi}
  \bigl(\rho-\rho_{\us}\bigr)\biggr\},\\
  \ue^{2g}=&c_{1}\ue^{2\rho}\biggl\{1-\frac{g_{\us}N_{f}}{2\pi}
  \biggl[\bigl(\rho-\rho_{\us}\bigr)-\frac{1}{6}+c_{2}\ue^{-6\rho}
  \biggr]\biggr\}^{1/3},
  \\
  \ue^{2f}=&c_{1}\ue^{2\rho}
  \frac{1-\frac{g_{\us}N_{f}}{2\pi}\left(\rho-\rho_{\us}\right)}
  {\left\{1-\frac{g_{\us}N_{f}}{2\pi}
      \left[\left(\rho-\rho_{\us}\right)-\frac{1}{6}+c_{2}\ue^{-6\rho}\right]
    \right\}^{2/3}},\\
  \ue^{-4A}=&-16\pi g_{\us}N\int \ud \rho\,\ue^{-4g\left(\rho\right)}+c_{3}.
\end{align}
\end{subequations}
The solution becomes singular as we take $\rho\to-\infty$, but the
divergence can be made more mild by setting $c_{2}=0$; this can be
thought of as the result of imposing IR singularity in the massive
case and then taking the limit in which the mass
vanishes~\cite{Bigazzi:2008zt}.  The solution exhibits a Landau pole
at $\rho=\rho_{\mathrm{LP}}=\rho_{\us}+\frac{2\pi}{g_{\us}N_{f}}$
where $\ue^{-\phi}\to 0$ and hence the coupling in the dual gauge
theory diverges.  We choose $c_{3}=0$ so that, as
in~\cite{Bigazzi:2008zt}, $\ue^{-4A}=0$ at the Landau pole.  Finally,
$c_{1}$ rescales the warp factor and so can be set to $1$ by rescaling
the Minkowski directions\footnote{These conditions are a little
  arbitrary.  In principle, they would be fixed after the geometry is
  completed in the UV to resolve the Landau pole.  In any case, we
  will mostly be interested in physics at much lower scales where the
  corrections from a more precise set of boundary conditions will be
  exponentially small.}.  With these choices, the solution describing
the backreaction of smeared $\uD 7$ branes in $AdS^{5}\times S^{5}$ is
\begin{subequations}
\label{eq:flat_flavored_space_solution}
\begin{align}
  \ue^{-\phi}=&\frac{1}{g_{\us}}\biggl\{1-\frac{g_{\us}N_{f}}{2\pi}
  \bigl(\rho-\rho_{\us}\bigr)\biggr\},\\
  \ue^{2g}=&\ue^{2\rho}\biggl\{1-\frac{g_{\us}N_{f}}{2\pi}
  \biggl[\bigl(\rho-\rho_{\us}\bigr)-\frac{1}{6}\biggr]\biggr\}^{1/3},
  \\
  \ue^{2f}=&\ue^{2\rho}
  \frac{1-\frac{g_{\us}N_{f}}{2\pi}\left(\rho-\rho_{\us}\right)}
  {\left\{1-\frac{g_{\us}N_{f}}{2\pi}
      \left[\left(\rho-\rho_{\us}\right)-\frac{1}{6}\right]
    \right\}^{2/3}},\\
  \ue^{-4A}=&-16\pi g_{\us}N\int_{\rho_{\mathrm{LP}}}^{\rho}
  \ud x\,\ue^{-4g\left(x\right)}.
\end{align}
\end{subequations}

Because the above configuration is supersymmetric, we are guaranteed
that the supergravity and $\uD 7$-brane equations of motion are
satisfied.  However, when supersymmetry is broken the Killing spinor
analysis that lead to the above solution does not apply and the
equations of motion must be explicitly considered.  We begin by
recasting the equations of motion~\eqref{eq:IIB_eom}.  The charge
distribution $\Omega$ is generically not decomposable but we can
express it as a sum of decomposable pieces.  In particular, since $Q$
is a constant we have, for our smeared approximation,
\begin{equation}
  \Omega=2Q\cJ,
\end{equation}
where $\cJ$ is the K\"ahler form on $CP^{2}$.  Still
following~\cite{Benini:2006hh,Bigazzi:2009bk} we define a local frame
$\bigl\{e^{\ul{\rho}},e^{\ul{0}},e^{\ul{a}}\bigr\}$ via the vielbein
\begin{equation}
  e_{\rho}^{\phantom{\rho}\ul{\rho}}=\alpha'^{1/2}\ue^{f},\quad
  e_{\psi}^{\phantom{\psi}\ul{0}}=\alpha'^{1/2}\ue^{f},\quad
  e_{a}^{\phantom{a}\ul{0}}=\alpha'^{1/2}\ue^{f}\cA_{a},\quad
  e_{a}^{\phantom{a}\ul{b}}=\alpha'^{1/2}\ue^{g}
  \tilde{e}_{a}^{\phantom{a}\ul{b}},
\end{equation}
in which we have written the coordinates on $CP^{2}$ as $\ud x^{a}$
and have chosen a frame $\tilde{e}^{\ul{a}}$ such that
\begin{equation}
  \cJ=\tilde{e}^{1}\wedge\tilde{e}^{2}+\tilde{e}^{3}\wedge\tilde{e}^{4}.
\end{equation}
Hence we can write
\begin{equation}
  \Omega=\Omega_{1}+\Omega_{2},\quad\Omega_{1}=
  2Q\,\tilde{e}^{1}\wedge\tilde{e}^{2},\quad
  \Omega_{2}=2Q\,\tilde{e}^{3}\wedge\tilde{e}^{4}.
\end{equation}

In this local frame, we find that the components of the Ricci tensor
built from $g_{mn}$ are
\begin{subequations}
\begin{align}
  R_{\ul{\rho}\ul{\rho}}=&\frac{1}{\alpha'}\ue^{-2f}
  \bigl\{-\partial_{\rho}^{2}f-4\partial_{\rho}^{2}g
  -4\bigl(\partial_{\rho}g\bigr)^{2}
  +4\partial_{\rho}f\partial_{\rho}g\bigr\},\\
  R_{\ul{0}\ul{0}}=&\frac{1}{\alpha'}\ue^{-2f}\bigl\{
  -\partial_{\rho}^{2}f-4\partial_{\rho}f\partial_{\rho}g
  +4\ue^{4\left(f-g\right)}\bigr\},\\
  R_{\ul{a}\ul{b}}=&\frac{1}{\alpha'}\ue^{-2f}
  \bigl\{-\partial_{\rho}^{2}g-4\bigl(\partial_{\rho}g\bigr)^{2}
  -2\ue^{4\left(f-g\right)}+6\ue^{2\left(f-g\right)}\bigr\}
  \delta_{\ul{a}\ul{b}}.
\end{align}
\end{subequations}
Then, using the above form for $\Omega_{i}$, the equations of
motion~\eqref{eq:IIB_eom} take the form (where again
$\Theta_{+}=\Phi_{+}^{-1}$)
\begin{subequations}
\begin{align}
  0=&\partial_{\rho}^{2}\Phi_{-}+
  4\partial_{\rho}g\partial_{\rho}\Phi_{-}
  -\frac{2\Theta_{+}}{1+\Phi_{-}\Theta_{+}}
  \partial_{\rho}\Phi_{-}\partial_{\rho}\Phi_{-},\\
  0=&{\ue^{-2f}}
  \bigl\{\partial_{\rho}^{2}\ue^{-\phi}
  +4\partial_{\rho}g\partial_{\rho}\ue^{-\phi}\bigr\}
  +{\ue^{\phi-2f}}\bigl\{Q^{2}-
  \bigl(\partial_{\rho}\ue^{-\phi}\bigr)^{2}\bigr\}
  +{4Q}\ue^{-2g},\\
  0=&-\partial_{\rho}^{2}f-4\partial_{\rho}^{2}g
  -4\bigl(\partial_{\rho}g\bigr)^{2}+4\partial_{\rho}f\,\partial_{\rho}g
  +\frac{2}{\left(1+\Theta_{+}\Phi_{-}\right)^{2}}
  \partial_{\rho}\Theta_{+}\partial_{\rho}\Phi_{-}
  -\frac{1}{2}\ue^{2\phi}\bigl(\partial_{\rho}\ue^{-\phi}\bigr)^{2},\\
\intertext{\newpage}
  0=&-\partial_{\rho}^{2}f-4\partial_{\rho}f\partial_{\rho}g+
  4\ue^{4\left(f-g\right)}-\frac{1}{2}\ue^{2\phi}Q^{2},\\
  0=&-\partial_{\rho}^{2}g
  -4\bigl(\partial_{\rho}g\bigr)^{2}
  -2\ue^{4\left(f-g\right)}
  +6\ue^{2\left(f-g\right)}
  -Q\ue^{\phi}\ue^{2\left(f-g\right)},\\
  0=&\partial_{\rho}^{2}\Theta_{+}+
  4\partial_{\rho}g\partial_{\rho}\Theta_{+}
  -\frac{2\Phi_{-}}{1+\Phi_{-}\Theta_{+}}
  \partial_{\rho}\Theta_{+}\partial_{\rho}\Theta_{+},
\end{align}
\end{subequations}
where we have suppressed the appearance of explicit $\uD 3$-branes.
It is easy to confirm that~\eqref{eq:flat_flavored_space_solution}
satisfies these equations.

We now perturb the geometry by the addition of $p$ $\uD
3$-$\overline{\uD 3}$ pairs.  The equations are difficult to solve,
even treating $\frac{p}{N}$ as a perturbation.  To make progress, we
will consider the case in which both the backreaction of $\uD 7$s and
the backreaction of the $\overline{\uD 3}$s can be treated as
comparable perturbations.

First, largely following~\cite{Bigazzi:2009bk}, we will argue that a
$\uD 7$ satisfying $z^{3}=0$ will still be a solution to the $\uD 7$
equations of motion.  To do so, we consider a brane filling the
$\rho$, $\theta$, $\vp$, and $\eta$ directions (where we
use~\eqref{eq:C3_coords} and~\eqref{eq:flat_radial_coord_2}) and take
$\ga$ and $\psi$ to be functions of $\rho$.  The $z^{3}=0$ embedding
is $\ga=\psi=0$.  From our analysis in
section~\ref{subsec:D7_probe_in_flat_space} we expect the Lagrangian
describing the $\uD 7$ fluctuations to be a function of
$\bigl(\partial_{\rho}\psi\bigr)^{2}$,
$\bigl(\partial_{\rho}\eta\bigr)^{2}$, the worldvolume angles, and, in
general, $\psi$ and $\ga$ (as well as the worldvolume gauge fields,
but those can again be consistently set equal to zero).  To show that
$\psi=\eta=0$ is a solution to the equations of motion, it then
suffices to show that the $\frac{\delta
  \cL}{\delta\psi}=\frac{\delta\cL}{\delta\ga}=0$ at these points.
The $\uD 7$ action again has two contributions, the DBI and CS piece.
Setting $\psi$ and $\ga$ to be constant, the induced metric on the
$\uD 7$ is
\begin{align}
  \ud s_{4}^{2}=\alpha'\biggl\{&\ue^{2f}\bigl[\ud\rho^{2}+
  \frac{1}{4}\cos^{4}\frac{\ga}{2}\bigl(\ud\eta+\cos\theta\ud\vp\bigr)^{2}
  \bigr]\notag\\
  &+\ue^{2g}\bigl[\frac{1}{4}\cos^{2}\frac{\ga}{2}
  \bigl(\ud\theta^{2}+\sin^{2}\theta\ud\vp^{2}\bigr)
  +\frac{1}{4}\cos^{2}\frac{\ga}{2}\sin^{2}\frac{\ga}{2}
  \bigl(\ud\eta+\cos\theta\ud\vp\bigr)^{2}\bigr]\biggr\}.
\end{align}
All of the components are independent of $\psi$ and have vanishing
derivatives with respect to $\ga$ at $\ga=0$.  Furthermore, from the
isometries the dilaton will depend only on $\rho$. Hence we have that
$\frac{\delta\cL^{\mathrm{DBI}}}{\delta\psi}=\frac{\delta\cL^{\mathrm{DBI}}}{\delta\ga}=0$
at $\psi=\ga=0$.

Next we consider the CS contribution to the $\uD 7$ action. If
$z^{3}=0$ is a solution even after the $\uD 7$s and $\overline{\uD
  3}$s backreact, then the smearing of such branes will produce an
$F_{1}$ given by
\begin{equation}
  F_{1}=-\frac{N_{f}}{2\pi}\bigl(\ud\psi+\cA\bigr).
\end{equation}
The field strength of the magnetic dual is
\begin{equation}
  F_{9}\sim \frac{N_{f}}{2\pi}\ue^{2\phi+4g}\ud\vol_{R^{3,1}}
  \wedge\ud\rho\wedge\cJ\wedge\cJ.
\end{equation}
Since $\cJ=\frac{1}{2}\ud\cA$ we can write
\begin{equation}
  C_{8}\sim \frac{N_{f}}{2\pi}\ue^{2\phi+4g}\ud\vol_{R^{3,1}}
  \wedge\ud\rho\wedge\cJ\wedge\cA.
\end{equation}
The CS action is proportional $\int \mathrm{P}\bigl[C_{8}\bigr]$ and
so the corresponding Lagrangian behaves as
\begin{equation}
  \cL^{\mathrm{CS}}\sim Q\ue^{2\phi+4g}\sin\theta\cos^{4}\frac{\ga}{2},
\end{equation}
and hence
$\frac{\delta\cL^{\mathrm{CS}}}{\delta\psi}=\frac{\delta\cL^{\mathrm{CS}}}{\delta\ga}=0$
at $\ga=\psi=0$.  From this, we conclude that $\ga=\psi=0$ is a
solution to the $\uD 7$ equations of motion even once the backreaction
of such branes and $\overline{\uD 3}$s is taken into account.  From
the isometries of the problem, we immediately have that all embeddings
of the type $a_{i}z^{i}=0$ solve the $\uD 7$ equations of motion.

Treating the $\overline{\uD 3}$s and $\uD 7$s as comparable
perturbations, the equations of
motion~\eqref{eq:D7_backreaction_in_flat_space} can be simply solved
by the method of undetermined coefficients.  The result is
\begin{subequations}
\label{eq:flat_flavored_nonsusy_solution}
\begin{align}
  \Phi_{-}=&-8\pi g_{\us}N\frac{p}{N}\ue^{-4\rho}
  -\frac{16}{3}\pi g_{\us}N\frac{g_{\us}N_{f}}{2\pi}\frac{p}{N}
  \bigl(\rho-\rho_{\us}+\frac{1}{12}\bigr)\ue^{-4\rho}\notag\\
  &+\frac{128}{5}\bigl(\pi g_{\us}N\bigr)^{3}\frac{p^{2}}{N^{2}}\ue^{-12\rho}
  +\Phi_{-}^{\mathrm{LP}},\\
  \ue^{-\phi}=&\frac{1}{g_{\us}}
  \biggl\{1-\frac{g_{\us}N_{f}}{2\pi}\bigl(\rho-\rho_{\us}\bigr)
  +\frac{8}{5}\bigl(\pi g_{\us}N\bigr)^{2}
  \frac{g_{\us}N_{f}}{2\pi}\frac{p}{N}\ue^{-8\rho}\biggr\},\\
  \ue^{2g}=&\ue^{2\rho}\biggl\{
  1-\frac{1}{3}\frac{g_{\us}N_{f}}{2\pi}
  \biggl(\rho-\rho_{\us}-\frac{1}{6}\biggr)
  -\frac{16}{5}\bigl(\pi g_{\us}N\bigr)^{2}
  \frac{p}{N}\ue^{-8\rho}\notag\\
  &\phantom{\ue^{2\rho}\biggl\{}
  -\frac{1}{9}\left(\frac{g_{\us}N_{f}}{2\pi}\right)^{2}
  \biggl[\bigl(\rho-\rho_{\us}\bigr)^{2}
  -\frac{1}{3}\bigl(\rho-\rho_{\us}\bigr)
  +\frac{1}{36}\biggr]\notag\\
  &\phantom{\ue^{2\rho}\biggl\{}
  -\frac{16}{5}
  \bigl(\pi g_{\us}N\bigr)^{2}\frac{g_{\us}N_{f}}{2\pi}
  \frac{p}{N}\biggl(\rho-\rho_{\us}+\frac{1}{6}\biggr)\ue^{-8\rho}
  -\frac{128}{25}\bigl(\pi g_{\us}N\bigr)^{4}\left(\frac{p}{N}\right)^{2}
  \ue^{-16\rho}\biggr\},\\
  \ue^{2f}=&\ue^{2\rho}\biggl\{
  1-\frac{1}{3}\frac{g_{\us}N_{f}}{2\pi}
  \biggl(\rho-\rho_{\us}+\frac{1}{3}\biggr)
  -\frac{16}{5}\bigl(\pi g_{\us}N\bigr)^{2}
  \frac{p}{N}\ue^{-8\rho}\notag\\
  &\phantom{\ue^{2\rho}\biggl\{}
  -\frac{1}{9}\left(\frac{g_{\us}N_{f}}{2\pi}\right)^{2}
  \biggl[\bigl(\rho-\rho_{\us}\bigr)^{2}
  +\frac{2}{3}\bigl(\rho-\rho_{\us}\bigr)
  -\frac{5}{36}\biggr]\notag\\
  &\phantom{\ue^{2\rho}\biggl\{}
  -\frac{16}{5}
  \bigl(\pi g_{\us}N\bigr)^{2}\frac{g_{\us}N_{f}}{2\pi}
  \frac{p}{N}\bigl(\rho-\rho_{\us}\bigr)\ue^{-8\rho}
  -\frac{128}{25}\bigl(\pi g_{\us}N\bigr)^{4}\left(\frac{p}{N}\right)^{2}
  \ue^{-16\rho}\biggr\},\\   
  \Phi_{+}^{-1}=&2\pi g_{\us}N\ue^{-4\rho}
  +\frac{4\pi}{3}g_{\us}N\frac{g_{\us}N_{f}}{2\pi}
  \biggl(\rho-\rho_{\us}+\frac{1}{12}\biggr)\ue^{-4\rho}
  -\frac{32}{5}\bigl(\pi g_{\us}N\bigr)^{3}\frac{p}{N}\ue^{-12\rho}\notag\\
  &+\frac{10}{9}\pi g_{\us}N\left(\frac{g_{\us}N_{f}}{2\pi}\right)^{2}
  \biggl[\bigl(\rho-\rho_{\us}\bigr)^{2}
  +\frac{1}{6}\bigl(\rho-\rho_{\us}\bigr)
  +\frac{5}{72}\biggr]\ue^{-4\rho}\notag\\
  &-\frac{64}{5}\bigl(\pi g_{\us}N\bigr)^{3}
  \frac{g_{\us}N_{f}}{2\pi}\frac{p}{N}
  \biggl[\rho-\rho_{\us}+\frac{1}{12}\biggr]\ue^{-8\rho}
  +\frac{2048}{125}\bigl(\pi g_{\us} N\bigr)^{5}\left(\frac{p}{N}\right)^{2}
  \ue^{-20\rho}+\Theta_{+}^{\mathrm{LP}}.
\end{align}
\end{subequations}
Integration constants were chosen so that when either $p=0$ or
$N_{f}=0$ the solution matches what was previously found.  This leaves
unfixed a constant in the $\frac{p}{N}\frac{g_{\us}N_{f}}{2\pi}$ term
which we chose by requiring $\Phi_{-}=-4\frac{p}{N}\Phi_{+}^{-1}$ at
$\cO\left(\frac{p}{N}\right)$ (since the $\uD 7$ effects enter the
equations for $\Phi_{-}$ and $\Phi_{+}^{-1}$ in the same way).  The
constants $\Phi_{-}^{\mathrm{LP}}$ and $\Theta_{+}^{\mathrm{LP}}$ are
introduced to impose that $\ue^{-4A}$ and $\omega^{-1}$ vanish at the
Landau pole.  However, since the Landau poles occur where the
perturbation treatment of the $\uD 7$s breaks down, the constants
cannot be reliably found.  However, they should scale as
$\ue^{-4\rho_{\mathrm{LP}}}$ and so will provide a negligible
correction.

\subsection{\label{subsec:backreacting_Kuperstein}Kuperstein branes on
  the conifold}

The backreaction of massive Kuperstein $\uD 7$s in $AdS^{5}\times
T^{1,1}$ was first considered in~\cite{Bigazzi:2008ie}, though the
massless case which to which we will specialize falls under the
general analysis of~\cite{Benini:2006hh}.

For the conifold, we again define a radial coordinate $\rho$
by~\eqref{eq:flat_radial_coord_2}.  The conifold metric is then
\begin{equation}
  \ud s_{6}^{2}=\alpha'\ue^{2\rho}\bigl\{\ud\rho^{2}
  +\bigl(\frac{1}{3}\ud\psi+\cA\bigr)^{2}+\ud s_{S^{2}\times S^{2}}^{2}\bigr\}.
\end{equation}
Our fiducial Kuperstein embedding is
then~\eqref{eq:Kuperstein_embedding}
\begin{equation}
  z^{4}=\left(\frac{2}{27}\right)^{1/4}\alpha'^{3/4}\ue^{3\rho_{\nu}/2}.
\end{equation}
A $\SO{4}\times\U{1}$ rotation of this embedding is
\begin{equation}
  \label{eq:generalized_Kuperstein}
  a_{A}z^{A}=\left(\frac{2}{27}\right)^{1/4}\alpha'^{3/4}\ue^{3\rho_{\nu}/2
    +\ui\alpha},
\end{equation}
in which again $\alpha\in\left[0,2\pi\right)$.  The $a_{A}$ are real
numbers satisfying
\begin{equation}
  a_{A}a_{A}=1,
\end{equation}
and so define an $S^{3}$ which we denote $\tilde{S}^{3}$.  We can
write
\begin{align}
  a_{1}=&\cos\frac{\tilde{\theta}}{2}\cos\frac{\tilde{\ga}+\tilde{\vp}}{2},&
  a_{2}=&\cos\frac{\tilde{\theta}}{2}\sin\frac{\tilde{\ga}+\tilde{\vp}}{2},
  \notag\\
  a_{3}=&\sin\frac{\tilde{\theta}}{2}\cos\frac{\tilde{\ga}-\tilde{\vp}}{2},&
  a_{4}=&\sin\frac{\tilde{\theta}}{2}\sin\frac{\tilde{\ga}-\tilde{\vp}}{2},
\end{align}
in which $\tilde{\theta}\in\left[0,\pi\right]$,
$\tilde{\vp}\in\left[0,2\pi\right)$, and
$\tilde{\ga}\in\left[0,4\pi\right)$.  The smeared charge distribution
\begin{equation}
  \rho_{\uD 7}=\frac{N_{f}}{4\pi^{3}}
\end{equation}
satisfies the property
\begin{equation}
  N_{f}=\int\ud\vol_{\tilde{S}^{3}}\ud\alpha\,\rho_{\uD 7}.
\end{equation}

The generalized embedding~\eqref{eq:generalized_Kuperstein}
corresponds to the vanishing of
\begin{subequations}
\begin{align}
  f_{1}=&\tilde{\psi}-\vp^{1}-\vp^{2}-\tilde{\ga}-\tilde{\vp}
  +2\arg\bigl(\Ga_{1}+\Ga_{2}\bigr)
  -2\alpha+4\pi n,\\
  f_{2}=&\ue^{3\rho}\abs{\Ga_{1}+\Ga_{2}}^{2}-\ue^{3\rho_{\nu}},
\end{align}
\end{subequations}
where $n\in Z$.  The Poincar{\'e} duals of the smeared $\uD 7$s again
follow from
\begin{equation}
  \Omega=\int\ud\vol_{\tilde{S}^{3}}\ud\alpha\,\rho_{\uD 7}\,
  \delta\bigl(f_{1}\bigr)\delta\bigl(f_{2}\bigr)
  \ud f_{1}\wedge f_{2}.
\end{equation}
Following steps as in the flat space case ($\theta^{i}=0$ is a
convenient point to evaluate the integral), we
find~\cite{Bigazzi:2008ie}
\begin{equation}
  Q=\begin{cases}
    0 & \rho<\rho_{\nu},\\
    \frac{3N_{f}}{4\pi}\left(1-\ue^{3\left(\rho_{\nu}-\rho\right)}\right)
    &\rho\ge\rho_{\nu}.
  \end{cases},
\end{equation}
where we have written
\begin{equation}
  F_{1}=-Q\bigl(\rho\bigr)\biggl(\frac{1}{3}\ud\psi+\cA\biggr),
\end{equation}
and used $\ud F_{1}=-\Omega$.

We again focus on the massless limit, $\rho_{\nu}\to-\infty$.  Writing
the perturbed metric as
\begin{equation}
  \ud s_{6}^{2}=\alpha'\bigl\{\ue^{2f}
  \bigl[\ud\rho^{2}+\bigl(\frac{1}{3}\ud\psi+\cA\bigr)^{2}\bigr]
  +\ue^{2g}\ud s_{S^{2}\times S^{2}}^{2}\bigr\},
\end{equation}
we find that the equations of motion are solved
by~\eqref{eq:flat_flavored_space_solution} with the
replacement~\cite{Benini:2006hh}
\begin{equation}
  \frac{g_{\us}N_{f}}{2\pi}\to\frac{3g_{\us}N_{f}}{4\pi}.
\end{equation}
The essentially identical solution is a consequence of the common
conical nature and Einstein-Sasaki base of $C^{3}$ and the
conifold~\cite{Benini:2006hh}.  Similarly, the backreaction of
$\overline{\uD 3}$s is once again given
by~\eqref{eq:flat_flavored_nonsusy_solution} with the above
replacement.

\bibliography{falling}

\providecommand{\href}[2]{#2}\begingroup\raggedright\begin{thebibliography}{10}

\bibitem{Maldacena:1997re}
J.~M. Maldacena, {\it The large {N} limit of superconformal field theories and
  supergravity},  {\em Adv.Theor.Math.Phys.} {\bf 2} (1998) 231--252,
  [\href{http://xxx.lanl.gov/abs/hep-th/9711200}{{\tt hep-th/9711200}}].

\bibitem{Witten:1998qj}
E.~Witten, {\it Anti de {S}itter space and holography},  {\em
  Adv.Theor.Math.Phys.} {\bf 2} (1998) 253--291,
  [\href{http://xxx.lanl.gov/abs/hep-th/9802150}{{\tt hep-th/9802150}}].

\bibitem{Gubser:1998bc}
S.~Gubser, I.~R. Klebanov, and A.~M. Polyakov, {\it Gauge theory correlators
  from noncritical string theory},  {\em Phys.Lett.} {\bf B428} (1998)
  105--114, [\href{http://xxx.lanl.gov/abs/hep-th/9802109}{{\tt
  hep-th/9802109}}].

\bibitem{Aharony:1999ti}
O.~Aharony, S.~S. Gubser, J.~M. Maldacena, H.~Ooguri, and Y.~Oz, {\it Large
  {$N$} field theories, string theory and gravity},  {\em Phys.Rept.} {\bf 323}
  (2000) 183--386, [\href{http://xxx.lanl.gov/abs/hep-th/9905111}{{\tt
  hep-th/9905111}}].

\bibitem{Randall:1999ee}
L.~Randall and R.~Sundrum, {\it A large mass hierarchy from a small extra
  dimension},  {\em Phys.Rev.Lett.} {\bf 83} (1999) 3370--3373,
  [\href{http://xxx.lanl.gov/abs/hep-ph/9905221}{{\tt hep-ph/9905221}}].

\bibitem{Verlinde:1999fy}
H.~L. Verlinde, {\it Holography and compactification},  {\em Nucl.Phys.} {\bf
  B580} (2000) 264--274, [\href{http://xxx.lanl.gov/abs/hep-th/9906182}{{\tt
  hep-th/9906182}}].

\bibitem{Giddings:2001yu}
S.~B. Giddings, S.~Kachru, and J.~Polchinski, {\it Hierarchies from fluxes in
  string compactifications},  {\em Phys.Rev.} {\bf D66} (2002) 106006,
  [\href{http://xxx.lanl.gov/abs/hep-th/0105097}{{\tt hep-th/0105097}}].

\bibitem{Baumann:2010sx}
D.~Baumann, A.~Dymarsky, S.~Kachru, I.~R. Klebanov, and L.~McAllister, {\it
  {D3}-brane potentials from fluxes in {AdS/CFT}},  {\em JHEP} {\bf 1006}
  (2010) 072, [\href{http://xxx.lanl.gov/abs/1001.5028}{{\tt
  arXiv:1001.5028}}].

\bibitem{Gandhi:2011id}
S.~Gandhi, L.~McAllister, and S.~{Sj\"ors}, {\it A toolkit for perturbing flux
  compactifications},  {\em JHEP} {\bf 1112} (2011) 053,
  [\href{http://xxx.lanl.gov/abs/1106.0002}{{\tt arXiv:1106.0002}}].

\bibitem{Grossman:1999ra}
Y.~Grossman and M.~Neubert, {\it Neutrino masses and mixings in
  non-factorizable geometry},  {\em Phys.Lett.} {\bf B474} (2000) 361--371,
  [\href{http://xxx.lanl.gov/abs/hep-ph/9912408}{{\tt hep-ph/9912408}}].

\bibitem{Cremades:2004wa}
D.~Cremades, L.~Ib{\'a\~n}ez, and F.~Marchesano, {\it Computing {Y}ukawa
  couplings from magnetized extra dimensions},  {\em JHEP} {\bf 0405} (2004)
  079, [\href{http://xxx.lanl.gov/abs/hep-th/0404229}{{\tt hep-th/0404229}}].

\bibitem{Acharya:2006mx}
B.~S. Acharya, F.~Benini, and R.~Valandro, {\it Warped models in string
  theory},  \href{http://xxx.lanl.gov/abs/hep-th/0612192}{{\tt
  hep-th/0612192}}.

\bibitem{Karch:2002sh}
A.~Karch and E.~Katz, {\it Adding flavor to {AdS/CFT}},  {\em JHEP} {\bf 0206}
  (2002) 043, [\href{http://xxx.lanl.gov/abs/hep-th/0205236}{{\tt
  hep-th/0205236}}].

\bibitem{Kachru:2009kg}
S.~Kachru, D.~Simi{\'c}, and S.~P. Trivedi, {\it Stable non-supersymmetric
  throats in string theory},  {\em JHEP} {\bf 1005} (2010) 067,
  [\href{http://xxx.lanl.gov/abs/0905.2970}{{\tt arXiv:0905.2970}}].

\bibitem{Dymarsky:2011ve}
A.~Dymarsky and S.~Kuperstein, {\it Non-supersymmetric conifold},  {\em JHEP}
  {\bf 1208} (2012) 033, [\href{http://xxx.lanl.gov/abs/1111.1731}{{\tt
  arXiv:1111.1731}}].

\bibitem{Klebanov:2000hb}
I.~R. Klebanov and M.~J. Strassler, {\it Supergravity and a confining gauge
  theory: {D}uality cascades and {$\chi$SB} resolution of naked singularities},
   {\em JHEP} {\bf 0008} (2000) 052,
  [\href{http://xxx.lanl.gov/abs/hep-th/0007191}{{\tt hep-th/0007191}}].

\bibitem{Kachru:2002gs}
S.~Kachru, J.~Pearson, and H.~L. Verlinde, {\it Brane/flux annihilation and the
  string dual of a nonsupersymmetric field theory},  {\em JHEP} {\bf 0206}
  (2002) 021, [\href{http://xxx.lanl.gov/abs/hep-th/0112197}{{\tt
  hep-th/0112197}}].

\bibitem{DeWolfe:2008zy}
O.~DeWolfe, S.~Kachru, and M.~Mulligan, {\it A gravity dual of metastable
  dynamical supersymmetry breaking},  {\em Phys.Rev.} {\bf D77} (2008) 065011,
  [\href{http://xxx.lanl.gov/abs/0801.1520}{{\tt arXiv:0801.1520}}].

\bibitem{Benini:2009ff}
F.~Benini, A.~Dymarsky, S.~Franco, S.~Kachru, D.~Simic, and H.~Verlinde, {\it
  Holographic gauge mediation},  {\em JHEP} {\bf 0912} (2009) 031,
  [\href{http://xxx.lanl.gov/abs/0903.0619}{{\tt arXiv:0903.0619}}].

\bibitem{McGuirk:2009am}
P.~McGuirk, G.~Shiu, and Y.~Sumitomo, {\it Holographic gauge mediation via
  strongly coupled messengers},  {\em Phys.Rev.} {\bf D81} (2010) 026005,
  [\href{http://xxx.lanl.gov/abs/0911.0019}{{\tt arXiv:0911.0019}}].

\bibitem{Fischler:2011xd}
W.~Fischler and W.~T. Garcia, {\it Hierarchies of susy splittings in
  holographic gauge mediation},  {\em JHEP} {\bf 1106} (2011) 046,
  [\href{http://xxx.lanl.gov/abs/1104.2078}{{\tt arXiv:1104.2078}}].

\bibitem{McGuirk:2011yg}
P.~McGuirk, {\it {Hidden-sector current-current correlators in holographic
  gauge mediation}},  {\em Phys.Rev.} {\bf D85} (2012) 045025,
  [\href{http://xxx.lanl.gov/abs/1110.5075}{{\tt arXiv:1110.5075}}].

\bibitem{Gabella:2007cp}
M.~Gabella, T.~Gherghetta, and J.~Giedt, {\it A gravity dual and {LHC} study of
  single-sector supersymmetry breaking},  {\em Phys.Rev.} {\bf D76} (2007)
  055001, [\href{http://xxx.lanl.gov/abs/0704.3571}{{\tt arXiv:0704.3571}}].

\bibitem{McGarrie:2010yk}
M.~McGarrie and D.~C. Thompson, {\it Warped general gauge mediation},  {\em
  Phys.Rev.} {\bf D82} (2010) 125034,
  [\href{http://xxx.lanl.gov/abs/1009.4696}{{\tt arXiv:1009.4696}}].

\bibitem{Skenderis:2012bs}
K.~Skenderis and M.~Taylor, {\it Holographic realization of gauge mediated
  supersymmetry breaking},  {\em JHEP} {\bf 1209} (2012) 028,
  [\href{http://xxx.lanl.gov/abs/1205.4677}{{\tt arXiv:1205.4677}}].

\bibitem{Argurio:2012cd}
R.~Argurio, M.~Bertolini, L.~Di~Pietro, F.~Porri, and D.~Redigolo, {\it
  Holographic correlators for general gauge mediation},  {\em JHEP} {\bf 1208}
  (2012) 086, [\href{http://xxx.lanl.gov/abs/1205.4709}{{\tt
  arXiv:1205.4709}}].

\bibitem{Argurio:2012bi}
R.~Argurio, M.~Bertolini, L.~Di~Pietro, F.~Porri, and D.~Redigolo, {\it
  Exploring holographic general gauge mediation},  {\em JHEP} {\bf 1210} (2012)
  179, [\href{http://xxx.lanl.gov/abs/1208.3615}{{\tt arXiv:1208.3615}}].

\bibitem{McGarrie:2012fi}
M.~McGarrie, {\it Holography for general gauge mediation},
  \href{http://xxx.lanl.gov/abs/1210.4935}{{\tt arXiv:1210.4935}}.

\bibitem{McGuirk:2009xx}
P.~McGuirk, G.~Shiu, and Y.~Sumitomo, {\it Non-supersymmetric infrared
  perturbations to the warped deformed conifold},  {\em Nucl.Phys.} {\bf B842}
  (2011) 383--413, [\href{http://xxx.lanl.gov/abs/0910.4581}{{\tt
  arXiv:0910.4581}}].

\bibitem{Bena:2009xk}
I.~Bena, M.~Gra{\~n}a, and N.~Halmagyi, {\it On the existence of meta-stable
  vacua in {K}lebanov-{S}trassler},  {\em JHEP} {\bf 1009} (2010) 087,
  [\href{http://xxx.lanl.gov/abs/0912.3519}{{\tt arXiv:0912.3519}}].

\bibitem{Bena:2010ze}
I.~Bena, G.~Giecold, M.~Gra{\~n}a, and N.~Halmagyi, {\it On the inflaton
  potential from antibranes in warped throats},  {\em JHEP} {\bf 1207} (2012)
  140, [\href{http://xxx.lanl.gov/abs/1011.2626}{{\tt arXiv:1011.2626}}].

\bibitem{Bena:2011hz}
I.~Bena, G.~Giecold, M.~Gra{\~n}a, N.~Halmagyi, and S.~Massai, {\it On
  metastable vacua and the warped deformed conifold: {A}nalytic results},
  \href{http://xxx.lanl.gov/abs/1102.2403}{{\tt arXiv:1102.2403}}.

\bibitem{Bena:2011wh}
I.~Bena, G.~Giecold, M.~Gra{\~n}a, N.~Halmagyi, and S.~Massai, {\it The
  backreaction of anti-{D3} branes on the {K}lebanov-{S}trassler geometry},
  \href{http://xxx.lanl.gov/abs/1106.6165}{{\tt arXiv:1106.6165}}.

\bibitem{Massai:2012jn}
S.~Massai, {\it A comment on anti-brane singularities in warped throats},
  \href{http://xxx.lanl.gov/abs/1202.3789}{{\tt arXiv:1202.3789}}.

\bibitem{Bena:2012bk}
I.~Bena, M.~Gra{\~n}a, S.~Kuperstein, and S.~Massai, {\it {$\overline{D3}$}'s -
  {S}ingular to the bitter end},  \href{http://xxx.lanl.gov/abs/1206.6369}{{\tt
  arXiv:1206.6369}}.

\bibitem{Dymarsky:2011pm}
A.~Dymarsky, {\it On gravity dual of a metastable vacuum in
  {K}lebanov-{S}trassler theory},  {\em JHEP} {\bf 1105} (2011) 053,
  [\href{http://xxx.lanl.gov/abs/1102.1734}{{\tt arXiv:1102.1734}}].

\bibitem{Blaback:2011nz}
J.~Bl{\r a}b{\" a}ck, U.~H. Danielsson, D.~Junghans, T.~Van~Riet, T.~Wrase, and
  M.~Zagermann, {\it The problematic backreaction of {SUSY}-breaking branes},
  {\em JHEP} {\bf 1108} (2011) 105,
  [\href{http://xxx.lanl.gov/abs/1105.4879}{{\tt arXiv:1105.4879}}].

\bibitem{Blaback:2011pn}
J.~Bl{\r a}b{\"a}ck, U.~H. Danielsson, D.~Junghans, T.~Van~Riet, T.~Wrase, and
  M.~Zagermann, {\it {(Anti-)Brane} backreaction beyond perturbation theory},
  {\em JHEP} {\bf 1202} (2012) 025,
  [\href{http://xxx.lanl.gov/abs/1111.2605}{{\tt arXiv:1111.2605}}].

\bibitem{Blaback:2012nf}
J.~Bl{\r a}b{\"a}ck, U.~H. Danielsson, and T.~Van~Riet, {\it Resolving
  anti-brane singularities through time-dependence},
  \href{http://xxx.lanl.gov/abs/1202.1132}{{\tt arXiv:1202.1132}}.

\bibitem{Bena:2012tx}
I.~Bena, D.~Junghans, S.~Kuperstein, T.~Van~Riet, T.~Wrase, and M.~Zagermann,
  {\it Persistent anti-brane singularities},  {\em JHEP} {\bf 1210} (2012) 078,
  [\href{http://xxx.lanl.gov/abs/1205.1798}{{\tt arXiv:1205.1798}}].

\bibitem{Kachru:2003aw}
S.~Kachru, R.~Kallosh, A.~D. Linde, and S.~P. Trivedi, {\it De {S}itter vacua
  in string theory},  {\em Phys.Rev.} {\bf D68} (2003) 046005,
  [\href{http://xxx.lanl.gov/abs/hep-th/0301240}{{\tt hep-th/0301240}}].

\bibitem{Kachru:2003sx}
S.~Kachru, R.~Kallosh, A.~D. Linde, J.~M. Maldacena, L.~P. McAllister, and
  S.~Trivedi, {\it Towards inflation in string theory},  {\em JCAP} {\bf 0310}
  (2003) 013, [\href{http://xxx.lanl.gov/abs/hep-th/0308055}{{\tt
  hep-th/0308055}}].

\bibitem{Polchinski:1998rr}
J.~Polchinski, {\em String theory. {V}ol. 2: {S}uperstring theory and beyond}.
\newblock Cambridge University Press, 1998.

\bibitem{Benini:2006hh}
F.~Benini, F.~Canoura, S.~Cremonesi, C.~N{\'u\~n}ez, and A.~V. Ramallo, {\it
  Unquenched flavors in the {K}lebanov-{W}itten model},  {\em JHEP} {\bf 0702}
  (2007) 090, [\href{http://xxx.lanl.gov/abs/hep-th/0612118}{{\tt
  hep-th/0612118}}].

\bibitem{Nunez:2010sf}
C.~N{\'u\~n}ez, {\'A}.~Paredes, and A.~V. Ramallo, {\it Unquenched flavor in
  the gauge/gravity correspondence},  {\em Adv.High Energy Phys.} {\bf 2010}
  (2010) 196714, [\href{http://xxx.lanl.gov/abs/1002.1088}{{\tt
  arXiv:1002.1088}}].

\bibitem{Brax:2000cf}
P.~Brax, G.~Mandal, and Y.~Oz, {\it Supergravity description of {non-BPS}
  branes},  {\em Phys.Rev.} {\bf D63} (2001) 064008,
  [\href{http://xxx.lanl.gov/abs/hep-th/0005242}{{\tt hep-th/0005242}}].

\bibitem{Zhou:1999nm}
B.~Zhou and C.-J. Zhu, {\it The complete black brane solutions in
  {D}-dimensional coupled gravity system},
  \href{http://xxx.lanl.gov/abs/hep-th/9905146}{{\tt hep-th/9905146}}.

\bibitem{Bigazzi:2009bk}
F.~Bigazzi, A.~L. Cotrone, J.~Mas, A.~Paredes, A.~V. Ramallo, and
  J.~Tarr{\'i}o, {\it {D3-D7} quark-gluon plasmas},  {\em JHEP} {\bf 0911}
  (2009) 117, [\href{http://xxx.lanl.gov/abs/0909.2865}{{\tt
  arXiv:0909.2865}}].

\bibitem{Bigazzi:2009gu}
F.~Bigazzi, A.~L. Cotrone, A.~Paredes, and A.~V. Ramallo, {\it Screening
  effects on meson masses from holography},  {\em JHEP} {\bf 0905} (2009) 034,
  [\href{http://xxx.lanl.gov/abs/0903.4747}{{\tt arXiv:0903.4747}}].

\bibitem{Candelas:1989js}
P.~Candelas and X.~C. de~la Ossa, {\it Comments on conifolds},  {\em
  Nucl.Phys.} {\bf B342} (1990) 246--268.

\bibitem{Klebanov:1998hh}
I.~R. Klebanov and E.~Witten, {\it Superconformal field theory on three-branes
  at a {C}alabi-{Y}au singularity},  {\em Nucl.Phys.} {\bf B536} (1998)
  199--218, [\href{http://xxx.lanl.gov/abs/hep-th/9807080}{{\tt
  hep-th/9807080}}].

\bibitem{Kuperstein:2004hy}
S.~Kuperstein, {\it Meson spectroscopy from holomorphic probes on the warped
  deformed conifold},  {\em JHEP} {\bf 0503} (2005) 014,
  [\href{http://xxx.lanl.gov/abs/hep-th/0411097}{{\tt hep-th/0411097}}].

\bibitem{Ouyang:2003df}
P.~Ouyang, {\it Holomorphic {D7} branes and flavored {$\mathcal{N}=1$} gauge
  theories},  {\em Nucl.Phys.} {\bf B699} (2004) 207--225,
  [\href{http://xxx.lanl.gov/abs/hep-th/0311084}{{\tt hep-th/0311084}}].

\bibitem{Benini:2007kg}
F.~Benini, {\it A chiral cascade via backreacting {D7}-branes with flux},  {\em
  JHEP} {\bf 0810} (2008) 051, [\href{http://xxx.lanl.gov/abs/0710.0374}{{\tt
  arXiv:0710.0374}}].

\bibitem{Chen:2008jj}
H.-Y. Chen, P.~Ouyang, and G.~Shiu, {\it On supersymmetric {D7}-branes in the
  warped deformed conifold},  {\em JHEP} {\bf 1001} (2010) 028,
  [\href{http://xxx.lanl.gov/abs/0807.2428}{{\tt arXiv:0807.2428}}].

\bibitem{Klebanov:2000nc}
I.~R. Klebanov and A.~A. Tseytlin, {\it Gravity duals of supersymmetric
  {$\SU{N}\times\SU{N+M}$} gauge theories},  {\em Nucl.Phys.} {\bf B578} (2000)
  123--138, [\href{http://xxx.lanl.gov/abs/hep-th/0002159}{{\tt
  hep-th/0002159}}].

\bibitem{Minasian:1999tt}
R.~Minasian and D.~Tsimpis, {\it On the geometry of non-trivially embedded
  branes},  {\em Nucl.Phys.} {\bf B572} (2000) 499--513,
  [\href{http://xxx.lanl.gov/abs/hep-th/9911042}{{\tt hep-th/9911042}}].

\bibitem{McGuirk:2012sb}
P.~McGuirk, G.~Shiu, and F.~Ye, {\it Soft branes in {supersymmetry-breaking}
  backgrounds},  {\em JHEP} {\bf 1207} (2012) 188,
  [\href{http://xxx.lanl.gov/abs/1206.0754}{{\tt arXiv:1206.0754}}].

\bibitem{Herzog:2001xk}
C.~P. Herzog, I.~R. Klebanov, and P.~Ouyang, {\it Remarks on the warped
  deformed conifold},  \href{http://xxx.lanl.gov/abs/hep-th/0108101}{{\tt
  hep-th/0108101}}.

\bibitem{Camara:2004jj}
P.~G. C{\'a}mara, L.~Ib{\'a\~n}ez, and A.~Uranga, {\it Flux-induced
  {SUSY}-breaking soft terms on {D7-D3} brane systems},  {\em Nucl.Phys.} {\bf
  B708} (2005) 268--316, [\href{http://xxx.lanl.gov/abs/hep-th/0408036}{{\tt
  hep-th/0408036}}].

\bibitem{Lust:2004fi}
D.~L{\"u}st, S.~Reffert, and S.~Stieberger, {\it Flux-induced soft
  supersymmetry breaking in chiral type {IIB} orientifolds with
  {D3/D7}-branes},  {\em Nucl.Phys.} {\bf B706} (2005) 3--52,
  [\href{http://xxx.lanl.gov/abs/hep-th/0406092}{{\tt hep-th/0406092}}].

\bibitem{Lust:2004dn}
D.~L{\"u}st, S.~Reffert, and S.~Stieberger, {\it {MSSM} with soft {SUSY}
  breaking terms from {D7}-branes with fluxes},  {\em Nucl.Phys.} {\bf B727}
  (2005) 264--300, [\href{http://xxx.lanl.gov/abs/hep-th/0410074}{{\tt
  hep-th/0410074}}].

\bibitem{Lust:2008zd}
D.~L{\"u}st, F.~Marchesano, L.~Martucci, and D.~Tsimpis, {\it Generalized
  non-supersymmetric flux vacua},  {\em JHEP} {\bf 0811} (2008) 021,
  [\href{http://xxx.lanl.gov/abs/0807.4540}{{\tt arXiv:0807.4540}}].

\bibitem{Kruczenski:2003be}
M.~Kruczenski, D.~Mateos, R.~C. Myers, and D.~J. Winters, {\it Meson
  spectroscopy in {AdS/CFT} with flavour},  {\em JHEP} {\bf 0307} (2003) 049,
  [\href{http://xxx.lanl.gov/abs/hep-th/0304032}{{\tt hep-th/0304032}}].

\bibitem{Marchesano:2008rg}
F.~Marchesano, P.~McGuirk, and G.~Shiu, {\it Open string wavefunctions in
  warped compactifications},  {\em JHEP} {\bf 0904} (2009) 095,
  [\href{http://xxx.lanl.gov/abs/0812.2247}{{\tt arXiv:0812.2247}}].

\bibitem{Marchesano:2010bs}
F.~Marchesano, P.~McGuirk, and G.~Shiu, {\it Chiral matter wavefunctions in
  warped compactifications},  {\em JHEP} {\bf 1105} (2011) 090,
  [\href{http://xxx.lanl.gov/abs/1012.2759}{{\tt arXiv:1012.2759}}].

\bibitem{Cecotti:2009zf}
S.~Cecotti, M.~C. Cheng, J.~J. Heckman, and C.~Vafa, {\it {Y}ukawa couplings in
  {F}-theory and non-commutative geometry},
  \href{http://xxx.lanl.gov/abs/0910.0477}{{\tt arXiv:0910.0477}}.

\bibitem{Marchesano:2009rz}
F.~Marchesano and L.~Martucci, {\it Non-perturbative effects on seven-brane
  yukawa couplings},  {\em Phys.Rev.Lett.} {\bf 104} (2010) 231601,
  [\href{http://xxx.lanl.gov/abs/0910.5496}{{\tt arXiv:0910.5496}}].

\bibitem{Grana:2001xn}
M.~Gra{\~n}a and J.~Polchinski, {\it Gauge/gravity duals with holomorphic
  dilaton},  {\em Phys.Rev.} {\bf D65} (2002) 126005,
  [\href{http://xxx.lanl.gov/abs/hep-th/0106014}{{\tt hep-th/0106014}}].

\bibitem{Burrington:2004id}
B.~A. Burrington, J.~T. Liu, L.~A. Pando~Zayas, and D.~Vaman, {\it Holographic
  duals of flavored {$\cN=1$} super {Y}ang-{M}ills: {B}eyond the probe
  approximation},  {\em JHEP} {\bf 0502} (2005) 022,
  [\href{http://xxx.lanl.gov/abs/hep-th/0406207}{{\tt hep-th/0406207}}].

\bibitem{Ihl:2012bm}
M.~Ihl, A.~Kundu, and S.~Kundu, {\it Back-reaction of non-supersymmetric
  probes: {P}hase transition and stability},  {\em JHEP} {\bf 1212} (2012) 070,
  [\href{http://xxx.lanl.gov/abs/1208.2663}{{\tt arXiv:1208.2663}}].

\bibitem{Bigazzi:2008zt}
F.~Bigazzi, A.~L. Cotrone, and A.~Paredes, {\it {K}lebanov-{W}itten theory with
  massive dynamical flavors},  {\em JHEP} {\bf 0809} (2008) 048,
  [\href{http://xxx.lanl.gov/abs/0807.0298}{{\tt arXiv:0807.0298}}].

\bibitem{Bigazzi:2008ie}
F.~Bigazzi, A.~L. Cotrone, A.~Paredes, and A.~Ramallo, {\it Non chiral
  dynamical flavors and screening on the conifold},  {\em Fortsch.Phys.} {\bf
  57} (2009) 514--520, [\href{http://xxx.lanl.gov/abs/0810.5220}{{\tt
  arXiv:0810.5220}}].

\end{thebibliography}\endgroup

\end{document}